\documentclass[12pt]{article}

\usepackage{amsmath,amssymb,amsthm}
\usepackage[letterpaper,margin=1in]{geometry}
\usepackage{graphicx}
\usepackage{fourier}
\usepackage[abspath]{currfile} 

\usepackage{tikz}
\usetikzlibrary{decorations.pathreplacing,calligraphy}

\usepackage{xcolor}
\definecolor{lightgray}{rgb}{0.9, 0.9, 0.9}
\tikzstyle{underbrace text style}=[font=\tiny, below, pos=.5, yshift=-3mm]
\usetikzlibrary {patterns,patterns.meta}

\usepackage[]{natbib}
\usepackage{comment} 
\usepackage{setspace}
\usepackage{nicefrac}
\usepackage{enumitem}
\usepackage{refcount}
\usepackage{appendix} 
\usepackage{subcaption}

\newcommand{\cvgd}{\stackrel{d}{\rightarrow}}
\newcommand{\sumin}{\sum_{i=1}^{n}}
\allowdisplaybreaks

\newtheorem{theorem}{Theorem}
\newtheorem{proposition}[theorem]{Proposition}

\newtheorem{assumption}{Assumption}
\newtheorem{lemma}{Lemma}

\usepackage[utf8]{inputenc} 

\title{Asymptotic Representations for Sequential Decisions, Adaptive Experiments, and Batched Bandits\thanks{This research was supported by the U.S.~National Science Foundation under grants  SES-2117260 (Hirano) and SES-2117261 (Porter). Any opinions, findings and conclusions or recommendations expressed in this material are those of the authors and do not necessarily reflect the views of the U.S.~National Science Foundation. Kensuke Sakamoto and Han Xu provided excellent research assistance.}}
\author{Keisuke Hirano\thanks{Department of Economics, Pennsylvania State University. Email: \texttt{kuh237@psu.edu}} 
\and Jack R.~Porter\thanks{Department of Economics, University of Wisconsin-Madison. Email: \texttt{jrporter@ssc.wisc.edu}}}
\date{\today}

\begin{document}

\maketitle

\begin{abstract}
\noindent 
We develop asymptotic approximations that can be applied to sequential estimation and inference problems, adaptive randomized controlled trials, and related settings. In batched adaptive settings where the decision at one stage can affect the observation of variables in later stages, our asymptotic representation characterizes all limit distributions attainable through a joint choice of an adaptive design rule and statistics applied to the adaptively generated data. This facilitates local power analysis of tests, comparison of adaptive treatments rules, and other analyses of batchwise sequential statistical decision rules.
\end{abstract}

\section{Introduction}

Empirical problems that involve dynamic analysis of data, and adaptive choice of sampling design and treatment allocations, are challenging to analyze using standard large-sample characterizations of statistical decision rules and optimality theory. 
In this paper, we develop new asymptotic approximations that can be applied to 
sequential estimation and inference problems, adaptively randomized controlled trials,
and other statistical decision problems that involve multiple decision 
nodes with structured and possibly endogenous information sets.  
Our asymptotic representation theorem  
parsimoniously characterizes all limit distributions that can be attained by 
some choice of a dynamic, data-driven procedure in a multi-stage setting where the researcher 
can take actions after each stage, including possibly adjusting allocations of a multi-valued treatment. 
This representation generalizes classic limit experiment results, leading to an approximation of the sequential statistical decision problem by a limiting data environment that interacts with the choice of adaptive sampling and treatment allocation rules. 
Through this approximation we can characterize 
the asymptotic behavior of statistical procedures, 
perform valid inference, search for optimal policy rules, and obtain best adaptive sampling designs.

The empirical analysis of dynamic policies, for both individual-level policies and aggregate policy rules, is an important and long-standing problem in econometrics and a variety of other fields. 
Recently, there has been renewed interest in sequential statistical design problems, where the researcher can adapt the treatment assignment and data-collection rules in light of prior data. 
Applications of adaptively randomized experiments in economics and related fields include \citet{schw:brad:fade:2017} and \citet{caria:etal:2020}.

Sequential statistical decision problems can be more complex than conventional static statistical design and inference problems, and introduce a number of complications. 
For example, if treatment arms are randomized adaptively, conventional hypothesis tests that are appropriate for simple randomized experiments will not be valid, even in large samples. \citet{zhan:jans:murp:2020} and \citet{hada:etal:2020} have highlighted this problem and proposed alternative tests that are asymptotically correctly sized. 
However, the asymptotic power properties of these tests, and the form of optimal tests in these nonstandard settings, remain an open question. 
In addition to testing, some work including \citet{hahn:hira:karl:2011} and \citet{arms:2022} has  considered efficient estimation of average treatment effects under adaptive treatment assignment; 
however, those papers consider situations in which the conditional treatment rules are deterministic in the limit, whereas a key technical challenge here is to account for random limits. 
More broadly, a general approximation theory for statistical decision rules in such settings would facilitate analysis of a wider range of possible statistical decision problems and procedures. 

We provide such a framework for multi-stage settings where data arrive in stages (or batches) and choices can be made after each stage is realized. 
Stage-wise or batched adaptive experiments are sometimes termed group-sequential randomized experiments in the biostatistics literature, and are called multi-armed batched bandits in 
the bandit algorithm literature.\footnote{For general introductions to  bandits, see \citet{bube:cesa:2012} and \citet{latt:szep:2020}.}

Our results  extend classic asymptotic representation theorems in Le Cam's limits of experiments theory. 
The standard asymptotic representation theorem applies to a static decision setting, where the full sample of data is realized and can be used to make an inference (such as a hypothesis test or a point estimator). 
In those static settings, the representation theorem yields an approximation to the original statistical problem by a simpler one, which in turn provides a useful foundation for asymptotic power analysis of tests, optimal estimation theory, and other statistical decision problems such as the choice of a statistical treatment rule. 
\citet{hira:port:2020} review some applications of this type of asymptotic approximation in economics, but note that there are very few available results for 
sequential decision problems. 
In this paper, we show that the limiting behavior of statistical procedures in sequential and adaptive problems can be characterized to a similar degree as in the static case. 
In particular, a single limiting decision rule represents the asymptotic distributions of a decision rule in the original problem under every local parameter. 
However, our representation is constructed not as a single limit experiment, but rather as a limiting data environment, or limit bandit, that reflects the sequential informational restrictions of the original decision problem, and the interaction between the choice of the adaptive sampling rules and the data generated through the decision process. 
Some recent papers, including \citet{chen:andr:2023} and \citet{adus:2025}, have used our main representation result to study optimal testing under different criteria in batched bandit experiments. 

Our work is closely related to recent, powerful approximation results for multi-armed bandits obtained  
by \citet{wage:xu:2021}, 
\citet{fan:glyn:2021}, \citet{kalv:zeev:2021}, and \citet{adus:2021}. 
Those papers, like ours, use $n^{-1/2}$ local parametrizations in the spirit of 
\citet{hira:port:2009} under which the the optimal treatment arm cannot be perfectly determined in limit. 
\citet{wage:xu:2021} and \citet{fan:glyn:2021} obtain continuous-time approximations 
for a number of specific bandit algorithms such as Thompson sampling, under specific distributional assumptions for the original data, and \citet{kalv:zeev:2021} develop results on the properties of the upper confidence bound algorithm under general arm distributions. 
\citet{adus:2021} builds on these results to solve for an approximately optimal continuous-time bandit algorithm. 
In contrast to these papers, which allow arm allocations to change at every new observation, we consider a batched setting where adjustments can only occur at a fixed, discrete set of times. 
However, in our most general result we allow the underlying parameters to change over time from batch to batch, a feature that \citet{zhan:jans:murp:2020} highlight as being empirically relevant in some adaptive experiments. 
\citet{adus:2021} obtains a very powerful limiting notion of Bayes risk and uses this to obtain asymptotically optimal rules in the classic bandit problem. 
Our representation differs from his in that it covers rules outside the class of Bayes rules, which in stationary environments can be expressed as functions of a current state. 

In the next section, we discuss some examples of sequential statistical decision problems
in economics and other fields, which are difficult to fully analyze except in special cases. These examples motivate the large-sample distributional approximations that we will develop in this paper. 
Section \ref{sec:artfixed} briefly 
reviews the classic limits of experiments theory for non-sequential decision problems, and 
develops an extension of the asymptotic representation theorem for locally asymptotically normal parametric statistical models to sequential settings. 
Here, a finite number of decisions are made at different times based on different subsets of the full sample of data. In this section, the information sets available are exogenous, in the sense that they are not influenced by prior choices made by the decision-maker. 
Section \ref{sec:artrand} contains our most general result. 
It considers the case where choices made by the decision-maker at one stage can affect what is observed in later stages, and other actions can be taken at every stage. 
For this setup, we obtain an asymptotic representation through a limiting Gaussian bandit environment. 
Section \ref{sec:applications} illustrates how our ``limiting bandit'' framework can be used for local asymptotic power analysis for data obtained from batched bandit experiments, and for batched sequential treatment choice problems. 

\section{Motivation}\label{subsec:examples}

We begin by introducing some stylized examples that motivate our approach and illustrate the potential 
scope of our projected results. 
In the first example, we consider a setting with exogenous, sequential arrival of data. 
The second example illustrates the additional complications that can arise when the sampling design is chosen adaptively as data arrives.

{\bf Example 1: Sequential Estimation, Forecasting, and Policy Choice}

Consider a sequence of decisions that must be made on the basis of gradually arriving information. 
For example, one may be interested in estimating a parameter, forecasting some quantity, or choosing among some set of policies, with the opportunity to revise the choice as additional data observations become available. 
Moreover, in empirical analysis of economic time series, it is common to base inference on a restricted 
{\em window} of time periods, to avoid issues that could arise from model instability or structural breaks. 
There is a large literature in econometric time series on estimation, forecasting, and testing 
using rolling and recursive windows, and on window selection. 
Many such problems can be cast as sequential decision problems, with multiple decision points that have corresponding information sets. 
		  
A simple example of such a problem is a sequential estimation problem with adjustment costs. 
Suppose we wish to estimate a scalar parameter $\theta$ on the basis of some observations $i=1,\ldots,n$. 
We can form an initial estimate $S_1$ with observations $1,\dots,n_1 < n$, and then form an updated estimate $S_2$ at the end of the sampling process based on the full $n$ observations. 
Suppose that we measure the performance of the estimates by compound squared error loss, with a penalty for adjustment. Then we can write the loss function as 
\[L(\theta,s_1,s_2) = (s_1-\theta)^2 + (s_2-\theta)^2 + c(|s_1-s_2|),\]
where $c(\cdot)$ is some (typically convex) cost function. 
In conventional single-stage estimation problems, classic results establish the optimality of the maximum likelihood estimator of $\theta$ under standard regularity conditions. 
To investigate optimality for the problem outlined here, an important first step is to 
characterize the possible limiting distributions of estimators $(S_1,S_2)$ in a way that reflects the 
information structure inherent in the setup, while being simple enough to be tractable. 
Our sequential representation result in Section \ref{sec:artfixed} provides such a 
characterization, in a form that facilitates comparisons 
of the limiting distributions of estimators in terms of their asymptotic 
risk.

{\bf Example 2: Inference from Adaptive Randomized Trials}

Adaptive randomized trials (as surveyed, for example, in \citealt{hu:rose:2006}) and related sequential methods in statistics have a long history, 
but they have gained renewed attention recently due to their increasing use in medicine, economics, education, and industry. 
In adaptive experiments, the treatment randomization probabilities or other aspects
of the data collection are modified after collecting the results from initial batches 
or pilot experiments. 
These methods have a number of potential benefits, including improving the precision of
measuring certain effects and improving the overall allocation of treatments to patients. 
The multi-armed bandit model is one framework for designing such adaptive trials that 
has been the focus of recent theoretical and applied work, as discussed in \citet{vill:bowd:waso:2015}.

Classic work on bandit algorithms, such as \citet{lai:robb:1985}, focus on their long-run allocation implications, typically 
studied with tools from large-deviations analysis.  Recent work on bandits 
includes \citet{per:rig:cha:sno:2016}, \citet{kock:thyr:2017}, \citet{kock:prei:veli:2020},
and \citet{kasy:saut:2021}. 
From this perspective, bandit algorithms have considerable appeal.
Batched bandits, which group observations into batches and revise treatment arm probabilities 
batchwise, have seen renewed interest due to their practical 
relevance for clinical trials and other applications. 
However, bandits typically induce a type of sample selection that causes conventional estimators and test statistics to have nonstandard properties.  
A number of recent papers, including 
\citet{vill:bowd:waso:2015}, \citet{bowd:trip:2017},
and \citet{shin:ramd:rina:2019}, point out that sample means and other conventional estimators and associated tests 
do not have their usual distributions when applied to data from multi-armed bandits.

Under the null hypothesis that two (or more) treatments are equally effective, 
standard bandit algorithms do not converge to deterministic allocations.  
As a result, conventional test statistics such as the t-test have non-normal limiting 
distributions, rendering standard inference methods asymptotically invalid. 
To date most work on large-sample theory in these problems has focused 
on large-deviations analysis or pointwise asymptotic theory, and does not address uniform validity or local power properties of inference procedures.

\citet{zhan:jans:murp:2020},
\citet{hada:etal:2020}, and other authors have proposed a number of alternative testing procedures to address this problem. 
Most of these proposed tests reweight observations to counteract the sample selection effect induced by the adaptive algorithm, thus 
restoring asymptotic normality under the null hypothesis. 
While such methods control asymptotic size, their power properties are not fully 
understood. 
Unlike the standard setting where conventional tests can be shown to have certain optimality properties, little is known about the form of optimal tests in this setting. 

In order to study the power properties of tests in this problem, and to develop optimality results, we need to characterize the set of feasible limiting power functions of tests 
under local alternatives. 
In Section \ref{sec:artrand}, we develop a novel limit experiment theory that fills in this theoretical gap. In particular, we derive a local asymptotic power envelope for the problem and compare it to the asymptotic power curve of one of the leading tests. 
We then discuss our proposed work to extend such results to more general adaptive randomized trials and other sequential statistical problems. 

\section{Asymptotic Representation for Sequential Decisions}\label{sec:artfixed}\

A powerful tool in asymptotic statistics is the Asymptotic Representation Theorem, which 
establishes the approximation of statistical problems by simpler ones, often with a Gaussian form. 
Pioneering work on representation theorems and their associated limit experiments include
\cite{leca:1972}, \cite{haje:1970}, and \cite{vdv:1991a}.
These works form the basis for many results about asymptotic efficiency of estimators and 
asymptotic optimality of tests and confidence intervals. 
In this section, we set up some basic notation and review the classic representation result for locally asymptotically normal parametric models. 
Then we give an extension of the classic result to multi-stage sequential statistical decision problems. 

Suppose that the data $z_1,\dots,z_n$ are i.i.d.~with distribution $P_\theta$ and corresponding density $p_{\theta}(z)$ for 
$\theta \in \Theta \subset \mathbb{R}^m$. 
We assume this parametric statistical model is smooth in the following sense. 

\begin{assumption}
\label{asm:dqm}
(a) Differentiability in quadratic mean (DQM):  there exists a function
 $s:\mathcal{Z}\to \mathbb{R}^m$, the score function, such
that
\[
\int \left[ dP_{\theta_0+h}^{1/2}(z) - dP_{\theta_0}^{1/2}(z)
-\frac{1}{2} h^\prime \cdot s(z) dP_{\theta_0}^{1/2}(z) \right]^2 = o(\|h\|^2)
\quad \mbox{as }h\to 0;
\]
\end{assumption}
This is a standard condition used to show, for example, the  asymptotic normality the maximum likelihood estimator. 
More generally, the DQM condition implies that the statistical experiment is locally asymptotically normal (LAN). 
Fix $\theta_0$ in the interior of $\Theta$  and consider local parameter sequences of the form 
\begin{equation}\label{eq:localpar}
\theta_n(h) = \theta_0 + \frac{h}{\sqrt{n}}, \quad h \in \mathbb{R}^m.
\end{equation}
Every local parameter $h$ generates a sequence of parameters approaching $\theta_0$ such that $\theta_n(h)$ is statistically difficult to distinguish from $\theta_0$. 
The classic asymptotic representation theorem shows that this situation can be approximated by a simple shifted normal model.
More precisely, following (\cite{vdv:1998}, Theorem 7.10), suppose that the DQM assumption holds and that the Fisher information matrix $J_0 = E_{\theta_0}[s s^\prime]$ is nonsingular.\footnote{The requirement that the Fisher information is nonsingular can be dropped at the cost of a slightly more complicated asymptotic representation. In our main result below, Theorem \ref{thm:bandit}, we will not impose this condition.} 
Let $S_n$ be {\em any} sequence of real-valued statistics (based on the data $z_1,\dots,z_n$) that possesses limit distributions under every local parameter $h$: 
\[ S_n \stackrel{h}{\leadsto} \mathcal{L}_h,\]
where $\stackrel{h}{\leadsto}$ indicates convergence in distribution (weak convergence) under $\theta_n(h)$.
Then the sequence $S_n$ can be asymptotically represented by a randomized statistic $T(Z,U)$, where $Z \sim N\left(h,J_0^{-1}\right)$ and $U$ is an independent uniform randomizing device, such that 
\[ T(Z,U) \sim \mathcal{L}_h \quad \text{ for every } h.\]
By analyzing the simple shifted normal  ``limit experiment'' where $Z$ is a single multivariate normal draw with unknown mean and known variance matrix, we can derive performance bounds on the asymptotic 
properties of procedures in the original problem, and in some cases we can directly deduce the form of optimal procedures. 

Asymptotic representation theorems have been developed for many other problems, including 
semiparametric models \citep[e.g.][]{vdv:1991a}, 
nonstationary times series \citep[e.g.][]{jega:1995},
and nonregular models with parameter-dependent support \citep[e.g.][]{hira:port:2003}.
They have been used to study inference with weak identification, as in 
\cite{catt:crum:jans:2012}, \cite{hp5}, \cite{andr:arms:2017}, and \cite{andr:miku:2020},
and inference on nonsmooth parameters in partially identified settings, such as 
\cite{hp4}, \cite{fang:sant:2014}, and \cite{fang:2016}.

However, most of the existing results of this type only apply to single-stage statistical decision problems, where the decision-maker chooses the action based on all $n$ observations. 
With the exception of results in \citet[chap.~13]{leca:1986} for a class of optimal stopping problems, limit experiment theory for sequential statistical decision problems is relatively underdeveloped. 
As a first step towards our main result, we extend the classic asymptotic representation theorem to multi-stage sequential decision settings. 

Suppose that the observations $z_1,\dots,z_n$ arrive sequentially, and we have a collection of ``times''  $\lambda_1,\lambda_2,\dots,\lambda_B$, 
where $0 < \lambda_1 < \lambda_2 < \dots < \lambda_B \le 1$. 
Each $\lambda_b$ defines a data window consisting of observations $ i=1,\dots, \lfloor \lambda_b n\rfloor$, the first $\lambda_b$ fraction of the data set. 
At each decision note $b$, the decision-maker can make a decision based on the observations available up to that point in time. 
The following result is a mild extension of the classic asymptotic representation theorem for LAN models that represents the limits of such a collection of sequential decisions by statistics based on a sequence of shifted Gaussian random variables. 

\begin{proposition} \label{thm:artfixed} 
Suppose Assumption \ref{asm:dqm} (DQM) holds with nonsingular Fisher information matrix $J_0$. 
    Let $\{\lambda_1,\lambda_2,\dots,\lambda_B\}$ satisfy $0 < \lambda_1 < \lambda_2 < \cdots < \lambda_B\le 1$. 
    Suppose that the real-valued statistics $S_n(\lambda_b)$ are adapted to $z_1,\dots,z_{\lfloor \lambda_b n\rfloor}$ and suppose that, for every $h \in \mathbb{R}^m$:
	\[ \left( S_n(\lambda_1), \dots,  S_n(\lambda_B)\right)^\prime  \stackrel{h}{\leadsto} \mathcal{L}_h. \] 
	Let
	\[ Z(\lambda) = \lambda h + J_0^{-1/2} W(\lambda), \quad \lambda\in [0,1],\]
	where $W(\lambda)$ is an $m$-dimensional standard Brownian motion, and let $U \sim \text{Unif}[0,1]$, independently of $Z(\cdot)$.
	Then there is a collection of functions $T_1,\ldots,T_B$ such that, for all $h$,
	\[ \begin{pmatrix} T_1(Z(\lambda_1),U) \\ T_2(Z(\lambda_1),Z(\lambda_2),U) \\ \vdots \\ T_B(Z(\lambda_1),\dots, Z(\lambda_B),U) \end{pmatrix}  \sim \mathcal{L}_h.
	\]	
\end{proposition}

All proofs are contained in the Appendix. 

We can relate the proposition to the classic asymptotic representation result for LAN models as follows. 
For any decision time $\lambda_b$, the term $Z(\lambda_b)$ is distributed $N\left(\lambda_b h, J_{0}^{-1} \lambda_b\right)$, which is equivalent to observing the shifted Gaussian $N\left(h, J_{0}^{-1} \lambda_b^{-1}\right)$. 
The latter is the usual LAN limit experiment for a sample of size $\lfloor \lambda_b  n\rfloor$. 
If we only wish to represent a single statistic $S_n(\lambda_b)$, this can be based simply on $Z(\lambda_b)$ and an independent randomization $U$. 
However, to jointly represent $S_n(\lambda_1),\dots,S_n(\lambda_b)$ for any $S_n(\cdot)$ and all $h$, we generally need to include the additional terms $Z(\lambda_1),\dots, Z(\lambda_{b-1})$ as arguments in the limiting statistic $T_b$. 
Alternatively, we can express the asymptotic representations of $S_n(\lambda_b)$ in the form 
$\tilde{T}_b(Z(\lambda_b), U, V_b)$, where $V_b$ is a vector independent of $h$ but dependent on $Z(\lambda)$ for $\lambda < \lambda_B$; we give this construction in Appendix \ref{app:altprop}.

Proposition \ref{thm:artfixed}, like the classic asymptotic representation theorem for LAN models, shows that essentially any procedure in the problem of interest is asymptotically equivalent to a procedure in a certain Gaussian limit 
experiment. 
Here, the limit experiment inherits its variance structure from the information structure in the 
original problem in a natural way. 
This in turn will allow us to maintain the informational restrictions on the procedures $S_n$ 
when developing large-sample performance bounds and optimality results.  
The theorem is not only useful for analyzing estimators and test statistics; it applies more generally to 
any (real-valued) action spaces, such as point forecasts, multi-stage policy choices, or allocation decisions. 

\section{Asymptotic Representations for Batched Adaptive Experiments}\label{sec:artrand}

Proposition \ref{thm:artfixed} in Section \ref{sec:artfixed} cover cases where the information sets are fixed, 
in the sense that 
they do not depend on the unknown parameters, nor on the realization of the data. 
While this can apply to a number of interesting applications, including sequential estimation and forecasting problems, and some multi-stage policy and treatment assignment problems, it does not directly handle settings where 
sampling may be adapted in light of current data, or other situations where there is feedback from initial actions of the decision-maker to later revelation of information. 
Specifically, in adaptive batched randomized experiments, the realized outcomes from earlier batches are used to determine the 
experimental allocations in later batches. 
This makes the information in the later batches random, and endogenous with respect to the overall experimental frame. 

In this section, we develop a new asymptotic representation for adaptive experiments. Our setup covers group-sequential randomized experiments (i.e.,  multi-armed batched bandits), where treatment arm probabilities can be modified in later waves based on realized data from earlier waves. 
The setup can handle settings with more than two arms, and the parameters and even the form of the data may be batch- and arm-specific.
We provide a joint asymptotic representation for a  collection of statistical decision rules applied at various stages of the experiment, {\em and} the choice of the adaptive treatment assignment rules which affect the information gained from later batches. 
In this representation, a limiting Gaussian bandit environment concisely expresses all attainable limit distributions from any choice of statistics and dynamic allocation rules that satisfy certain convergence conditions. 

\subsection{Intuition in a Single-Armed Bandit Setting}

We first give a heuristic overview of our main findings and proof technique in a simplified setting, with only a single treatment arm and two batches or waves. 
We will give a more general formal result in the next subsection. 

There will be two batches of data, $b=1,2$. 
Let the observations from batch $b$ be denoted $z_{b,i}$, where $z_{b,i}$ are i.i.d.~across batches and across individuals with density $p_{\theta}(z)$ and support $\mathcal{Z}$,
where $\theta$ is the parameter. 
In batch 1, we observe $n$ observations: $z_{1,1},\dots,z_{1,n}$.
After observing the batch 1 data, we can choose the size of batch 2 as 
\[ n_2 = \lfloor n \lambda \rfloor, \quad 
\lambda \in [0,\bar{\lambda}],\]  
where $0 < \bar{\lambda} < \infty$. 
Note that $\lambda=1$ implies that the batches are of equal size. 
The decision rule $\Lambda_{n}$ maps the batch 1 data into a choice for $\lambda$: 
\[ \Lambda_{n}: \mathcal{Z}^n \to \left[ 0,\bar{\lambda}\right]. \] 
Then, the data for batch 2 are realized with (random) sample size $\Lambda_n$, and a real-valued statistic
$S_{2,n}$ is calculated using both batches of data. 
The statistic $S_{2,n}$ could be a test statistic, for example, or a point estimator of the parameter $\theta$. 

Our goal is to characterize the attainable limit distributions for the pair $(\Lambda_n,S_{2,n})$. 
This is complicated by the fact that the sample size of batch 2 is random and depends on batch 1. 
One immediate implication is that the statistic $S_{2,n}$ must in fact be a collection of statistics indexed by the batch 2 sample size. We express this through a stochatic process notation: 
\[ \tilde{S}_{2,n}(\lambda): \mathcal{Z}^n \times \mathcal{Z}^{\lfloor n\lambda \rfloor} \to \mathbb{R}, \quad \lambda \in [0,\bar{\lambda}]. \]
Here $\tilde{S}_{2,n}(\lambda)$ indicates the statistical decision rule to be applied if the second-batch sample size is $\lfloor n\lambda \rfloor$. 
This notation suppresses the inputs $z_{1,1},\ldots$ of the function $\tilde{S}_{2,n}(\lambda)(z_{1,1},\ldots)$.
The ``realized'' end-of-sample statistic is the randomly stopped process 
$S_{2,n} = \tilde{S}_{2,n}(\Lambda_n)$. 

Our goal is to characterize the possible limiting distributions of both $\Lambda_n$ and $S_{2,n}$
under similar smoothness conditions as in Section \ref{sec:artfixed}.  
Suppose that the pair of decision rules converges jointly under every value of $h$:
\[ 
  \left( \Lambda_n, S_{2,n} \right) \stackrel{h}{\leadsto} \left( \Lambda, S_2 \right).
\]  
We seek a parsimonious specification of a data environment, and a pair of decision rules adapted to that environment whose distributions match those of $(\Lambda,S_2)$ under every $h$.

Consider the likelihood ratio processes for the data under $\theta_n(h)$ vs.~$\theta_0$. For batch 1, the likelihood ratio process takes the usual form 
\[ \ell_{1,n}(h)=\prod_{i=1}^n \frac{ p_{\theta_n(h)}(z_{1,i})}{p_{\theta_0}(z_{1,i})}.\]
For batch 2, it will again be convenient to introduce a process  notation with respect to $\lambda$: 
\[ \ell_{2,n}(h,\lambda) = \prod_{i=1}^{\lfloor n\lambda \rfloor} \frac{ p_{\theta_n(h)}(z_{2,i})}{p_{\theta_0}(z_{2,i})}.\] 
A standard argument shows that the pair of likelihood ratio processes converges jointly; 
for any alternative $h$ and taking limits under $\theta_0$, we have  
\[ \left\{ \ell_{1,n}(h), \ell_{2,n}(h,\cdot) \right\} 
\stackrel{0}{\leadsto} \left\{ \exp\left( h'\Delta_1 - \frac{1}{2}h'J_0 h\right), \exp\left( h'\Delta_2(\cdot) - \frac{(\cdot)}{2} h'J_0 h\right)   \right\},
\] 
where $\Delta_1 \sim N(0,J_0)$, and $\Delta_2(\cdot)$ is a Gaussian process independent of $\Delta_1$, where $\Delta_2(\cdot)$ has independent increments with $\Delta_2(\lambda) \sim N(0,\lambda J_0)$. In particular, $\Delta_2(\bar{\lambda})$ corresponds to the limiting likelihood ratio for batch 2 in the case where all of the potential data are realized. 
The limits of the likelihood ratios can be shown to match the likelihood ratios that would obtain in the statistical model of observing a single draw for a shifted multivariate normal $Z_1 \sim N(h,J_0^{-1})$,  and a single realization of the shifted Gaussian process $Z_2(\lambda) = \lambda h + J_0^{-1}W(\lambda)$, where $W(\lambda)$ is a $k$-dimensional standard Brownian motion independent of $Z_1$.  

The full likelihood ratio $\ell_{2,n}(h,\bar{\lambda})$ for batch 2, and its limiting counterpart above, corresponds to a latent, maximally informative potential data set. Depending on the 
form of $\Lambda_n$, the tail portion of this process may be discarded, but working with the likelihood ratios in this form allows us to handle any choice for $\Lambda_n$ and apply asymptotic change-of-measures arguments to obtain limiting distributions under alternative local parameters. 

Next we construct representations for the limits of $\Lambda_n$ and $S_{2,n}$. Under $\theta_0$ we have joint weak convergence 
\[ \left(\Lambda_n,S_{2,n}\right)\stackrel{0}{\leadsto} \left(\Lambda^0, S_2^0\right).\]
We expect that the limit distribution of $\Lambda_n$ will be independent of the second batch, so we consider its conditional distribution given only $\Delta_1$. Let $q_{\Lambda^0}(u|\delta_1)$ be the conditional $u$-quantile of $\Lambda^0$ given $\Delta_1 = \delta_1$, where $u\in [0,1]$. Let $U_1 \sim \text{Unif}[0,1]$ independently of $\Delta_1$, and let $Q_{\Lambda^0} = q_{\Lambda^0}(U_1|\Delta_1)$. Then 
\[ ( Q_{\Lambda^0}, \Delta_1) \sim (\Lambda^0,\Delta_1),
\] so we have constructed a statistic that matches $\Lambda^0$. This step mirrors the argument used in Theorem \ref{thm:artfixed} for fixed information sets. 

Matching the limit of $S_{2,n}$ is more delicate, because it is restricted to use only the observed portion of the potential data in batch 2, which in turn depends on the realization of $\Lambda_n$. 
In the limit, the likelihood ratio associated with the realized data in batch 2 corresponds to a randomly stopped portion of the process $\Delta_2(\cdot)$, with stopping time $\Lambda^0$. We represent $S_{2,n}$ by taking the conditional distribution of $S_2^0$ given $\Delta_1$, $\Delta_2(\Lambda^0)$, and $\Lambda^0$. Using $q_{S_2^0}(u|\cdot)$ to denote the quantile function of this conditional distribution, we construct 
\[ Q_{S_2^0}(U_2 | \Delta_1, \Delta_2(Q_{\Lambda^0}),Q_{\Lambda^0}), 
\]
where $U_2$ is another independent standard uniform random variable. Due to the various conditional independencies (for example, of the increments of $\Delta_2(\cdot)$), this can be shown to give a suitable representation in the sense that 
\[ \left(\Delta_1, \Delta_2(\bar{\lambda}), Q_{\Lambda^0},Q_{S_2^0} \right) 
\sim \left(\Delta_1, \Delta_2(\bar{\lambda}), \Lambda^0, S_2^0 \right). \] 
At this point, we have constructed a representation of the limit of $\Lambda_n$ and $S_{2,n}$ with the desired form under $h=0$. 
The representation is constructed jointly with the full potential likelihood ratios for the two batches, which is key for the remainder of the argument. 

The final step is to show that the {\em same} construction matches the limit distributions under alternative local parameter values, when we change $\Delta_1$ and $\Delta_2$ to their shifted counterparts $Z_1$ and $Z_2$ in the constructions of $Q_{\Lambda^0}$ and $Q_{S_2^0}$.
To do this, we can appeal to Le Cam's third lemma \citep[Theorem 6.6]{vdv:1998}, which provides an asymptotic change-of-measures formula to obtain the limit distributions under local alternatives. 
The formula uses the limits of the likelihood ratios, which depend on $\Delta_1$ and $\Delta_2$. 
However, the realized likelihood ratio for batch 2 will depend on the rule $\Lambda_n$, whose 
limit distribution will change if we vary $h$. As a result, it is difficult to directly apply Le Cam's third lemma in this adaptive sampling setting. 
Instead, we use the full or potential likelihood ratio $\Delta_2(\bar{\lambda})$. This 
has a well defined limit under every $h$ that does not depend on the choice of the adaptive rule $\Lambda_n$. Moreover, the statistics $Q_{\Lambda^0}$ and $Q_{S_2^0}$ were constructed relative to the full potential likelihood ratios. This device allows us to apply Le Cam's third lemma, and with some work, verify that our constructions do indeed have the correct limit distributions under every value of the local parameter. 

This sketch of an argument has resulted in the following representations for the decision rules. First, for $\Lambda_n$, its limits 
correspond to a statistic $T_{\Lambda}(Z_1,U_1) = q_{\Lambda^0}(U_1|Z_1)$, where $Z_1 \sim N(h,J_0^{-1})$ and $U_1$ is an independent uniform random variable. 
Next, the limit of $S_{2,n}$, can be represented as a function of $Z_1$, $T_{\Lambda}$, another independent uniform, and the stopped process $W_2(T_{\Lambda})$. This stopped process is equivalent to observing a 
normal with mean $h$ and variance $J_0^{-1}/T_{\Lambda}$. We can therefore represent the limit of $S_{2,n}$ alternatively as a function of $Z_1, Z_2, U_1, U_2$, where 
$Z_2 \sim N(h,J_0^{-1}/T_{\Lambda})$. The dependence of $Z_2$ on the rule $\Lambda$ makes this representation somewhat more complex than the usual form of a limit experiment, such as those we have expressed in Section \ref{sec:artfixed}.
Here, $Z_1$ has a fixed distribution (for every $h$), but the distribution of $Z_2$ will depend on $\Lambda$ which may be random in the limit. 
This interaction between the decision rule and the data is a special case of a bandit environment, so we can view this result as providing a ``limiting bandit'' rather than a conventional limit experiment. 

\subsection{Multi-Armed Batched Bandits}

Building on the intuition above, we now consider multi-armed batched bandit settings
where, at each stage, the decision-maker can choose the arm probabilities for the next stage, and carry out other statistical inferences based on the data realized from all prior batches. We will allow the form and underlying distributions of the data to vary across batches, although an important special case is the standard stationary bandit in which the potential outcome distributions remain the same over time. 

We assume there are $K$ treatment arms denoted $k=0,\dots,K-1$, and $B$ batches $b=1,\dots,B$. Here $B$ and $K$ are both finite and fixed in advance\footnote{In principle the number of periods $B$ could be arbitrarily large, but the asymptotic approximations are based on having an increasing number of potential observations per batch.}, but we can allow the adaptive sampling scheme to eventually rule out certain arms or stop sampling early. 

In every batch $b$ and observation index $i$ there are potential outcomes $z_{b,i}(k)$, which are random variables over $i$ with marginal densities $p_{\theta,k}^{(b)}(\cdot)$  on some support $\mathcal{Z}_{k}^{(b)}$. We allow the distributions, and the supports of the variables, to differ across batches. 
The parameter $\theta \in \Theta \subset \mathbb{R}^m$ is a joint parameter of dimension $m$, characterizing all of the batch/arm distributions, and we fix a centering value $\theta_0 \in \text{int}(\Theta)$. In addition to the usual potential outcomes, we can handle covariates within this setup by defining, for example, $z_{b,i}(k) = (y_{b,i}(k),x_{b,i})$ where $x_{b,i}$ are covariates that are not ``affected'' by the choice of treatment arm. The variables contained in $x_{b,i}$ could also differ across batches. 
We assume that the variables $z_{b,i}(0),\dots,z_{b,i}(K-1)$ are independent across batches $b$ and individuals $i$ while allowing for arbitrary dependence across arms. 
With this setup and at most one observed arm for each observation, 
the sampling distributions of any decision rules, and the realized likelihood functions, only depend on the marginal distributions of the potential outcomes specified by the densities $p_{\theta,k}^{(b)}(\cdot)$.

\begin{figure}[h!]
  \caption{Setup and Notation for Batched Bandits}\label{fig:bandit}
  \framebox{
\begin{tikzpicture}[ultra thick,scale=0.9]
  \tikzset{every node}=[font=\small\sffamily]
  \node at (-2,1.5) {\Large Batch};
  \node at (-2,0) {\Large $1$};

  
  \draw (0, 0) node[above=7pt] {$\lambda_{1,0} = 0$};

  \draw (2, 0) node[above=7pt] {$\lambda_{1,2}$};

  \draw (6, 0) node[above=7pt] {$\lambda_{1,3}$};

  \draw[] (6,0) -- (10,0);
  \draw (10, 0) node[above=7pt] {$1$};
  \draw (10, 0) node[right=1pt] {sample size $n$};

  \foreach  \l/\x/\c/\w in {A/0/{arm 1}/2, B/2/{arm 2}/4, C/6/{arm 3}/4}
  {\draw[fill=lightgray] (\x,0-.4) rectangle (\x+\w,0+.2)node[midway]{\c};}

  \draw [decorate, 
          decoration = {brace,mirror,raise=5pt,amplitude=10pt}] (0,-.5) --  (10,-.5) 
          node[black,midway,yshift=-0.8cm] {};

   \draw[->] (5, -1) -- (5, -1.5);
   \node[draw, red] at (5,-2+0.1) {$\Lambda_{2,n} \quad S_{1,n}$};

   \node at (-2,-4) {\Large $2$};

   \foreach  \l/\x/\c/\w/\p in {A/0/{arm 1}/2/{}, B/2/{arm 2}/3/{}, C/5/{arm 3}/4/{}, D/9/{}/2/{north east lines}}
   {\draw[fill=lightgray, pattern = \p] (\x,-4-.4) rectangle (\x+\w,-4+.2)node[midway]{\c};}

   \draw (0, -4) node[above=7pt] {$\lambda_{2,0} = 0$};
   \draw (2, -4) node[above=7pt] {$\lambda_{2,1}$};
   \draw (5, -4) node[above=7pt] {$\lambda_{2,2}$};
   \draw (9, -4) node[above=7pt] {$\lambda_{2,3}$};
   \draw (11, -4) node[above=7pt] {$\bar{\lambda}_{2}$};

   \draw [red] (4.5, -2.1) -- (4.5, -2.25);
   \draw [->,red] (4.5, -2.25) .. controls (4.5,-2.5) and (2+0.1,-2.5) .. (2+0.1,-3-0.2);
   \draw [->,red] (4.5, -2.25) .. controls (4.5,-2.5) and (5,-2.5) .. (5,-3-.2);
   \draw [->,red] (4.5, -2.25) .. controls (6,-2.5) and (9,-2.5) .. (9,-3-.2);

   \draw [decorate, 
          decoration = {brace,mirror,raise=5pt,amplitude=10pt}] (0,-4-.5) --  (9,-4-.5) 
          node[black,midway,yshift=-0.8cm] {};
   \node[draw, red] at (4.5,-6) {$\Lambda_{3,n} \quad S_{2,n}$};
   \draw[->] (4.5, -5) -- (4.5, -5.5);

   \draw [->, red] (5,-1) .. controls (14,-2) and (15,-5) .. (4.5+1.5,-6);

   \draw [ ->, dashed, blue] (6,-2) -- (10-0.3, -2.05);
\end{tikzpicture}
}
\end{figure}

Batch $b=1$ has sample size $n$. The relative numbers of observations from each arm in batch 1 are determined by $\lambda_1 = (\lambda_{1,0},\dots,\lambda_{1,K})$ where $0 = \lambda_{1,0} \le \lambda_{1,1}\le \cdots \le \lambda_{1,K-1} \le \lambda_{1,K}=1$, with the interpretation that observations $\lfloor n \lambda_{1,k}\rfloor + 1,\dots, \lfloor n \lambda_{1,k+1} \rfloor$ are assigned to arm $k$. The allocation $\lambda_1$ to the 
arms in batch $b=1$ is treated as fixed (nonstochastic).  

For batches $b=2,\dots,B$, let $0<\bar{\lambda}_b < \infty$ be fixed numbers representing the maximal sample size in batch $b$ relative to batch 1. 
For notational convenience, also define $\bar{\lambda}_1 = 1$. 
The number of observations (relative to $n$) from each arm are determined by $\lambda_b = (\lambda_{b,0},\dots,\lambda_{b,K})$, where 
\[ 0 = \lambda_{b,0} \le \lambda_{b,1} \le \cdots \le \lambda_{b,K} \le \bar{\lambda}_b. \] 
The statistical decision rule will include the choices of arm sizes in batches $2,\dots,B$ based on prior realized data. For each $b \in \{2,\dots,B\}$, let 
\[\tilde{\Lambda}_{b,n}\left(\lambda_1,\lambda_2,\dots,\lambda_{b-1}\right) \] 
be a collection of mappings from data (whose domain depends on prior batch/arm sizes $\lambda_1,\dots,\lambda_{b-1}$) to vectors of the form $\lambda_b$. Then the realized batch/arm sizes will be determined recursively through 
\[ \Lambda_{b,n} = \tilde{\Lambda}_{b,n}\left(\lambda_1,\Lambda_{2,n},\dots,\Lambda_{b-1,n}\right). \] 
After each batch is realized, we may also take some other action that depends on data up to that batch. For $b=1,\dots,B$, let 
\[ S_{b,n} = \tilde{S}_{b,n}\left(\lambda_1,\Lambda_{2,n},\dots,\Lambda_{b,n}\right),\] 
where, as before,  $\tilde{S}_{b,n}$ represents a collection of functions for each possible set of prior batch/arm sizes, indexed by $\lambda_1,\dots,\lambda_b$. 
Figure \ref{fig:bandit} summarizes the notation and the relationships between different components of this sequential decision rule. 
The full collection of sequential decision rules is given by 
\[ 
  \left( S_{1,n},\Lambda_{2,n}, S_{2,n},\dots, \Lambda_{B,n},S_{B,n}\right).  
\]  

In the statement of the theorem below, we assume that the underlying parametric models corresponding to the $p_{\theta,k}^{(b)}$ satisfy differentiability in quadratic mean, as in 
Assumption \ref{asm:dqm}(a). However, we do  not require their Fisher information matrices $J_k^{(b)}$ to be invertible. 
We also use the convention that $N(0,0)$ is a point mass at zero. 

\begin{theorem}\label{thm:bandit}  
Suppose that Assumption \ref{asm:dqm}(a) (DQM) holds for each $p_{\theta,k}^{(b)}$ at $\theta_0$ with Fisher information $J_{k}^{(b)}$. 
Suppose that, for every $h \in \mathbb{R}^m$, we have joint weak convergence 
under $\theta_n(h) = \theta_0 + h/\sqrt{n}$: 
\[ 
  \left( S_{1,n},\Lambda_{2,n}, S_{2,n}, \dots, \Lambda_{B,n},S_{B,n} \right) \stackrel{h}{\leadsto} \left( S_{1},\Lambda_{2}, S_{2},\dots, \Lambda_{B}, S_{B} \right).
\]
Then there exist statistics $T_{S_1}, T_{\Lambda_2}, T_{S_2},\dots, T_{\Lambda_B}, T_{S_B}$ and random variables $Z_{b} = \left( Z_{b,0}, \dots, Z_{b,K-1}\right)$ for $b=1,\dots,B$ and $U$, where: 

\begin{enumerate}
  \item $Z_{1,0},\dots,Z_{1,K-1}$ and $U$ are independent with $U \sim \text{Unif}[0,1]$ and 
  \[ Z_{1,k} \sim N\left( (\lambda_{1,k+1}-\lambda_{1,k}) J_k^{(1)} h, (\lambda_{1,k+1}-\lambda_{1,k}) J_k^{(1)} \right). \] 
   The statistics $T_{S_1}(Z_1,U)$ and $T_{\Lambda_2}(Z_1,U)$ are based on the realizations of $Z_1$ and $U$. 
  \item For $b=2,\dots,B$, the variables $Z_b$ and statistics $T_{S_b}$, $T_{\Lambda_{b+1}}$ are generated as follows. 
  The distribution of $Z_b$ conditional on $Z_1,\dots,Z_{b-1}, U$, has conditionally independent components 
  \[ Z_{b,k} \mid Z_1,\dots,Z_{b-1},U  \sim   N\left( (\lambda_{b,k+1}-\lambda_{b,k}) J_k^{(b)}h , (\lambda_{b,k+1}-\lambda_{b,k})J_k^{(b)}\right),
  \] 
  where 
  \[ \left( \lambda_{b,0},\dots,\lambda_{b,K}\right) = T_{\Lambda_b}(Z_1,\dots,Z_{b-1},U).\]
  The statistics $T_{S_b}$ and $T_{\Lambda_{b+1}}$ (the latter defined only for $b=2,\dots,B-1$) are functions of variables up to stage $b$: 
  \[ T_{S_b}\left( Z_1,\dots,Z_b,U\right), \quad T_{\Lambda_{b+1}}\left( Z_1,\dots,Z_b,U\right). \] 
\end{enumerate}
The statistics $T_{S_1}, T_{\Lambda_2},\dots$, in conjunction with this recursive specification for the $Z_1,\dots,Z_B$, and $U$, have the property that
\[ \left(T_{S_1}, T_{\Lambda_2}, T_{S_2},\dots,T_{\Lambda_B}, T_{S_B}\right)
\stackrel{d}{=} \left( S_1, \Lambda_2, S_2, \dots, \Lambda_B, S_B \right),\] 
where the equality in distribution holds for every value of $h$. 
\end{theorem}

\vspace{.75cm}
{\bf Remarks on the Theorem:}

At each stage $b=2,\dots,B$, the conditional distribution of the $Z_{b,k}$ depend on $T_{\Lambda_b}$, which in turn depends on the realizations of the variables in prior batches.
This captures the interacting structure of the original problem, where the rules for adaptively modifying the treatment arm probabilities affects the information gained in later stages. The resulting data will be nonstationary in a manner that is determined by the choice of the allocation rules. 

The representation given in the theorem has the form of a $B$-horizon Gaussian bandit {\em environment}, where the Gaussian observations have a specific shift form reflecting the local parameter $h$ and the asymptotic form of the allocation rule. 
Any decision rules (including the dynamic rules for selecting treatment probabilities) in the original batched problem that satisfies the convergence conditions can be represented by some choice of statistics $T_{S_1},T_{\Lambda_2},\dots$ in this limiting environment. 

The matrices $J_{k}^{(b)}$, which are Fisher information matrices relative to the full parameter vector $\theta$ at the value $\theta_0$, are not required to be invertible in the theorem. 
This is useful for handling cases where some components of $\theta$ correspond to specific treatment arms (so that observations from a different arm may not be informative about that sub-component of $\theta$). 
There may also be some components of $\theta$ that are common across arms, and some components of $\theta$ could be nonidentified even with data on all the arms. 
The theorem also allows arms to be adaptively assigned zero weight 
(leading to $\lambda_{b,k+1}-\lambda_{b,k} =0$), and allows for decision rules that can adaptively stop the experiment early (leading to zero weights for all arms in later batches). 
Handling these different possible settings, with possibly different distributions across both arms and batches, leads to a somewhat complicated notation. 
In the next section, we show how the result can be specialized to a simpler and more intuitive 
form for two-armed batched bandits where each arm has its own vector of parameters. 

\section{Applications}\label{sec:applications}

\subsection{Two-Arm Adaptive Experiments}\label{sec:experiment}

We now illustrate how Theorem \ref{thm:bandit} can be applied to the benchmark case of an 
adaptive randomized experiment with two treatments arms (treatment vs.~control, $K=2$) and two or more 
equally-sized batches.

Suppose that individuals are i.i.d.~(within and across batches), with potential outcomes $y_{b,i}(0)$ and $y_{b,i}(1)$ for individual $i=1,\dots,n$ in batch $b=1,\dots,B$.
Let $D_{b,i}=0,1$ denote the treatment received. 
We observe $D_{b,i}$ and $Y_{b,i} = D_{b,i} y_{b,i}(1) + (1-D_{b,i}) y_{b,i}(0)$. 
In the first batch ($b=1$), we assign an equal number of individuals to each of the two treatment arms. 
In the notation of Section \ref{sec:artrand}, we have $\lambda_{1,1}=0.5$, and 
\[ D_{1,i} = \left\{ \begin{array}{ll} 
  0 &\quad i=1,\dots, \lfloor 0.5 n \rfloor \\
  1 &\quad i=\lfloor 0.5n \rfloor+1,\dots,n. \end{array}\right.
\] 
In each later batch $b\ge 2$, the treatment allocations can be based on the data $\{ D_{c,i}, Y_{c,i}\}_{i=1}^n$ from batches $c\le b-1$. 
The treatment allocation rules $\Lambda_{b,n}$
indicate the fraction of units in batch $b$ that are assigned to treatment arm 0. 
We set $\lambda_{b,1} = \Lambda_{b,n}$ and allocate the first $\lfloor \lambda_{b,1} n\rfloor$ units in batch $b$ to treatment arm 0.

We assume a parametric model for the potential outcomes. 
The marginal distributions of $y_{b,i}(0)$ and $y_{b,i}(1)$ have densities $p_{\beta_0}(y)$ and $p_{\beta_1}(y)$, with parameters $\beta_0, \beta_1 \in \mathcal{B}\subset \mathbb{R}^{d_\beta}$. 
We fix centering values $\beta_{0,0}$ and $\beta_{0,1}$ for the two parameters (where both $\beta_{0,0}$ and $\beta_{0,1}$ are in the interior of $\mathcal{B}$), and impose Assumption \ref{asm:dqm} (DQM) for each of the two potential outcomes distributions. 
Let $\Sigma_{0}$ and $\Sigma_1$ denote the Fisher information matrices for the two potential outcomes models at $\beta_{0,0}$ and $\beta_{0,1}$.\footnote{These information matrices are invertible by Assumption \ref{asm:dqm}(b), which simplifies the notation to follow, but we could drop the invertibility requirement as discussed in Section \ref{sec:artrand}.}
Let $h_0$ and $h_1$ denote arm-specific local parameters, and consider local alternatives $\beta_0 = \beta_{0,0}+h_0/\sqrt{n}$ and $\beta_1 = \beta_{0,1}+h_1/\sqrt{n}$. We can write these more compactly with the joint parameter $\theta = (\beta_0,\beta_1)$ and joint local parameter $h=(h_0,h_1)$ as 
$\theta_n(h) = \theta_0 + h/\sqrt{n} = (\beta_{0,0}, \beta_{0,1}) +  (h_0,h_1)/\sqrt{n}$. 
Suppose that the average treatment effect $\tau_n$ depends on the parameters as follows: 
\begin{align*}
  \tau_n(\theta) 
  &= E_{\theta}\left[ y_{b,i}(1)\right] - E_{\theta}\left[ y_{b,i}(0)\right] \\
  &= g_1(\beta_1) - g_0(\beta_0).
\end{align*}
Assume that $\theta_0$ is such that $\tau_n(\theta_0) = g_1(\beta_{0,1})-g_0(\beta_{0,0}) = 0$, and that $g_1$ and $g_0$ are  differentiable with total derivatives $\dot{g}_1(\cdot)$ and 
$\dot{g}_0(\cdot)$ so that 
\[ \tau_n(\theta_0 + h/\sqrt{n}) \approx \frac{1}{\sqrt{n}}( \dot{g}_1'h_1 - 
 \dot{g}_0'h_0), \]  
where $\dot{g}_1 := \dot{g}_1(\beta_{0,1})$ and $\dot{g}_0 :=\dot{g}_0(\beta_{0,0})$. 

Then, after reducing some of the variables by sufficiency, the limiting bandit environment of Theorem \ref{thm:bandit} can be expressed in the following, more intuitive form. At each stage $b=1,\dots,B$, we observe
\begin{equation} 
\label{eqn:simple_limitexp}
Z_b = \begin{pmatrix}
  Z_{b,0} \\ Z_{b,1}
\end{pmatrix} \mid h, Z_{1},\dots,Z_{b-1} \sim N\left( \begin{pmatrix}
  h_0 \\ h_1 
\end{pmatrix}, \begin{pmatrix}
  \frac{1}{\lambda_{b,1}} \Sigma_0^{-1} & 0 \\ 
  0 & \frac{1}{1-\lambda_{b,1}} \Sigma_1^{-1}
\end{pmatrix} \right),
\end{equation}
where $\lambda_{1,1}=0.5$ and, for $b=2,\dots,B$, we set $\lambda_{b,1} = \Lambda_{b}(Z_1,\dots,Z_{b-1},U)$, where $\Lambda_b(\cdot)$ is the limiting representation of $\Lambda_{b,n}$.

The variables $Z_{b,0}$ and $Z_{b,1}$ are shifted Gaussians, with means $h_0$ and $h_1$, and variance scaled by the relative sampling proportions (propensity scores) of the two arms in that batch. 
Thus, the asymptotic representation reduces the problem to a $B$-stage bandit with a pair of Gaussian observations at each stage. 
The choice of the adaptive treatment assignment rules $\Lambda_{b,n}\leadsto \Lambda_b$ then induces a dynamic conditional variance structure on the two observations in the second and later stages. 

Importantly, the limiting treatment allocation rules $\Lambda_b$ may be random. 
For example, consider the widely used Thompson sampling algorithm. 
In the setup considered here, the Thompson sampling rule for batch 2 can be expressed as follows. 
Let $\pi(\beta_0,\beta_1)$ be a prior distribution for the model parameters, and let 
the corresponding posterior after batch 1 be 
\[ \pi \left(\beta_0,\beta_1 | \left\{ D_{1,i},Y_{1,i}\right\}_{i=1}^n \right) \quad \propto \quad  
\pi(\beta_0,\beta_1) \cdot \prod_{i=1}^n p_{\beta_1}(Y_{1,i})^{D_{1,i}} \cdot p_{\beta_0}(Y_{1,i})^{(1-D_{1,i})}. \] 
We assign units in batch 2 to treatment arms based on the posterior probability that arm 0 is better than arm 1: 
\begin{align}
  \Lambda_{2,n} &= \text{Pr}\left( g(\beta_0)>g(\beta_1) | \left\{ D_{1,i},Y_{1,i}\right\}_{i=1}^n \right) 
 \nonumber  \\
  &= \iint {\bf 1}\left( g(\beta_0)>g(\beta_1)\right)  \pi \left(\beta_0,\beta_1 | \left\{ D_{1,i},Y_{1,i}\right\}_{i=1}^n \right) d\beta_0 d\beta_1. 
  \label{eqn:thom} 
\end{align}

Heuristically, the posterior $\pi \left(\beta_0,\beta_1 | \left\{ D_{1,i},Y_{1,i}\right\}_{i=1}^n \right)$ will be approximately normal, centered at the realizations of the Batch 1 limit experiment $Z_1=(Z_{1,0},Z_{1,1})$, and with variance equal to the Fisher information for Batch 1. 
This suggests that the Thompson sampling rule will converge to the probability that treatment 0 has higher mean than treatment 1 under this normal posterior. 

Proposition \ref{prop:bvm} in Appendix \ref{sec:TSlimit} formalizes this intuition, and shows that under relatively mild conditions, the Thompson sampling rule has the following limiting distribution under every local parameter $h$:

  \[ \Lambda_{2,n} \leadsto \Lambda_{2}(Z_1) = \text{Pr}\left( \dot{g}_1'h_1 < \dot{g}_0'h_0 \mid Z_1 \right) = 
  \Phi\left( \frac{\dot{g}_0'Z_{1,0}-\dot{g}_1'Z_{1,1} }{\sqrt{2 \dot{g}_0'\Sigma_0^{-1}\dot{g}_0+ 2 \dot{g}_1'\Sigma_1^{-1}\dot{g}_1}}\right).\]
 Then, by Theorem \ref{thm:bandit}, the second stage limiting bandit observation is 
  \[ Z_2 \mid h, Z_1 \sim N\left( \begin{pmatrix}
    h_0 \\ h_1 
  \end{pmatrix}, \begin{pmatrix}
    \frac{1}{\Lambda_{2}(Z_1)} \Sigma_0^{-1} & 0 \\ 
    0 & \frac{1}{1-\Lambda_{2}(Z_1)} \Sigma_1^{-1}
  \end{pmatrix} \right).
  \]

Pointwise limits such as these are sufficient for our purposes in the following subsections, which focus on obtaining local asymptotic power envelopes and risk bounds.
In other applications, stronger results may be needed to ensure that a given combination of adaptive sampling rule and other statistics satisfies the desired limiting behavior. 
\citet{chen:andr:2023} show that a slightly modified version of Thompson sampling converges uniformly to its limiting Gaussian version, which they use to obtain results on uniform asymptotic coverage of confidence interval procedures. 

  For some of the numerical work below, we also consider settings with more than two batches. 
In those cases, we will assume that the treatment allocation rules $\Lambda_{b,n}$ have limits corresponding to Bayesian posterior means in the Gaussian model with known variances, updated to reflect the information in $Z_1,\dots,Z_{b-1}$.

Given a specific adaptive treatment rule, such as the Thompson sampling rule, 
Theorem \ref{thm:bandit} yields the conclusion that any statistical procedure that uses the data generated by the two-stage experiment, if it has limits, must be asymptotically matched by some rule of the form $T_2(Z_1,Z_2,U)$ where the $(Z_1,Z_2)$ have the normal distributions above, for every local parameter value. 
However, the theorem also allows us to characterize the joint choice of adaptive treatment rule and any statistical procedures based on the realized data: any adaptive treatment rule with limits must be matched by some rule of the form $\Lambda_2(Z_1,U)$ which in turn generates the $Z_2$ used by the limiting statistic $T_2(Z_1,Z_2,U)$.

\subsection{Asymptotic Power Envelopes for Batched Thompson Sampling}\label{subsec:powerenv}

Using the asymptotic representation, we first study the power of hypothesis tests in adaptive experimental designs.  
To keep the notation and analysis simple, we retain the simple two-arm setup of Section \ref{sec:experiment}, and further specialize it to the case 
with scalar arm-specific parameters $\beta_0$ and $\beta_1$, 
where $\tau_n = \beta_1 - \beta_0$.  

To examine local asymptotic power, we take the  
average treatment effect or social welfare contrast to be  local to zero, i.e.~$\tau_n = (h_1-h_0)/\sqrt{n}$. 
We use the representation in Equation \eqref{eqn:simple_limitexp},
and write $\sigma_j^2 = \Sigma_j^{-1}$ for notational simplicity.
In this limiting bandit environment, we consider one-sided tests of the null 
$H_0: \tau = h_1-h_0 = 0$, 
 taking the adaptive treatment rules $\Lambda_b$ to be the limits of Thompson sampling as discussed in the previous subsection. 

We will compare two tests based on two  difference-in-means statistics.  
The pooled difference-in-means statistic combines the data from all batches:
\[
 \hat{S}^{PDM}_n = \hat{\beta}_{1} - \hat{\beta}_{0} 
 =  \frac{ \sum_{b=1}^B \sum_{i=1}^n Y_{b,i} D_{b,i}}{ \sum_{b=1}^B \sum_{i=1}^n  D_{b,i} } - 
 \frac{ \sum_{b=1}^B \sum_{i=1}^n Y_{b,i} (1-D_{b,i})}{ \sum_{b=1}^B \sum_{i=1}^n (1- D_{b,i}) }.
 \] 
This can be represented within our limiting Gaussian bandit environment as 
\begin{equation}  
\label{eqn:appt} 
  {T}^{PDM} = 
  \frac{ \sum_{b=1}^B \left(1-\Lambda_b\right)Z_{b,1} }{ \sum_{b=1}^B \left(1-\Lambda_b\right)  }
  -   \frac{ \sum_{b=1}^B \Lambda_b Z_{b,0} }{ \sum_{b=1}^B \Lambda_b  },
\end{equation} 
where we set $\Lambda_1 = 1/2$ and  $\Lambda_b = \Lambda_b(Z_1, \ldots, Z_{b-1})$ for $b\ge2$.
Due to the randomness in the arm probabilities for batches $b\ge 2$ induced by the 
sampling rule, the null sampling distribution of ${T}^{PDM}$ will not, in general, be normal. Further discussion and intuition for the nonstandard limiting distribution of the naive difference estimator are given in \citet{zhan:jans:murp:2020} and \citet{hada:etal:2020}. 

While the limiting null distribution of the pooled difference-in-means estimator is nonnormal, it is fully characterized by the  simple Gaussian bandit  representation given in 
\eqref{eqn:simple_limitexp} 
and \eqref{eqn:appt}, along with a particular limiting sampling rule $\Lambda_b$.  
To obtain a test based on the pooled difference-in-means estimator, 
the nonstandard null distribution of ${T}^{PDM}$ can be used to find a critical value for a one-sided test (in the direction of the alternative) that controls asymptotic size.  Below, we simulate this null distribution to obtain a critical value using mean zero Gaussian draws based on Equation
\eqref{eqn:simple_limitexp}.\footnote{The standardized version of the pooled difference-in-means statistic could also be used as a test statistic, and the critical values for its non-normal null limiting distribution are likewise available from simulating the limiting bandit asymptotic representation.  Here, we present the pooled difference-in-means statistic without standardization for simplicity. 
The performance of the two test statistics is very similar.}

\citet{hada:etal:2020} and \citet{zhan:jans:murp:2020} propose alternative testing procedures whose null distributions are asymptotically standard normal, so that conventional critical values can be used to obtain valid inference. 
While these new tests control asymptotic size, it is not clear to what extent these modified procedures sacrifice power to obtain a simple null distribution. 

We consider a batched difference-in-means test, which is a mild generalization of 
the test proposed by \cite{zhan:jans:murp:2020} and 
is in the class of tests proposed by \citet{hada:etal:2020}.  
The basic idea is to combine approximately normal batchwise difference-in-means statistics to 
obtain a single approximately normal statistic.  
The difference in means using only the first batch can be represented as 
$W_1 = Z_{1,1}-Z_{1,0}$. 
Under the null hypothesis, $W_1$
is normally distributed with zero mean and variance
$V(W_1) = 2(\sigma_{0}^2 + \sigma_{1}^2)$. 
Similarly, consider the batch $b$ statistic
$W_b = Z_{b,1}-Z_{b,0}$.   
Under the null, the conditional distribution of $W_b$ given $Z_1, \ldots, Z_{b-1}$ is normal with mean zero and conditional variance 
\[ V(W_b|Z_1, \ldots, Z_{b-1}) = \frac{\sigma_{0}^2}{\Lambda_b(Z_1, \ldots, Z_{b-1})} + \frac{\sigma_{1}^2}{1-\Lambda_b(Z_1, \ldots, Z_{b-1})}.\] 
As a result, the test statistic
\[ 
   {T}^{BDM} = \left[ 
 \sum_{b=1}^B 
 \frac{W_b}{\sqrt{\frac{\sigma_{0}^2}{\Lambda_b(Z_1, \ldots, Z_{b-1})} + \frac{\sigma_{1}^2}{1-\Lambda_b(Z_1, \ldots, Z_{b-1})}}}
\right] \bigg/ \sqrt{B} 
\] 
has a standard normal distribution under the null hypothesis.
This is the asymptotic analog of taking a linear combination of the batchwise differences-in-means, with weights inversely proportional to their estimated standard deviations.\footnote{A concrete formula for the finite sample statistic is given in Appendix \ref{app:bdm}.} 

We consider tests of the null hypothesis that $\tau=0$ against the one-sided alternative $\tau>0$, at the 5\% significance level. 
The two difference-in-means tests reject the null hypothesis if their test statistic exceeds a critical value. 
For the BDM test, the asymptotic critical value is simply 
the 0.95 quantile of the standard normal distribution.  
For the PDM test, we obtain an asymptotically size-corrected critical value 
by simulation.
We then simulate the local asymptotic power functions of these tests in the limiting Gaussian bandit environment. 
The asymptotic representations of both test statistics and the Thompson sampling rule depend on the local parameters only through the difference $\tau = h_1-h_0$.  
This invariance property means that we can express the local asymptotic power of both tests as a function of the scalar $\tau$ (for a given $\sigma_1^2$ and $\sigma_2^2$).  

We also obtain an envelope that gives a pointwise  upper bound on the local asymptotic power of any test.  
Applying the Neyman-Pearson lemma to the asymptotic representation 
provided by Theorem~\ref{thm:bandit} 
immediately yields a 
 pointwise bound for all alternative 
values $h_1-h_0\ne 0$.  
In the figures, we use 
this result 
to plot a least favorable power envelope in $\tau$ by finding the minimal power over all  $(h_0^1, h_1^1)$ such that $\tau = h_1^1-h_0^1$.\footnote{We can fix the null at $h^0=0$ without loss of generality, because  the likelihood ratio test appearing in the Neyman-Pearson Lemma  and the asymptotic representations of the PDM and BDM tests are invariant to additive shifts to the parameters under both the null and alternative hypotheses.}
\citet{adus:2025} 
develops an extended version of this envelope for semiparametric settings.

\begin{figure}[ht]
 \caption{Local Asymptotic Power Curves under Thompson Sampling: Equal Variances} 
 \label{fig:equal}
     \begin{subfigure}[]{0.5\textwidth}
             \caption{Two Batches ($B=2$): $\sigma_0=\sigma_1=\frac{1}{2}$}
       \includegraphics[width=.95\linewidth]{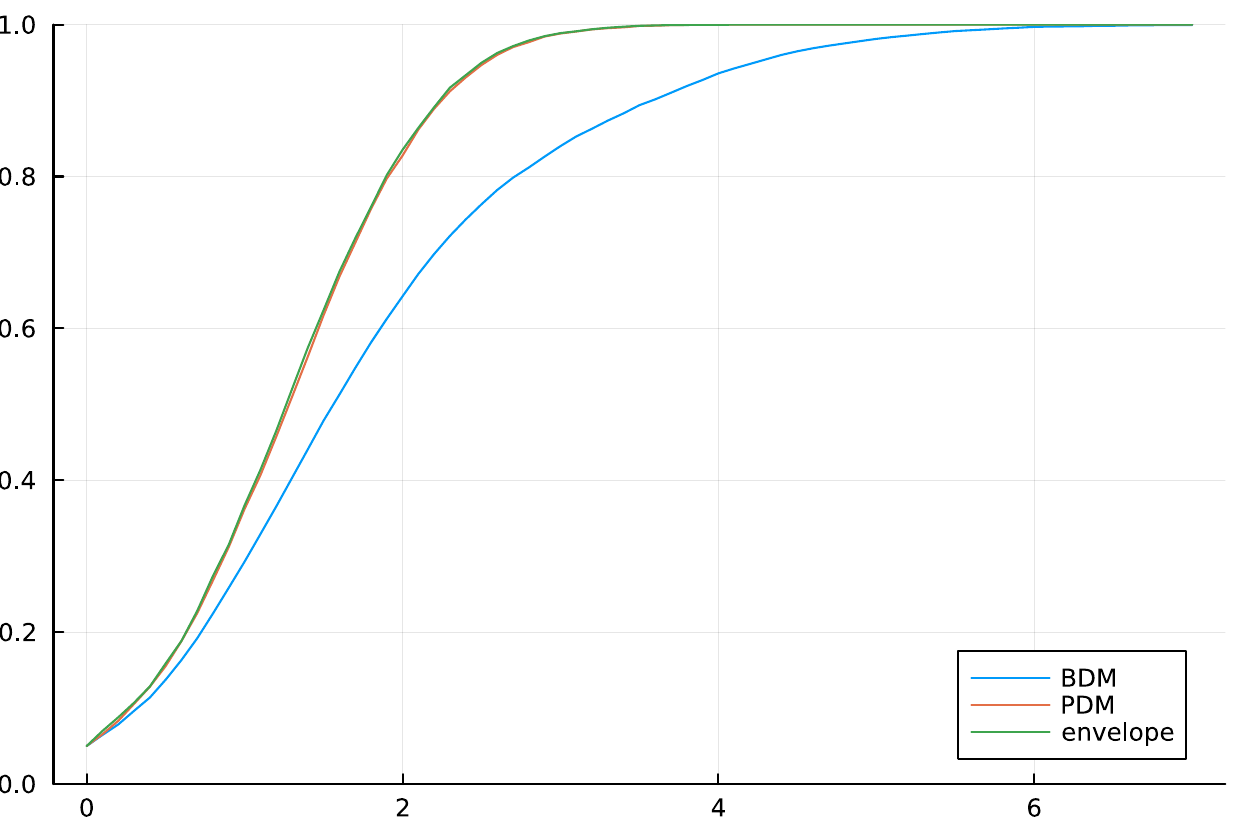}
    \end{subfigure}%
         \begin{subfigure}[]{0.5\textwidth}
             \caption{Five Batches ($B=5$): $\sigma_0=\sigma_1=\frac{1}{2}$}
       \includegraphics[width=.95\linewidth]{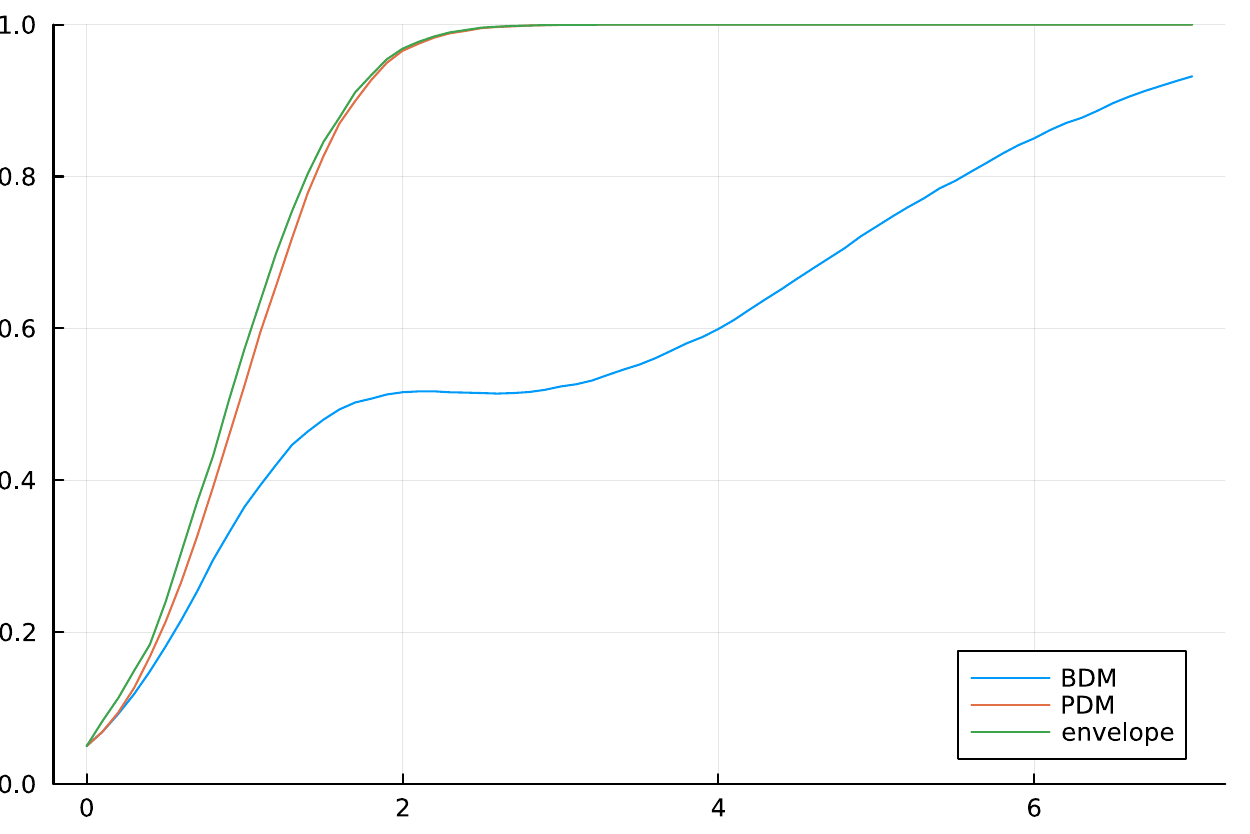}
    \end{subfigure}
\end{figure}

Figure~\ref{fig:equal} 
considers the case where the potential outcomes for the two treatment arms are homoskedastic with $\sigma_0=\sigma_1=\frac{1}{2}$, for experiments with two and five batches ($B=2,5$). 
The figure shows the local asymptotic power curves for the batched difference-in-means test and the pooled difference-in-means test (using the size-corrected critical value).  
The batched difference-in-means test has lower power than the pooled difference-in-means test, especially with more batches. 

The power envelope described above is also shown on each graph.   
Since this envelope is based on a pointwise optimality criterion, it is not necessarily attainable. 
However, in both cases, the pooled difference-in-means test nearly achieves the envelope.

\begin{figure}[ht]
 \caption{Local Asymptotic Power Curves under Thompson Sampling: Unequal Variances} 
 \label{fig:unequal}
     \begin{subfigure}[]{0.5\textwidth}
             \caption{Two Batches ($B=2$): $\sigma_0=\frac{1}{2}$, $\sigma_1=\frac{1}{16}$}
       \includegraphics[width=.95\linewidth]{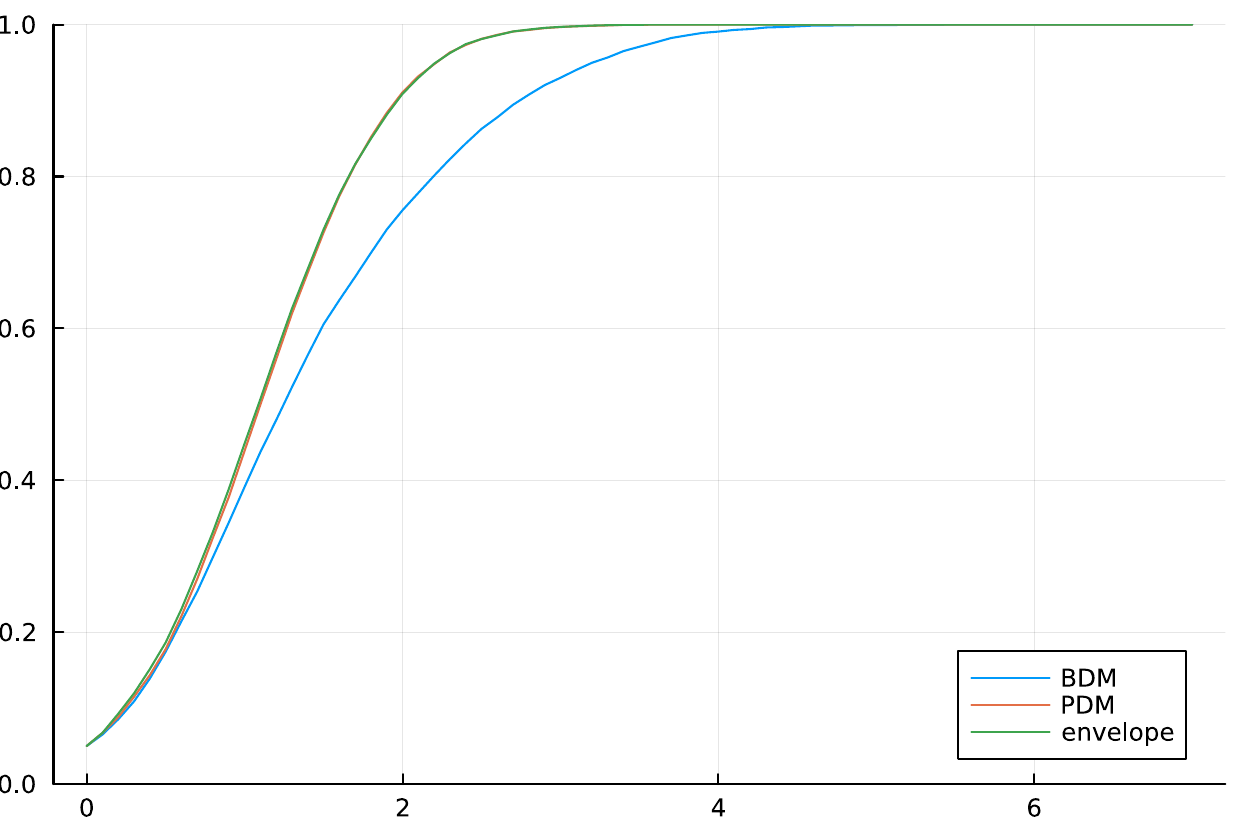}
    \end{subfigure}%
         \begin{subfigure}[]{0.5\textwidth}
             \caption{Five Batches ($B=5$): $\sigma_0=\frac{1}{2}$, $\sigma_1=\frac{1}{16}$ }
       \includegraphics[width=.95\linewidth]{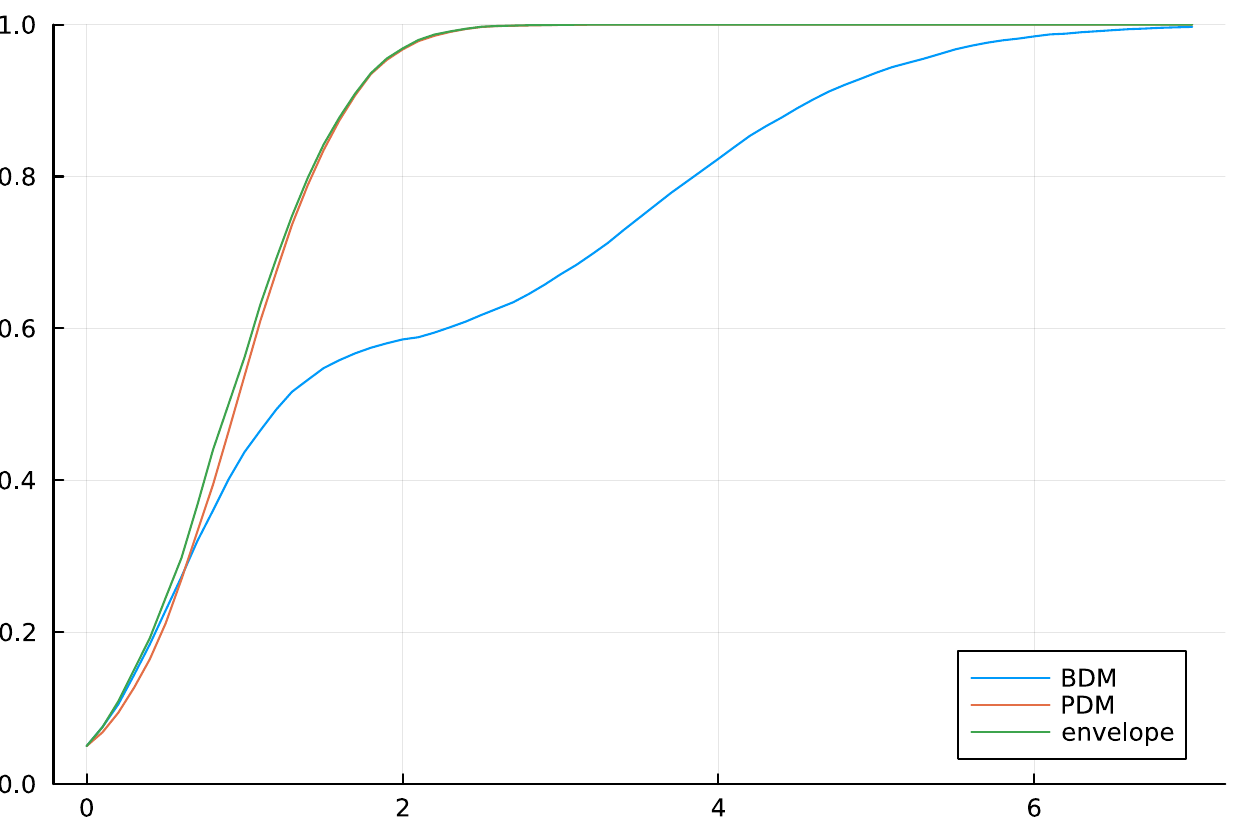}
    \end{subfigure} 
    \\ 
    \medskip 
    
         \begin{subfigure}[]{0.5\textwidth}
             \caption{Two Batches ($B=2$): $\sigma_0=\frac{1}{16}$, $\sigma_1=\frac{1}{2}$}
       \includegraphics[width=.95\linewidth]{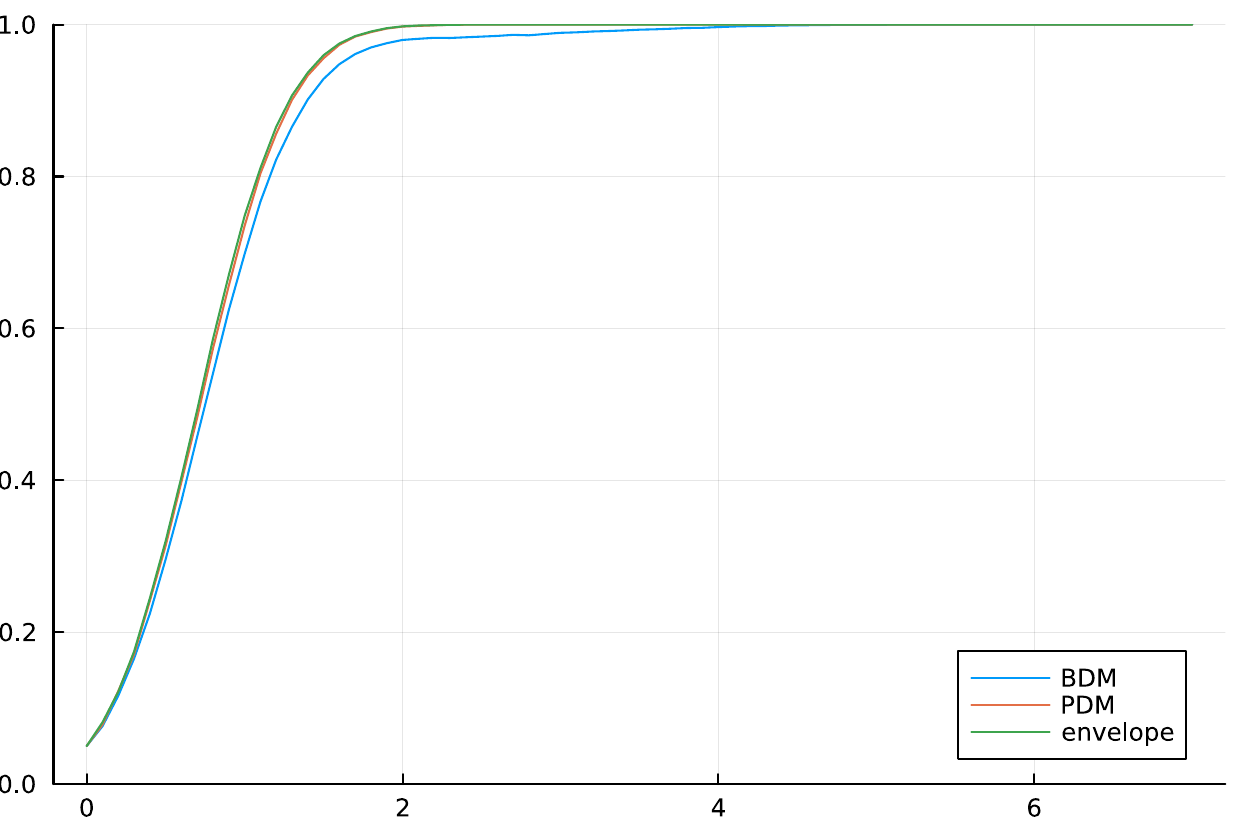}
    \end{subfigure}%
         \begin{subfigure}[]{0.5\textwidth}
             \caption{Five Batches ($B=5$): $\sigma_0=\frac{1}{16}$, $\sigma_1=\frac{1}{2}$ }
       \includegraphics[width=.95\linewidth]{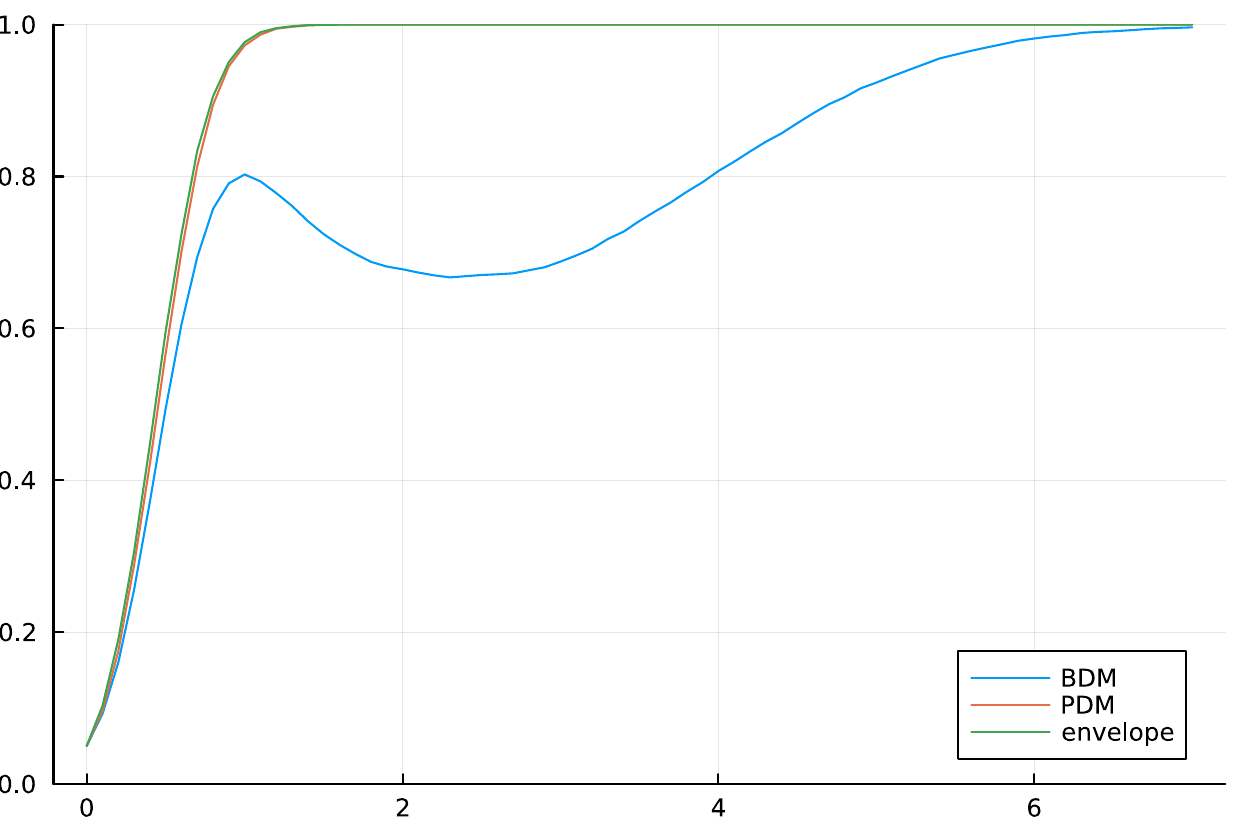}
    \end{subfigure}
\end{figure}

In Figure~\ref{fig:unequal} we consider cases with unequal arm variances.  
Panels (a) and (b) are qualitatively similar to the previous results with equal variances. 
When the more effective treatment arm has larger variance, as in panels (c) and (d), some distinctive features emerge.  For $B=2$, the relative power loss of the batched difference-in means test is quite small.  However, when $B=5$, the power curve for the batched difference-in-means test is noticeably worse and nonmonotonic in $\tau$.  
This nonmonotonicity arises because the value of $\tau$ affects not only the difference-in-means terms that appear in the numerator of the expression for $T^{BDM}$, but also the standard deviation terms in the denominator through the sampling rules $\Lambda_b$. 

In summary, these power comparisons reveal that the modification of tests to achieve pivotalness under the null can 
lead to significant loss of power that does not diminish with more batches.  
The conventional pooled difference-in-means test has a nonnormal null distribution, but can be size-corrected using its 
 asymptotic representation. It appears to perform quite well, even when compared to a conservative asymptotic power envelope.

\subsection{Batched Treatment Assignment}\label{subsec:treatmentassignment}

We next illustrate how the limiting bandit representation can be used to study batchwise treatment assignment rules, where the goal is to maximize the welfare of experimental and post-experimental subjects.
This can be viewed as a batchwise version of the classic two-armed bandit problem, with the addition of a post-sample group of individuals for whom treatment arms will also be assigned. 
We consider a simple, parametrized class of treatment assignment rules and compare their asymptotic welfare implications under different local parameters. 
We leave more extensive analysis of optimal decision rules for future work. 

We continue to use the simplified setup of Section \ref{subsec:powerenv}, with a scalar parameter for each arm and with the welfare contrast being local to zero: $\tau_n = \tau_n(h) = (h_1-h_0)/\sqrt{n}$. Subjects arrive and are assigned in two batches, followed by an out-of-sample treatment assignment choice for a group of future individuals drawn from the same population.
We continue to use $\Lambda_{b,n}$ to denote the fraction of subjects assigned to treatment 0 in stage $b$, where $\Lambda_{1,n} = 0.5$ and $\Lambda_{2,n}$ can be based on the data collected from the first batch. 
At the conclusion of the experiment, a treatment allocation $\Lambda_{3,n}$ is selected for the future group of individuals, based on the complete experimental data. 
We assume that $\Lambda_{2,n}$ and $\Lambda_{3,n}$ converge jointly in distribution under every $h=(h_0,h_1)$. By Theorem \ref{thm:bandit}, they can be represented asymptotically by rules $\Lambda_2(Z_1,U)$ and $\Lambda_{3}(Z_1,Z_2,U)$, where $Z_1$ and $Z_2$ have the distributions given in Section \ref{subsec:powerenv} and $U$ is an independent randomization. 

 Suppose that the decision-maker evaluates treatment allocations $\lambda \in [0,1]$ through their welfare regret loss, as in \citet{mans:2004}. The welfare regret under allocation $\lambda$ and parameter $\tau_n(h)$ is
\begin{align*} L^R(h,\lambda) &= \tau_n(h)\left[ {\bf 1}(\tau_n(h)>0) - (1-\lambda)\right]\\
   &= -\tau_n(h) \left[ {\bf 1}(\tau_n(h)\le 0) - \lambda\right].
\end{align*}
While the treatment allocation in batch 1 is fixed in advance, the allocation $\Lambda_{2,n}$ in batch 2 is under the control of the decision-maker, who considers the welfare of the subjects in batch 2 as well as the benefits from experimentation for making a better choice $\Lambda_{3,n}$ of the allocation for the future subjects.\footnote{Because we are focusing on regret loss and the decision rule for the first batch is fixed, we drop loss terms associated with batch 1 in the following analysis.}
We consider a weighted regret loss where, for some $\omega \in [0,1]$,  
\[ L^{\omega}(h,\lambda_2,\lambda_3) = (1-\omega) L^R(h,\lambda_2) + \omega L^R(h,\lambda_3). \] 
The weighted regret risk of $(\Lambda_{2,n}, \Lambda_{3,n})$ is
\[ R^{\omega}(h, \Lambda_{2,n},\Lambda_{3,n}) = E_h\left[ L^{\omega}(h, \Lambda_{2,n},\Lambda_{3,n})\right].\]
Here the expectation is with respect to the data in both batches, and the distribution of the data in the second batch depends on the choice of $\Lambda_{2,n}$.   
Following \cite{hira:port:2009}, we rescale the loss and risk functions by the factor $\sqrt{n}$ and consider their limits as $n\to\infty$. 
These limits can be characterized through a limiting version of the loss function. 
Since the basic regret loss has the limiting form
\[ \sqrt{n} L^R(h,\lambda) \to L_{\infty}^R(h,\lambda) := (h_0-h_1)\left[ {\bf 1}(h_0 \ge h_1) - \lambda \right], \] 
we can take the limit of the weighted loss to be 
\begin{align*} L_{\infty}^\omega (h,\lambda_2,\lambda_3) &= (1-\omega) L_{\infty}^R(h,\lambda_2) + \omega L_{\infty}^R(h,\lambda_3)\\ 
  &= (h_0-h_1) \left\{ {\bf 1}(h_0 \ge h_1) - (1-\omega) \lambda_2 - \omega \lambda_3 \right\}.
\end{align*}
The corresponding limiting risk, using the asymptotic representations for the decision rules, is
\[ R^\omega(h,\Lambda_2,\Lambda_3) = (h_0-h_1) \left\{ {\bf 1}(h_0\ge h_1) - (1-\omega) E_{h}\left[ \Lambda_2(Z_1,U) \right] - \omega E_{h,\Lambda_2} \left[ \Lambda_3(Z_1,Z_2,U) \right]\right\},
\] 
where the first expectation is taken with respect to the marginal distribution of $(Z_1, U)$ under $h$, and the second expectation is taken with respect to the marginal distribution of $(Z_1, U)$ and the conditional distribution of $Z_2$ given $\Lambda_2(Z_1,U)$. 

The limiting Gaussian bandit representation holds for any treatment assignment rules that possess limits. 
We consider a class of ``scaled'' Thompson rules. 
For $c \ge 0$, let $\Lambda_{2c,n} = \Phi\left(c \cdot \Phi^{-1}(\Lambda_{2,n})\right)$  where $\Lambda_{2,n}$ is the Thompson sampling rule
(and analogously for $\Lambda_{3c,n}$).
Since $\Lambda_{2,n} \leadsto \Lambda_2(Z_1)$, the limit of the scaled Thompson rule is 
\[ 
\Lambda_{2,c}(Z_1) =  \Phi\left( c \cdot \frac{\dot{g}_0'Z_{1,0}-\dot{g}_1'Z_{1,1} }{\sqrt{2 \dot{g}_0'\Sigma_0^{-1}\dot{g}_0+ 2 \dot{g}_1'\Sigma_1^{-1}\dot{g}_1}} \right).
\]
For $c=0$, the scaled Thompson rule assigns half of the subjects to each arm, regardless of the realization of the first batch. 
As $c \to \infty$, the rule assigns all individuals to the arm with the higher estimated mean; this is an asymptotic analog of Manski's empirical success rule 
\citep{mans:2004}.

\begin{figure}[ht]
  \caption{Thompson Sampling in Batch 2, Compare Options for $\Lambda_3$}\label{fig:assignment1}
  \begin{center}
    \includegraphics[width=10cm]{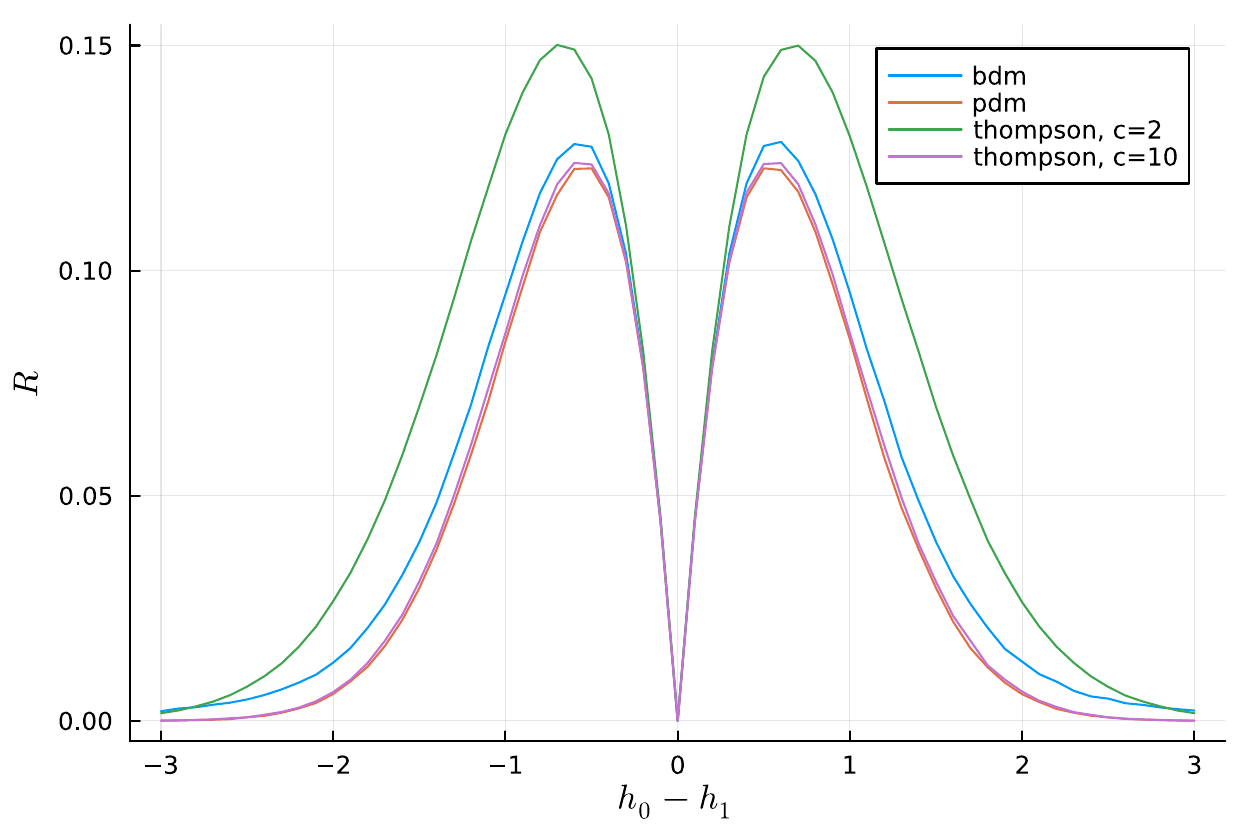}
  \end{center}
\end{figure}

\begin{figure}[ht]
  \caption{Scaled Thompson Sampling in Batch 2, Compare Options for $\Lambda_3$}\label{fig:assignment2}
  \begin{center}
    \includegraphics[width=10cm]{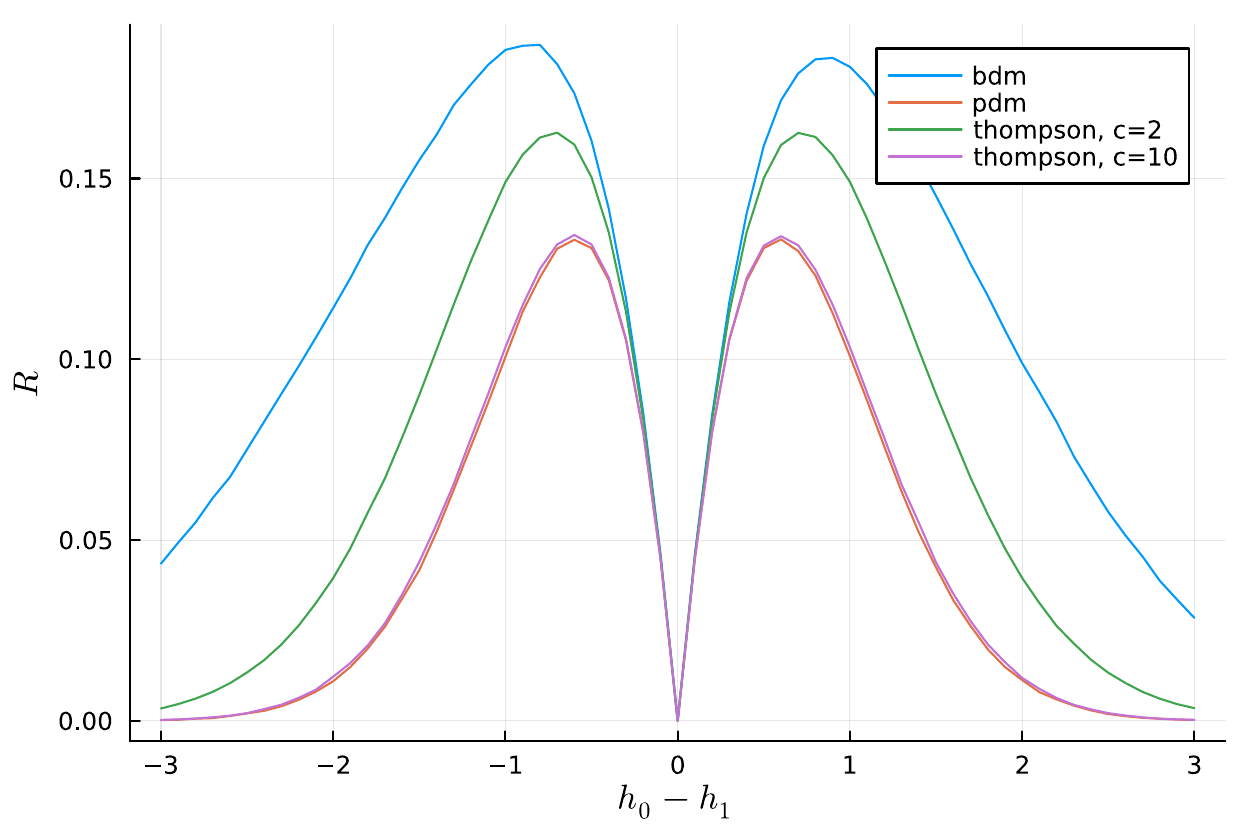}
  \end{center}
\end{figure}

We first consider cases where $\omega=1$, which puts all of the welfare weight on the future (out of sample) subjects. There is a fixed second stage rule $\Lambda_2$, and we compare alternative choices for the future group allocation $\Lambda_3$. 
In Figure \ref{fig:assignment1}, the two treatment arms have equal variances ($\sigma_1 = \sigma_0 = 0.5$), and $\Lambda_2$ is the conventional Thompson sampling rule. 
For $\Lambda_3$, we consider the pooled difference-in-means rule, the batched difference-in-means rule, and scaled Thompson sampling with $c=2$ and $c=10$. 
For all of these rules, their risk function is symmetric around zero, and tends to peak at moderate values of the treatment effect, where sampling uncertainty interacts with the welfare loss from choosing the inferior treatment. 
The PDM rule has uniformly lowest risk among the four rules. 
PDM is the large sample analog of the Manski's empirical success rule, which is known to be asymptotically optimal for non-adaptive treatment assignment problems \citep{hira:port:2009}. 
It also is the Bayes decision in the adaptive limiting Gaussian problem under a flat prior, so it is asymptotically average risk optimal in this adaptive setting with equal variances and a fixed choice for $\Lambda_2$.
The scaled Thompson rule with $c=10$ is similar to PDM, as expected. 

Figure \ref{fig:assignment2} shows the same comparison when  the second-stage allocation has $c=5$, which allocates units more aggressively  towards the arm with higher mean in the first batch. This leads to higher welfare regret for the third batch of individuals, but the PDM rule for batch 3 continues to dominate the other rules. 

\begin{figure}[ht]
  \caption{PDM in Batch 3, Compare Options for $\Lambda_2$}\label{fig:assignment3}
  \begin{center}
    \includegraphics[width=10cm]{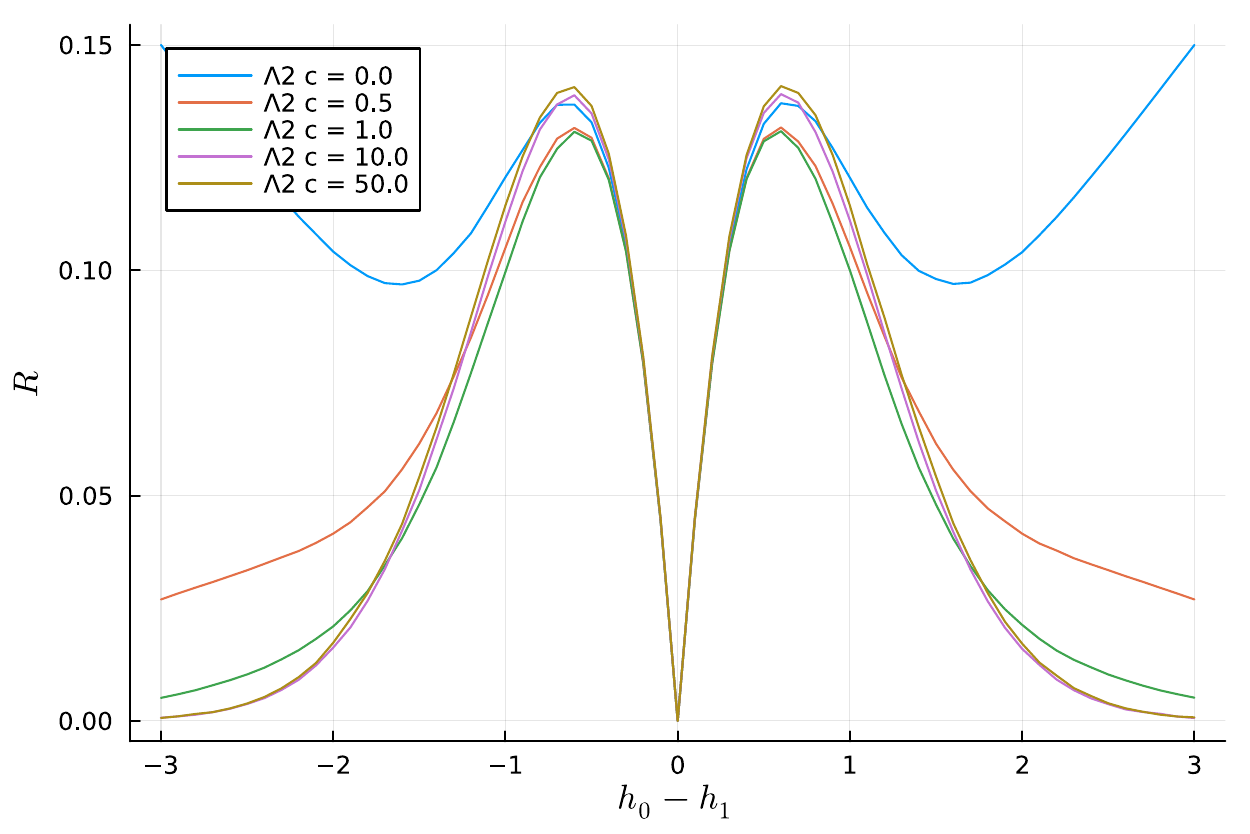}
  \end{center}
\end{figure}

The  PDM rule seems like a reasonable choice for $\Lambda_3$, at least in the case with equal arm variances. In Figure \ref{fig:assignment3}, we set $\Lambda_3$ to the PDM rule and consider a range of scaled Thompson rules for $\Lambda_2$. 
The welfare weighting of the future subject group is $\omega=0.9$. 
When $c=0$, the welfare regret risk diverges. This reflects the welfare regret from not adapting the treatment choice for subjects in batch 2, which has a 0.1 weight in the loss function. 
This problem disappears when there is a moderate to high degree of adaptation in batch 2. 
No single choice of $c$ dominates all the others. 
However, among the choices considered, $c=1$ (conventional Thompson sampling) has the lowest worst-case risk.

\section{Conclusion}

The representation results we have developed in this paper characterize all attainable limit distributions of procedures in a rich class of dynamic statistical decision problems where the agent's actions can include the choice of the adaptive experimentation rule.  
They do not require restricting attention to a specific class of rules, such as Bayes rules or empirical analog (plug-in) rules, and can be combined with any solution concept such as minmax regret or robust preferences. 
Decision problems that are intractable in their original form can be dramatically simplified through these approximations. 
However, whether or not the asymptotic approximation of the original problem by a limiting Gaussian bandit is simple enough to yield an explicit solution will depend on the specific problem at hand. 
For more complex problems, it may be useful to first apply our representation theorem, and then use numerical methods to obtain an approximate solution to the simplified decision problem.

\clearpage
\bibliography{seq}

\clearpage 

\appendix
\appendixpage
\section{Proofs}
\label{sec:app_proofs}

{\bf Proof of Proposition \ref{thm:artfixed}:}

This proposition follows from Theorem \ref{thm:bandit} with mild adjustments in notation.  
There is a single treatment arm $k=0$. 
Set $B$ (in Theorem \ref{thm:bandit}) equal to the number of decision nodes $B$ in Proposition \ref{thm:artfixed}. 
The allocation rules $\Lambda_{b,n} = \lambda_{b,1}$ are deterministic, with $\lambda_{1,1}$ in Theorem \ref{thm:bandit} equal to $\lambda_1$ in the Proposition, and $\lambda_{b,1}$ in the Theorem equal to $\lambda_b-\lambda_{b-1}$ for $b \ge 2$ in the Proposition. 
We can map the observation indices $(b,i)$ in Theorem \ref{thm:bandit} to equivalent indices $i$ that are used in Proposition \ref{thm:artfixed}.
Thus, in each successive batch $b$, we observe $\lfloor n (\lambda_b - \lambda_{b-1})\rfloor$ observations of the outcome under ``treatment 0.''

The probability density function of the observations $z_i$ is not time-varying in Proposition \ref{thm:artfixed}, so we can drop the $(b)$ superscript on $p_{\theta,k}^{(b}(\cdot)$. 
We assume that the Fisher information $J_0^{(b)} = J_0$ is fully invertible, which allows us to simplify some of the expressions in the Proposition using standard properties of the multivariate normal distribution. 
This combined with the properties of Brownian motion allow us to relate the increments of the process $Z(t)$ in Proposition \ref{thm:artfixed} to the corresponding shifted normal expressions in Theorem \ref{thm:bandit}.

{\bf Proof of Theorem \ref{thm:bandit}:}

The first step is to characterize the asymptotic behavior of the likelihood ratio process.  For each sample observation, we have both 
the observed arm data and the ``potential'' data on unobserved arms.  It will be useful to jointly characterize both the likelihood ratio process corresponding to the observed data and the ``potential'' likelihood ratio process corresponding 
to data from all arms including the unobserved potential arm data.  Without loss of generality, we specify an auxiliary model of 
unobserved arm data that will exactly preserve the finite sample behavior of the observed data and maintain the assumptions of the 
theorem.  In the auxiliary model, the marginal distribution of data associated with each arm, $p^{(b)}_{\theta,k}(z_{b,i}(k))$, will remain as provided in the 
supposition of the theorem, and we will additionally take $z_{b,i}(0), \ldots, z_{b,i}(K-1)$ to be independent across arms. 
By construction, both the auxiliary potential likelihood and the ``actual'' potential likelihood will generate 
identical likelihoods for the observable data.
In the remainder of the proof,  the ``potential'' likelihood will be a reference to the likelihood of observed and unobserved data for 
the auxiliary model. 

Given $\lambda_b = (\lambda_{b,0},\ldots,\lambda_{b,K})$ for batch $b \in \{1, \ldots, B\}$, the (observed data) likelihood ratio process of 
  $\theta_n(h)=\theta_0 + h/\sqrt{n}$ versus $\theta_0$  is given by 
\[ \ell_{b,n}(h; \lambda_b) = \prod_{k=0}^{K-1} 
\prod_{i= \lfloor n \lambda_{b,k}\rfloor + 1}^{\lfloor n \lambda_{b,k+1} \rfloor} 
\frac{p^{(b)}_{\theta_n(h),k}(z_{b,i}(k))}{p^{(b)}_{\theta_0,k}(z_{b,i}(k))}\] 
 for all $h \in \mathbb{R}^m$.
For batch  $b \in \{2, \ldots, B\}$, we take the 
 the auxiliary (unobserved data) likelihood ratio process of 
  $\theta_n(h)=\theta_0 + h/\sqrt{n}$ versus $\theta_0$  to be given by 
\[ \ell_{b,n}^-(h; \lambda_b) = \prod_{k=0}^{K-1} 
\prod_{\substack{ 1 \le i \le  \lfloor n \lambda_{b,k}\rfloor \\  \lfloor n \lambda_{b,k+1} \rfloor  + 1\le i \le  \lfloor n \bar{\lambda}_{b} \rfloor}}
\frac{p^{(b)}_{\theta_n(h),k}(z_{b,i}(k))}{p^{(b)}_{\theta_0,k}(z_{b,i}(k))}\] 
 for all $h \in \mathbb{R}^m$.  Note that, for observation $i$, this specification imposes independence 
 across arms.  Also, we focus only on batches $\{2, \ldots, B\}$ where the observed arms are determined 
 adaptively.

Jointly over batches, these likelihood ratio processes 
 form stochastic processes in $(\lambda_1, \ldots, \lambda_B)$: 
 \begin{align*}
  \ell_{n}(h)(\lambda_1, \ldots, \lambda_B) &= 
  (\ell_{1,n}(h; \lambda_1), \ldots, \ell_{B,n}(h; \lambda_B)); \text{ and }
  \\
  \ell_{n}^-(h)(\lambda_2, \ldots, \lambda_B) &= 
  (\ell_{2,n}^-(h; \lambda_2), \ldots, \ell_{B,n}^-(h; \lambda_B)).
  \end{align*}
  
Under the DQM assumption and Donsker's Theorem for partial-sum processes \citep[e.g.][Theorem 2.12.6]{vdv:well:1996}, the pair $(\ell_{n}(h), \ell_{n}^-(h))$ converges weakly to a tight random element: 
   \begin{equation} 
   \label{eqn:lrc} 
   (\ell_{n}(h), \ell_{n}^-(h)) \stackrel{0}{\leadsto} (\ell^0(h), \ell^{0-}(h)),
   \end{equation} 
where 
\begin{align} 
  \ell^0( h & )(\lambda_1, \ldots, \lambda_B)  
  \label{eqn:ell0}
  \\ = & 
  \left( \prod_{k=0}^{K-1} 
  \exp\left[ h^\prime
   (\Delta_{1,k}(\lambda_{1,k+1}) - \Delta_{1,k}(\lambda_{1,k})) 
   - \frac{(\lambda_{1,k+1}-\lambda_{1,k})}{2} h^\prime J_k^{(1)} h \right],  \ldots,  
  \right. 
   \nonumber 
  \\ 
  & \left. \ \ \ 
    \prod_{k=0}^{K-1} 
   \exp\left[ h^\prime ( \Delta_{B,k}(\lambda_{B,k+1}) - \Delta_{B,k}(\lambda_{B,k})) 
   - \frac{(\lambda_{B,k+1}-\lambda_{B,k})}{2} h^\prime J_k^{(B)} h \right]
    \right), 
     \nonumber 
\\[2ex] 
   \ell^{0-} & (h)({\lambda}_2, \ldots, \lambda_B) 
     \label{eqn:ell0minus}
   \\  =  
  & \left( \prod_{k=0}^{K-1} 
  \exp\left[ h^\prime
  [ (\Delta_{2,k}(\bar{\lambda}_{2}) - \Delta_{2,k}(\lambda_{2,k+1})) + (\Delta_{2,k}(\lambda_{2,k}) - \Delta_{2,k}(\lambda_{2,0})) ]
   - \frac{(\bar{\lambda}_{2}-\lambda_{2,k+1}+ \lambda_{2,k} )}{2} h^\prime J_k^{(2)} h \right],    \ldots, 
  \right. 
   \nonumber 
  \\ 
  & \left. \ \ \ 
   \prod_{k=0}^{K-1} 
  \exp\left[ h^\prime
  [ (\Delta_{B,k}(\bar{\lambda}_{B}) - \Delta_{B,k}(\lambda_{B,k+1})) + (\Delta_{B,k}(\lambda_{B,k}) - \Delta_{B,k}(\lambda_{B,0})) ]
   - \frac{(\bar{\lambda}_{B}-\lambda_{B,k+1}+ \lambda_{B,k} )}{2} h^\prime J_k^{(B)} h \right] 
    \right),   
    \nonumber 
\end{align} 
and the $\Delta_{b,k}(\cdot)$ are Gaussian processes with independent increments 
such that $\Delta_{b,k}(t) \sim N(0,tJ_k^{(b)})$ for $t\ge 0$, with independence 
across $b$ and $k$.

Putting together the observed and unobserved likelihood ratios, we can denote 
the  ``potential'' likelihood ratio for batch $b \in \{2, \ldots, B\}$ by 
\[ \bar{\ell}_{b,n}(h) = \prod_{k=0}^{K-1} 
\prod_{i= 1}^{\lfloor n \bar{\lambda}_{b} \rfloor} 
\frac{p^{(b)}_{\theta_n(h),k}(z_{b,i}(k))}{p^{(b)}_{\theta_0,k}(z_{b,i}(k))},\] 
and form the joint likelihood  over these batches:
$\bar{\ell}_{n}(h) = 
  (\bar{\ell}_{2,n}(h), \ldots, \bar{\ell}_{B,n}(h))$.   Note that 
  $\bar{\ell}_{b,n}(h) =  \ell_{b,n}(h; \lambda_b) \cdot \ell_{b,n}^-(h; \lambda_b)$, 
  but $\bar{\ell}_{b,n}(h)$ does not depend on $ \lambda_b$.  
  The weak limit of $\bar{\ell}_{n}(h)$ will be used for Le Cam's third lemma below, 
  and this limit follows by (\ref{eqn:lrc}):
  \[ 
    \bar{\ell}_{n}(h)  \stackrel{0}{\leadsto} \bar{\ell}^0(h),
   \] 
where 
\begin{equation} 
  \bar{\ell}^0(h) =  
  \left( \prod_{k=0}^{K-1} 
  \exp\left[ h^\prime
   \Delta_{2,k}(\bar{\lambda}_{2}) 
   - \frac{\bar{\lambda}_{2}}{2} h^\prime J_k^{(2)} k \right], 
   \ldots, 
   \prod_{k=0}^{K-1} 
   \exp\left[ h^\prime \Delta_{B,k}(\bar{\lambda}_{B})  
   - \frac{\bar{\lambda}_{B}}{2} h^\prime J_k^{(B)} h \right]
    \right). 
    \label{eqn:ell0pot} 
\end{equation}

For batch 1, $\lambda_1$ determines the observations from each arm. Let 
\[ 
  \Delta_1(\lambda_1)  \, = \, 
  \left( \Delta_{1,0}(\lambda_{1,1}) - \Delta_{1,0}(\lambda_{1,0}), \ldots, 
  \Delta_{1,K-1}(\lambda_{1,K}) - \Delta_{1,K-1}(\lambda_{1,K-1})
    \right).  
\] 
For batches $2, \ldots, B$, the scores $\Delta_{b,k}(\bar{\lambda}_{b})$ from \eqref{eqn:ell0pot}, which 
represent the limiting potential information from 
batch $b$ and arm $k$, can be partitioned into 
``observed'' and ``unobserved'' scores (see \eqref{eqn:ell0} and  \eqref{eqn:ell0minus}) as determined by 
$(\lambda_{b,k}, \lambda_{b,k+1})$:
\begin{align*} 
\Delta_b(\lambda_b) & \, = \, 
\left( \Delta_{b,0}(\lambda_{b,1}) - \Delta_{b,0}(\lambda_{b,0}), \ldots, 
\Delta_{b,K-1}(\lambda_{b,K}) - \Delta_{b,K-1}(\lambda_{b,K-1})
  \right); 
  \\ 
  \Delta_b^-(\lambda_b) & \, = \, 
\left( \Delta_{b,0}(\bar{\lambda}_{b}) - \Delta_{b,0}(\lambda_{b,1}), 
\Delta_{b,1}(\lambda_{b,1}) - \Delta_{b,1}(\lambda_{b,0}), 
\Delta_{b,1}(\bar{\lambda}_{b}) - \Delta_{b,1}(\lambda_{b,2}), \ldots, \right. 
\\ & \hskip1.9cm \left.  
\Delta_{b,K-1}(\lambda_{b,K-1}) - \Delta_{b,K-1}(\lambda_{b,0}), 
\Delta_{b,K-1}(\bar{\lambda}_{b}) - \Delta_{b,K-1}(\lambda_{b,K})
  \right);
  \\ 
  \bar{\Delta}_b & \, = \, \left( \Delta_{b,0}(\bar{\lambda}_{b}), \ldots, 
  \Delta_{b,K-1}(\bar{\lambda}_{b})
    \right).
\end{align*} 
For any $\lambda_b$, the terms $\Delta_b(\lambda_b)$ and $\Delta_b^-(\lambda_b)$ are independent.

By supposition, the statistics $S_{b,n}$ and $\Lambda_{b,n}$ jointly converge weakly under $\theta_0$. For convenience, we use zero superscripts  to denote their limits, and write 
\begin{align*} 
  \left( S_{1,n},\Lambda_{2,n},  S_{2,n},\dots, \Lambda_{B,n},S_{B,n} \right) 
   \  \stackrel{0}{\leadsto} \ 
   \left( S_{1}^0,\Lambda_{2}^0,  S_{2}^0, \dots, \Lambda_{B}^0, S_{B}^0, \right).
\end{align*}
Prohorov's Theorem \citep[Theorem 1.3.9]{vdv:well:1996} and auxiliary results \citep[Lemmas 1.4.3 and 1.4.4]{vdv:well:1996} imply that along a subsequence we have 
joint convergence of the statistics and the likelihood ratio 
processes, 
\begin{align*} 
  \left( S_{1,n},\Lambda_{2,n},  S_{2,n},\dots, \Lambda_{B,n},S_{B,n}, \ell_{n}(h),  \bar{\ell}_{n}(h)  \right) 
   \  \stackrel{0}{\leadsto} \ 
   \left( S_{1}^0,\Lambda_{2}^0,  S_{2}^0, \dots, \Lambda_{B}^0, S_{B}^0,\ell^0(h), \bar{\ell}^0(h) \right), 
\end{align*}
for each $h \in \mathbb{R}^m$.

Next we construct representations for the weak limit of the statistics under 
$\theta_0$, using conditional vector quantile functions.  
Given two random vectors $Y$ and $X$ taking values on $\mathbb{R}^{d_Y}$ and 
$\mathbb{R}^{d_X}$, the conditional vector quantile function of $Y$ given $X=x$ 
can be defined as a map $u \mapsto q_{Y|X}(u|x)$ where $u$ takes values on the 
$\mathbb{R}^{d_Y}$ unit cube and $q_{Y|X}(U|x)$ has the distribution of 
$Y$ conditional on  $X=x$ when $U$ is uniformly distributed on the 
$\mathbb{R}^{d_Y}$ unit cube \citep{carl:cher:gali:2016}.  

Our construction proceeds recursively.  Let $U_1, \ldots, U_B$ be 
independent 
uniformly distributed random vectors on the appropriate unit cube that 
are also independent of $\left( \Delta_{b,k} \right)_{b=1,k=0}^{B,K-1}$.  
\begin{enumerate} 
\item[1(a)] Given the  conditional vector quantile function 
$q_{(S_{1}^0,\Lambda_{2}^0)|\Delta_1(\lambda_1)}(u|\delta_1)$, define 
$$(Q_{S_{1}^0}, Q_{\Lambda_{2}^0}) 
= q_{(S_{1}^0,\Lambda_{2}^0)|\Delta_1(\lambda_1)}(U_1| \Delta_1(\lambda_1)).$$ 
Then, 
\[ 
  (\Delta_1(\lambda_1), Q_{S_{1}^0}, Q_{\Lambda_{2}^0}) 
  \sim 
  (\Delta_1(\lambda_1), S_{1}^0,\Lambda_{2}^0). 
\] 
\item[1(b)] Note that $(\Delta_{2,0}, \ldots, \Delta_{2,K-1}) \perp (\Delta_1(\lambda_1), Q_{S_{1}^0}, Q_{\Lambda_{2}^0}) $ 
 and $(\Delta_{2,0}, \ldots, \Delta_{2,K-1})  \perp (\Delta_1(\lambda_1), S_{1}^0,\Lambda_{2}^0)$.  Hence, 
 for the randomly stopped processes, $\Delta_2(Q_{\Lambda_{2}^0}), \Delta_2^-(Q_{\Lambda_{2}^0}), 
 \Delta_2(\Lambda_2^0), 
 \Delta_2^-(\Lambda_2^0)$, we have 
 \begin{align*} 
  (\Delta_1(\lambda_1), Q_{S_{1}^0}, Q_{\Lambda_{2}^0}, \Delta_2(Q_{\Lambda_{2}^0})) 
 & \sim 
  (\Delta_1(\lambda_1), S_{1}^0,\Lambda_{2}^0, \Delta_2(\Lambda_2^0)) \ \ \mbox{and}
  \\ 
  (\Delta_1(\lambda_1), Q_{S_{1}^0}, Q_{\Lambda_{2}^0}, \Delta_2^-(Q_{\Lambda_{2}^0})) 
 & \sim 
  (\Delta_1(\lambda_1), S_{1}^0,\Lambda_{2}^0, \Delta_2^-(\Lambda_2^0)) 
 \end{align*} 
 Since $\Delta_2(Q_{\Lambda_{2}^0}) \perp \Delta_2^-(Q_{\Lambda_{2}^0}) | 
 \Delta_1(\lambda_1), Q_{S_{1}^0}, Q_{\Lambda_{2}^0}$ and 
 $\Delta_2(\Lambda_2^0) \perp \Delta_2^-(\Lambda_2^0) | 
 \Delta_1(\lambda_1), S_{1}^0,\Lambda_{2}^0$, it follows that 
 \[ 
  (\Delta_1(\lambda_1), Q_{S_{1}^0}, Q_{\Lambda_{2}^0}, 
  \Delta_2(Q_{\Lambda_{2}^0}), \Delta_2^-(Q_{\Lambda_{2}^0})) 
  \sim 
  (\Delta_1(\lambda_1), S_{1}^0,\Lambda_{2}^0, 
  \Delta_2(\Lambda_2^0), \Delta_2^-(\Lambda_2^0))
 \] 
 \item[2(a)] For $b = 2, \ldots, B-1$, given 
 $(\Delta_1(\lambda_1), Q_{S_{1}^0}, Q_{\Lambda_{2}^0}, \Delta_2(Q_{\Lambda_{2}^0}), 
 \ldots, Q_{S_{b-1}^0}, Q_{\Lambda_{b}^0}, \Delta_b(Q_{\Lambda_{b}^0})) $, 
 we construct $(Q_{S_{b}^0}, Q_{\Lambda_{b+1}^0})$ using 
 the conditional vector quantile function \\ 
 $q_{(S_{b}^0, \Lambda_{b+1}^0) | 
 \Delta_1(\lambda_1), S_{1}^0,\Lambda_{2}^0, 
 \Delta_2(\Lambda_2^0), \ldots, S_{b-1}^0,\Lambda_{b}^0, 
 \Delta_b(\Lambda_b^0)
 }(u | \delta_1, s_1, \lambda_2, \delta_2, \ldots, s_{b-1}, \lambda_{b}, \delta_b)$.  
 Define 
 \begin{align*} 
 &  (Q_{S_{b}^0}, Q_{\Lambda_{b+1}^0}) 
  = 
  \\ 
  & 
  q_{(S_{b}^0, \Lambda_{b+1}^0) | 
 \Delta_1(\lambda_1), S_{1}^0,\Lambda_{2}^0, 
 \Delta_2(\Lambda_2^0), \ldots, S_{b-1}^0,\Lambda_{b}^0, 
 \Delta_b(\Lambda_b^0)
 }(U_b| 
 \Delta_1(\lambda_1), Q_{S_{1}^0}, Q_{\Lambda_{2}^0}, \Delta_2(Q_{\Lambda_{2}^0}), 
 \ldots, Q_{S_{b-1}^0}, Q_{\Lambda_{b}^0}, \Delta_b(Q_{\Lambda_{b}^0})). 
 \end{align*}
Then, 
\[ 
  (\Delta_1(\lambda_1), Q_{S_{1}^0}, Q_{\Lambda_{2}^0}, \Delta_2(Q_{\Lambda_{2}^0}), 
  \ldots, Q_{S_{b}^0}, Q_{\Lambda_{b+1}^0})
  \,  \sim \, 
  (\Delta_1(\lambda_1), S_{1}^0,\Lambda_{2}^0, \Delta_2({\Lambda_{2}^0}), 
  \ldots, {S_{b}^0}, {\Lambda_{b+1}^0}). 
\] 

For $b\ge 3$,  note that 
$U_b \perp ( \Delta_2^-(Q_{\Lambda_{2}^0}), \ldots, \Delta_{b}^-(Q_{\Lambda_{b}^0}))  $ $| \, 
\Delta_1(\lambda_1), Q_{S_{1}^0}, Q_{\Lambda_{2}^0}, 
\Delta_2(Q_{\Lambda_{2}^0}), 
\ldots, Q_{S_{b-1}^0}, Q_{\Lambda_{b}^0}, \Delta_b(Q_{\Lambda_{b}^0})$, and 
$({S_{b}^0}, {\Lambda_{b+1}^0}) \perp 
( \Delta_2^-({\Lambda_{2}^0}), \ldots, \Delta_{b}^-({\Lambda_{b}^0})) $ $ | \,  
\Delta_1(\lambda_1), S_{1}^0,\Lambda_{2}^0, \Delta_2({\Lambda_{2}^0}), \ldots 
S_{b-1}^0,\Lambda_{b}^0, 
 \Delta_b(\Lambda_b^0)$. So, 
\begin{align*} 
  & (Q_{S_{b}^0}, Q_{\Lambda_{b+1}^0}) \, | \, \Delta_1(\lambda_1), Q_{S_{1}^0}, Q_{\Lambda_{2}^0}, 
  \Delta_2(Q_{\Lambda_{2}^0}), \Delta_2^-(Q_{\Lambda_{2}^0}), 
  \ldots, Q_{S_{b-1}^0}, Q_{\Lambda_{b}^0}, \Delta_b(Q_{\Lambda_{b}^0}), \Delta_{b}^-(Q_{\Lambda_{b}^0})
  \\ 
  & \sim \,  (Q_{S_{b}^0}, Q_{\Lambda_{b+1}^0}) \, | \, \Delta_1(\lambda_1), Q_{S_{1}^0}, Q_{\Lambda_{2}^0}, 
  \Delta_2(Q_{\Lambda_{2}^0}), 
  \ldots, Q_{S_{b-1}^0}, Q_{\Lambda_{b}^0}, \Delta_b(Q_{\Lambda_{b}^0})
  \\ 
  &\sim \, 
  (S_{b}^0, \Lambda_{b+1}^0)  \, | \,  
  \Delta_1(\lambda_1), S_{1}^0,\Lambda_{2}^0, 
  \Delta_2(\Lambda_2^0), \ldots, S_{b-1}^0,\Lambda_{b}^0, 
  \Delta_b(\Lambda_b^0)
  \\ 
  & \sim \,   (S_{b}^0, \Lambda_{b+1}^0)  \, | \,  
  \Delta_1(\lambda_1), S_{1}^0,\Lambda_{2}^0, 
  \Delta_2(\Lambda_2^0), \Delta_2^-(\Lambda_2^0), \ldots, S_{b-1}^0,\Lambda_{b}^0, 
  \Delta_b(\Lambda_b^0), \Delta_b^-(\Lambda_b^0)
\end{align*}
and 
\begin{align*} 
  &  ( \Delta_1(\lambda_1), Q_{S_{1}^0}, Q_{\Lambda_{2}^0}, 
  \Delta_2(Q_{\Lambda_{2}^0}), \Delta_2^-(Q_{\Lambda_{2}^0}), 
  \ldots, Q_{S_{b-1}^0}, Q_{\Lambda_{b}^0}, 
  \Delta_b(Q_{\Lambda_{b}^0}), \Delta_{b}^-(Q_{\Lambda_{b}^0}), 
  Q_{S_{b}^0}, Q_{\Lambda_{b+1}^0})
  \\ 
  & \sim \,   
  (\Delta_1(\lambda_1), S_{1}^0,\Lambda_{2}^0, 
  \Delta_2(\Lambda_2^0), \Delta_2^-(\Lambda_2^0), \ldots, S_{b-1}^0,\Lambda_{b}^0, 
  \Delta_b(\Lambda_b^0), \Delta_b^-(\Lambda_b^0), 
  S_{b}^0, \Lambda_{b+1}^0)
\end{align*}
\item[2(b)]  
Note that $(\Delta_{b+1,0}, \ldots, \Delta_{b+1,K-1}) 
\perp (\Delta_1(\lambda_1), Q_{S_{1}^0}, Q_{\Lambda_{2}^0}, 
\Delta_2(Q_{\Lambda_{2}^0}),  \Delta_2^-(Q_{\Lambda_{2}^0}), 
\ldots, Q_{S_{b}^0}, Q_{\Lambda_{b+1}^0})$ 
 and \\  $(\Delta_{b+1,0}, \ldots, \Delta_{b+1,K-1}) 
 \perp (\Delta_1(\lambda_1), S_{1}^0,\Lambda_{2}^0,
  \Delta_2({\Lambda_{2}^0}),  \Delta_2^-(\Lambda_2^0), 
 \ldots, {S_{b}^0}, {\Lambda_{b+1}^0})$. 
   Hence, 
 for the randomly stopped processes, $\Delta_{b+1}(Q_{\Lambda_{b+1}^0})$, 
 $\Delta_{b+1}^-(Q_{\Lambda_{b+1}^0})$, 
 $\Delta_{b+1}({\Lambda_{b+1}^0})$, $\Delta_{b+1}^-({\Lambda_{b+1}^0})$, we have 
 \begin{align*} 
 & (\Delta_1(\lambda_1), Q_{S_{1}^0}, Q_{\Lambda_{2}^0}, \Delta_2(Q_{\Lambda_{2}^0}), 
 \Delta_2^-(Q_{\Lambda_{2}^0}), 
  \ldots, Q_{S_{b}^0}, Q_{\Lambda_{b+1}^0},\Delta_{b+1}(Q_{\Lambda_{b+1}^0}))
 \\ 
&  \hskip1cm   \sim \, 
  (\Delta_1(\lambda_1), S_{1}^0,\Lambda_{2}^0, \Delta_2({\Lambda_{2}^0}), \Delta_2^-({\Lambda_{2}^0}), 
  \ldots, {S_{b}^0}, {\Lambda_{b+1}^0}, \Delta_{b+1}({\Lambda_{b+1}^0})),  \ \ \mbox{and} 
  \\[1ex]  
  & (\Delta_1(\lambda_1), Q_{S_{1}^0}, Q_{\Lambda_{2}^0}, \Delta_2(Q_{\Lambda_{2}^0}), 
  \Delta_2^-(Q_{\Lambda_{2}^0}), 
  \ldots, Q_{S_{b}^0}, Q_{\Lambda_{b+1}^0}, \Delta_{b+1}^-(Q_{\Lambda_{b+1}^0}))
 \\ 
&  \hskip1cm   \sim \, 
  (\Delta_1(\lambda_1), S_{1}^0,\Lambda_{2}^0, \Delta_2({\Lambda_{2}^0}),   \Delta_2^-({\Lambda_{2}^0}), 
  \ldots, {S_{b}^0}, {\Lambda_{b+1}^0}, \Delta_{b+1}^-({\Lambda_{b+1}^0}))
\end{align*}
 Since $\Delta_{b+1}(Q_{\Lambda_{b+1}^0}) \perp 
 \Delta_{b+1}^-(Q_{\Lambda_{b+1}^0}) | 
 \Delta_1(\lambda_1), Q_{S_{1}^0}, Q_{\Lambda_{2}^0}, \Delta_2(Q_{\Lambda_{2}^0}), 
 \Delta_2^-(Q_{\Lambda_{2}^0}), 
  \ldots, Q_{S_{b}^0}, Q_{\Lambda_{b+1}^0}$ 
  and \\ 
 $ \Delta_{b+1}({\Lambda_{b+1}^0}) \perp  \Delta_{b+1}^-({\Lambda_{b+1}^0}) | 
 \Delta_1(\lambda_1), S_{1}^0,\Lambda_{2}^0, \Delta_2({\Lambda_{2}^0}), \Delta_2^-({\Lambda_{2}^0}), 
 \ldots, {S_{b}^0}, {\Lambda_{b+1}^0}$, 
 it follows that 
 \begin{align*} 
 & (\Delta_1(\lambda_1), Q_{S_{1}^0}, Q_{\Lambda_{2}^0}, 
  \Delta_2(Q_{\Lambda_{2}^0}), \Delta_2^-(Q_{\Lambda_{2}^0}), \ldots, 
  Q_{S_{b}^0}, Q_{\Lambda_{b+1}^0},\Delta_{b+1}(Q_{\Lambda_{b+1}^0}),  
  \Delta_{b+1}^-(Q_{\Lambda_{b+1}^0}) ) 
  \\ 
  & \hskip.3cm \sim 
  (\Delta_1(\lambda_1), S_{1}^0,\Lambda_{2}^0, 
  \Delta_2(\Lambda_2^0), \Delta_2^-(\Lambda_2^0), \ldots, S_{b}^0,\Lambda_{b+1}^0, 
  \Delta_{b+1}({\Lambda_{b+1}^0}), \Delta_{b+1}^-({\Lambda_{b+1}^0}))
 \end{align*} 
 \item[2(c)] Iterate Steps 2(a) and 2(b) until $b = B-1$. 
 \item[3 \hskip2ex]  For $b=B$, we construct 
$Q_{S_{B}^0}$ as $(Q_{S_{b}^0}, Q_{\Lambda_{b+1}^0})$ is constructed in Step 2(a).  
Then, following the argument in Step 2(a) with 
$Q_{S_{B}^0}$ and $S_{B}^0$ replacing 
$(Q_{S_{b}^0}, Q_{\Lambda_{b+1}^0}) $ and 
$(S_{b}^0, \Lambda_{b+1}^0)$, we obtain 
\begin{align*} 
 & (\Delta_1(\lambda_1), Q_{S_{1}^0}, Q_{\Lambda_{2}^0}, \Delta_2(Q_{\Lambda_{2}^0}), \Delta_2^-(Q_{\Lambda_{2}^0}), 
  \ldots, Q_{\Lambda_{B}^0}, \Delta_B(Q_{\Lambda_{B}^0}), \Delta_B^-(Q_{\Lambda_{B}^0}), Q_{S_{B}^0})
 \\ 
  & \,  \sim \, 
  (\Delta_1(\lambda_1), S_{1}^0,\Lambda_{2}^0, \Delta_2({\Lambda_{2}^0}), \Delta_2^-({\Lambda_{2}^0}), 
  \ldots, \Lambda_{B}^0, 
  \Delta_B(\Lambda_B^0), \Delta_B^-(\Lambda_B^0), {S_{B}^0}). 
\end{align*}
\end{enumerate} 

\bigskip 

It follows immediately that 
\begin{align*} 
  & (\Delta_{1,0}({\lambda}_1), \ldots, \Delta_{1,K-1}({\lambda}_1), 
  Q_{S_{1}^0}, Q_{\Lambda_{2}^0},
  \Delta_{2,0}(\bar{\lambda}_2), \ldots, \Delta_{2,K-1}(\bar{\lambda}_2), 
   \ldots, Q_{\Lambda_{B}^0}, 
   \Delta_{B,0}(\bar{\lambda}_B), \ldots, \Delta_{B,K-1}(\bar{\lambda}_B), Q_{S_{B}^0})
  \\ 
   & \,  \sim \, 
   (\Delta_{1,0}({\lambda}_1), \ldots, \Delta_{1,K-1}({\lambda}_1), 
   S_{1}^0,\Lambda_{2}^0, 
   \Delta_{2,0}(\bar{\lambda}_2), \ldots, \Delta_{2,K-1}(\bar{\lambda}_2), 
   \ldots, \Lambda_{B}^0, 
   \Delta_B(\Lambda_B^0), 
   \Delta_{B,0}(\bar{\lambda}_B), \ldots, \Delta_{B,K-1}(\bar{\lambda}_B), {S_{B}^0}). 
 \end{align*}
We can now use this representation under $\theta_0$  in conjunction 
with Le Cam's third lemma to obtain 
an expression for the the limit law of 
$\left( S_{1,n},\Lambda_{2,n},  S_{2,n},\dots, \Lambda_{B,n},S_{B,n} \right)$ 
under any $h$ in terms of the constructed variables.  For any Borel sets 
$A_{1}, \ldots, A_B$ and $C_2, \ldots, C_B$, the limit law 
under $h$, denoted $\mathcal{L}_h$, is 
\begin{align} 
  &  \mathcal{L}_h(A_1 \times C_2  \times A_2  \times \cdots \times C_B \times A_B) 
  \label{eqn:L3}
   \\ 
   & = E\Bigg[ {\bf 1}\{ S_{1}^0\in A_1, \Lambda_{2}^0 \in C_2,  S_{2}^0 \in A_2, 
       \ldots,   \Lambda_{B}^0 \in C_B, S_{B}^0 \in A_B \} 
   \nonumber 
   \\ 
   & \hskip1cm 
   \prod_{k=0}^{K-1} 
    \exp\left[ h^\prime
     \Delta_{1,k}({\lambda}_{1}) 
     - \frac{(\lambda_{1,k+1}-\lambda_{1,k})}{2} h^\prime J_k^{(1)} h \right]  
    \prod_{k=0}^{K-1} 
    \exp\left[ h^\prime \Delta_{2,k}(\bar{\lambda}_{2})  
    - \frac{\bar{\lambda}_{2}}{2} h^\prime J_k^{(2)} h \right] 
    \nonumber 
    \\ 
    & \hskip1cm 
     \cdots  
     \prod_{k=0}^{K-1} 
     \exp\left[ h^\prime \Delta_{B,k}(\bar{\lambda}_{B})  
     - \frac{\bar{\lambda}_{B}}{2} h^\prime J_k^{(B)} h \right]
            \Bigg] 
        \nonumber 
        \\ 
        & = E\Bigg[ {\bf 1}\{ Q_{S_{1}^0} \in A_1, Q_{\Lambda_{2}^0} \in C_2, Q_{S_{2}^0} \in A_2, 
            \ldots,  Q_{\Lambda_{B}^0} \in C_B, Q_{S_{B}^0} \in A_B \} 
            \nonumber 
            \\ 
            & \hskip1cm   \cdot \prod_{k=0}^{K-1} 
         \exp\left[ h^\prime
          \Delta_{1,k}({\lambda}_{1}) 
          - \frac{(\lambda_{1,k+1}-\lambda_{1,k})}{2} h^\prime J_k^{(1)} h \right]  
       \cdot \prod_{k=0}^{K-1} 
       \exp\left[ h^\prime \Delta_{2,k}(\bar{\lambda}_{2})  
       - \frac{\bar{\lambda}_{2}}{2} h^\prime J_k^{(2)} h \right]
          \nonumber 
        \\ 
        & \hskip1cm  
          \cdots  
          \prod_{k=0}^{K-1} 
          \exp\left[ h^\prime \Delta_{B,k}(\bar{\lambda}_{B})  
          - \frac{\bar{\lambda}_{B}}{2} h^\prime J_k^{(B)} h \right] 
             \Bigg]. 
             \nonumber 
\end{align}

We now state a limiting bandit model that will provide the representation 
under any local parameter $h$.  Let $W_{b,k}(\cdot)$ be a Gaussian process 
with independent increments such that 
$W_{b,k}(t) \stackrel{h}{\sim} N\left(t J^{(b)}_k h, t J^{(b)}_k\right)$, and let $U_b$ be 
uniformly distributed on the unit cube as above.  Take all the Gaussian processes 
 $W_{b,k}$ across $b$ and $k$ and all the uniform random variables $U_b$ across  
$b$ to be independent.  
Further, for $k = 0, \ldots, K-1$, define  for $b=1, \ldots, B$, 
\begin{align*} 
  \tilde{Z}_{b,k}(\lambda_b) & =   W_{b,k}(\lambda_{b,k+1}) - 
  W_{b,k}(\lambda_{b,k}) 
  \\ 
 & \sim N\left( (\lambda_{b,k+1}-\lambda_{b,k}) J_k^{(b)} h, 
 (\lambda_{b,k+1}-\lambda_{b,k}) J_k^{(b)} \right), 
\end{align*} 
and for  \ $b=2, \ldots, B$, 
  \begin{align*}  
  \tilde{Z}^-_{b,k}(\lambda_1) & = W_{b,k}(\bar{\lambda}_{b}) 
  - W_{b,k}(\lambda_{b,k+1}) + 
  W_{b,k}(\lambda_{b,k}) 
  \\ 
  & \sim N\left( (\bar{\lambda}_{b}-\lambda_{b,k+1}+\lambda_{b,k}) J_k^{(b)} h, 
  (\bar{\lambda}_{b}-\lambda_{b,k+1}+\lambda_{b,k}) J_k^{(b)} \right). 
\end{align*} 

\begin{enumerate}
  \item 
 For batch 1, $\lambda_1$ is fixed, so we set 
 \begin{align*} 
  {Z}_{1,k} & = \tilde{Z}_{1,k}(\lambda_1) 
 \end{align*} 
  Let $Z_1 =  (Z_{1,0},\dots,Z_{1,K-1})$
   and 
  notice that, under $h=0$, $Z_1 \sim \Delta_1(\lambda_1)$.
  The statistics $T_{S_1}(Z_1,U_1)$ and $T_{\Lambda_2}(Z_1,U_1)$ are constructed as:  
   $$(T_{S_1}, T_{\Lambda_2}) 
   = q_{(S_{1}^0,\Lambda_{2}^0)|\Delta_1(\lambda_1)}(U_1| Z_1).$$ 
   \item For $b=2,\dots,B$, the variables $Z_b$, ${Z}^-_{b,k}$ and statistics $T_{S_b}$, $T_{\Lambda_{b+1}}$ are generated recursively: 
   \begin{align*} 
    {Z}_{b,k} & = \tilde{Z}_{b,k}(T_{\Lambda_{b}});
    \\ 
    {Z}^-_{b,k} & = \tilde{Z}^-_{b,k}(T_{\Lambda_{b}});
    \\ 
    \bar{Z} _{b,k} & = {Z}_{b,k} + {Z}^-_{b,k} \sim N\left( \bar{\lambda}_{b}  J_k^{(b)} h, \bar{\lambda}_{b} J_k^{(b)} \right). 
   \end{align*} 
  Let $Z_b =  (Z_{b,0},\dots,Z_{b,K-1})$, 
  $\bar{Z}_b = (\bar{Z}_{b,0}, \ldots, \bar{Z}_{b,K-1})$.
   Under $h=0$, $(Z_b, \bar{Z}_b) \sim (\Delta_b(\lambda_b), \bar{\Delta}_b)$. 
  
  For $b=2,\dots,B-1$, the statistics $T_{S_b}\left( Z_1,\dots,Z_b,U_1, \ldots, U_{b}\right)$ 
  and $T_{\Lambda_{b+1}}\left( Z_1,\dots,Z_b,U_1, \ldots, U_{b}\right)$ are constructed as follows:
\[ (T_{S_b}, T_{\Lambda_{b+1}}) 
     = 
     q_{(S_{b}^0, \Lambda_{b+1}^0) | 
    \Delta_1(\lambda_1), S_{1}^0,\Lambda_{2}^0, 
    \Delta_2(\Lambda_2^0), \ldots, S_{b-1}^0,\Lambda_{b}^0, 
    \Delta_b(\Lambda_b^0)
    }(U_b| 
    Z_1, T_{S_1}, T_{\Lambda_{2}}, Z_2, 
    \ldots, T_{S_{b-1}}, T_{\Lambda_{b}}, Z_b). 
\] 
For batch $B$,  the statistic $T_{S_B}\left( Z_1,\dots,Z_B,U_1, \ldots, U_{B}\right)$ 
is constructed as:
\[    T_{S_B}
     = 
     q_{S_{B}^0 | 
    \Delta_1(\lambda_1), S_{1}^0,\Lambda_{2}^0, 
    \Delta_2(\Lambda_2^0), \ldots, S_{B-1}^0,\Lambda_{B}^0, 
    \Delta_B(\Lambda_B^0)
    }(U_B| 
    Z_1, T_{S_1}, T_{\Lambda_{2}}, Z_2, 
    \ldots, T_{S_{B-1}}, T_{\Lambda_{B}}, Z_B). 
\] 
\end{enumerate}
The constructed variables $T_{S_b}$ and $T_{\Lambda_{b}}$ 
use uniformly distributed variables on unit cubes $U_1, \ldots, U_B$.  This 
form 
differs from the 
expressions for the corresponding constructed variables in the statement of the theorem, 
which include only a single uniform variable, $U$.  Since a single uniform random variable 
can be used to construct any finite number of independent uniform variables on unit cubes (e.g. 
$U_1, \ldots, U_B$), constructed variables of the form given in the statement of the theorem 
follow immediately from the constructed variables provided in this proof.  

When $h=0$, 
 \[
  (Z_1, T_{S_1}, T_{\Lambda_{2}}, \bar{Z}_2, 
  \ldots, T_{S_{B-1}}, T_{\Lambda_{B}}, \bar{Z}_B, T_{S_B}) 
  \sim 
  (\Delta_1(\lambda_1), S_{1}^0,\Lambda_{2}^0, 
      \bar{\Delta}_2, \ldots, S_{B-1}^0,\Lambda_{B}^0, 
      \bar{\Delta}_B, S_{B}^0) 
 \]
 which will be used below in \eqref{eqn:z0}. 

 Finally, we verify that the joint distribution $(T_{S_1}, T_{\Lambda_{2}},  T_{S_2}, 
 \ldots,  T_{\Lambda_{B}},  T_{S_B})$ matches the limiting distribution 
of $ \left( S_{1,n},\Lambda_{2,n},  S_{2,n},\dots, \Lambda_{B,n},S_{B,n} \right) $ 
under  any $h$ as given in \eqref{eqn:L3}.  
To do this we use the following lemma. 
\begin{lemma} 
  \label{lem:lie3}
  For any (measurable) function $g$, 
  \begin{enumerate} 
    \item[(a)] 
    \[ 
      E_h\left[ 
        g(Z_1, U_1) \right]  = E_0\left[ 
        g(Z_1, U_1)
        \prod_{k=0}^{K-1} 
        \exp\left( h^\prime Z_{1,k} - \frac{(\lambda_{1,k+1}-\lambda_{1,k})}{2} 
        h^\prime J^{(1)}_k h  \right) 
  \right]; 
      \]
      and 
    \item[(b)] 
   for $b \in \{2, \ldots, B\}$, 
  \begin{align*} 
& E_h\left[ 
g(Z_1, \ldots, Z_b, U_1, \ldots, U_b) \, | \, 
Z_1 = z_1, \ldots, Z_{b-1}=z_{b-1}, U_1 = u_1, \ldots, U_{b-1} = u_{b-1}  
\right] 
\\ 
& = 
E_0\bigg[ 
g(Z_1, \ldots, Z_b, U_1, \ldots, U_b) 
\prod_{k=0}^{K-1}  \exp\left( h^\prime \bar{Z}_{b,k} 
- \frac{\bar{\lambda}_b}{2} h^\prime J^{(b)}_k h 
  \right)  
  \\ 
  & \hskip1.5cm 
\, \bigg| \, 
Z_1 = z_1, \ldots, Z_{b-1}=z_{b-1}, U_1 = u_1, \ldots, U_{b-1} = u_{b-1}  
\bigg].  
  \end{align*} 
\end{enumerate} 
\end{lemma}

Lemma~\ref{lem:lie3}(b) is used below in \eqref{eqn:pfB}, 
and Lemma~\ref{lem:lie3}(a) is used in  \eqref{eqn:pf1}.

\begin{align} 
  & {\Pr}_h(T_{S_1} \in A_1, T_{\Lambda_{2}} \in C_2,T_{S_2}\in A_2, 
  \ldots,  T_{\Lambda_{B}}\in C_B, T_{S_B} \in A_B) 
  \nonumber 
  \\[.5ex] 
  & = 
  E_h[ {\bf 1}\{T_{S_1}(Z_1,U_1) \in A_1\} 
  {\bf 1}\{T_{\Lambda_{2}}(Z_1,U_1)  \in C_2\} 
  {\bf 1}\{T_{S_2}(Z_1,Z_2,U_1,U_2) \in A_2\} \cdots 
  \nonumber 
  \\ 
  & \hskip1.1cm  \cdot  
  {\bf 1}\{T_{\Lambda_{B}}(Z_1,\ldots,Z_{B-1},U_1,\ldots, U_{B-1}) \in C_B\} 
  \nonumber 
  \\ 
  & \hskip1.1cm  \cdot 
  E_h( {\bf 1}\{T_{S_B}(Z_1,\ldots,Z_{B},U_1,\ldots, U_{B}) \in A_B\}
  |Z_1,\ldots,Z_{B-1},U_1,\ldots, U_{B-1}) \, ]
  \nonumber 
  \\[.5ex]  
  & = 
  E_h\bigg[ {\bf 1}\{T_{S_1}(Z_1,U_1) \in A_1\} 
  {\bf 1}\{T_{\Lambda_{2}}(Z_1,U_1)  \in C_2\} 
  {\bf 1}\{T_{S_2}(Z_1,Z_2,U_1,U_2) \in A_2\} \cdots 
  \nonumber 
  \\ 
  & \hskip1.1cm  \cdot  
  {\bf 1}\{T_{\Lambda_{B}}(Z_1,\ldots,Z_{B-1},U_1,\ldots, U_{B-1}) \in C_B\} 
  \nonumber 
  \\ 
  & \hskip1.1cm  \cdot 
  E_0\bigg( {\bf 1}\{T_{S_B}(Z_1,\ldots,Z_{B},U_1,\ldots, U_{B}) \in A_B\} 
  \prod_{k=0}^{K-1}  \exp\left( h^\prime \bar{Z}_{B,k} 
- \frac{\bar{\lambda}_B}{2} h^\prime J^{(b)}_k h 
  \right)  
  \nonumber 
  \\ 
  & \hskip2.3cm 
  \bigg| \, Z_1,\ldots,Z_{B-1},U_1,\ldots, U_{B-1}\bigg) \, \bigg] 
\label{eqn:pfB}
  \\[.5ex]  
  & \vdots 
  \nonumber 
  \\[.5ex]  
  & = 
  E_0\bigg[ {\bf 1}\{T_{S_1}(Z_1,U_1) \in A_1\} 
  {\bf 1}\{T_{\Lambda_{2}}(Z_1,U_1)  \in C_2\} 
  \prod_{k=0}^{K-1}  \exp\left( h^\prime {Z}_{1,k} 
  - \frac{(\lambda_{1,k+1}-\lambda_{1,k})}{2} h^\prime J^{(1)}_k h 
    \right) 
  \nonumber 
  \\ 
  & \hskip1.1cm  \cdot  
  E_0\bigg( {\bf 1}\{T_{S_{2}}(Z_1,Z_{2},U_1, U_{2}) \in A_{2}\} 
  {\bf 1}\{T_{\Lambda_{3}}(Z_1, Z_{2},U_1, U_{2}) \in C_3\} 
  \cdot  
  \prod_{k=0}^{K-1}  \exp\left( h^\prime \bar{Z}_{2,k} 
  - \frac{\bar{\lambda}_{2}}{2} h^\prime J^{(2)}_k h 
    \right) 
    \nonumber 
  \\ 
  & \hskip2.1cm \cdots 
  \nonumber   
  \\
  & \hskip2.1cm \cdot  
  E_0\bigg( {\bf 1}\{T_{S_B}(Z_1,\ldots,Z_{B},U_1,\ldots, U_{B}) \in A_B\} 
  \prod_{k=0}^{K-1}  \exp\left( h^\prime \bar{Z}_{B,k} 
- \frac{\bar{\lambda}_B}{2} h^\prime J^{(B)}_k h 
  \right)  
  \nonumber 
  \\ 
  & \hskip3.3cm 
  \bigg| \, Z_1,\ldots,Z_{B-1},U_1,\ldots, U_{B-1}\bigg)  \, 
  \cdots 
  \bigg| \, Z_1,U_1\bigg) 
  \, \bigg] 
  \label{eqn:pf1}
  \\[.5ex]  
& = 
E_0\bigg[ {\bf 1}\{T_{S_1} \in A_1, T_{\Lambda_{2}} \in C_2,  
\cdots 
T_{S_{B-1}} \in A_{B-1}, T_{\Lambda_{B}} \in C_B, T_{S_B}\in A_B\}   
\nonumber 
\\ 
& \hskip1.1cm  \cdot 
\prod_{k=0}^{K-1}  \exp\left( h^\prime {Z}_{1,k} 
- \frac{(\lambda_{1,k+1}-\lambda_{1,k})}{2} h^\prime J^{(1)}_k h 
\right) 
\prod_{k=0}^{K-1}  \exp\left( h^\prime \bar{Z}_{2,k} 
- \frac{\bar{\lambda}_2}{2} h^\prime J^{(2)}_k h 
\right)  
\nonumber 
\\ 
& \hskip1.1cm  
\cdots  
\prod_{k=0}^{K-1}  \exp\left( h^\prime \bar{Z}_{B,k} 
- \frac{\bar{\lambda}_B}{2} h^\prime J^{(B)}_k h 
\right)  
\, \bigg] 
\nonumber 
\\ 
& = E\left[ {\bf 1}\{ Q_{S_{1}^0} \in A_1, Q_{\Lambda_{2}^0} \in C_2, Q_{S_{2}^0} \in A_2, 
    \ldots,  Q_{\Lambda_{B}^0} \in C_B, Q_{S_{B}^0} \in A_B \} 
    \right. 
\nonumber 
\\ 
& \hskip1cm  
 \,\prod_{k=0}^{K-1} 
 \exp\left[ h^\prime
  \Delta_{1,k}(\lambda_{1}) 
  - \frac{(\lambda_{1,k+1}-\lambda_{1,k})}{2} h^\prime J_k^{(1)} h \right]  
 \cdot  \,\prod_{k=0}^{K-1}  \exp\left[ h^\prime
 \Delta_{2,k}(\bar{\lambda}_{2}) 
 - \frac{\bar{\lambda}_{2}}{2} h^\prime J_k^{(2)} h \right]  
 \nonumber 
\\ 
& \hskip1cm  \left. 
 \cdots  
  \prod_{k=0}^{K-1} 
  \exp\left[ h^\prime \Delta_{B,k}(\bar{\lambda}_{B})  
  - \frac{\bar{\lambda}_{B}}{2} h^\prime J_k^{(B)} h \right]
     \right]
     \label{eqn:z0} 
     \\[.5ex]  
     & =      \mathcal{L}_h(A_1 \times C_2  \times A_2  \times \cdots \times C_B \times A_B).   
     \nonumber 
\end{align}
\hfill $\Box$

\bigskip 

\bigskip

\noindent 
\textbf{Proof of Lemma~\ref{lem:lie3}:} \\ 
We focus on proving part (b).  Part (a) follows from equation \eqref{eqn:lemA} below 
with $b=1$.  

Let $\lambda_b = T_{\Lambda_{b}}(z_1, \ldots, z_{b-1}, u_1, \ldots, u_{b-1})$.  And let $d_b$ denote the dimension of 
the unit cube support of $U_b$.  
\begin{align} 
  & E_h\left[ 
  g(Z_1, \ldots, Z_b, U_1, \ldots, U_b) \, | \, 
  Z_1 = z_1, \ldots, Z_{b-1}=z_{b-1}, U_1 = u_1, \ldots, U_{b-1} = u_{b-1}  
  \right] 
  \nonumber 
    \\ 
    & = \int_{[0,1]^{d_b}}  \int \cdots \int 
      g(z_1, \ldots, z_{b-1}, z_b, u_1, \ldots, u_{b-1}, u_b) 
      \nonumber 
      \\ 
      & \hskip2.4cm 
      \cdot 
      \prod_{k=0}^{K-1} \phi\left(z_{b,k} \, | \, (\lambda_{b,k+1}-\lambda_{b,k}) J_k^{(b)} h, 
     (\lambda_{b,k+1}-\lambda_{b,k}) J_k^{(b)} \right) \, 
     dz_{b,0} \cdots dz_{b,K-1} 
      du_b 
      \nonumber 
      \\ 
        & = \int_{[0,1]^{d_b}}  \int \cdots \int 
          g(z_1, \ldots, z_{b-1}, z_b, u_1, \ldots, u_{b-1}, u_b) 
          \cdot 
          \prod_{k=0}^{K-1} 
          \exp\left( h^\prime z_{b,k} - \frac{(\lambda_{b,k+1}-\lambda_{b,k})}{2} 
          h^\prime J^{(b)}_k h  \right) 
          \nonumber 
         \\ 
         & \hskip2.4cm  
         \cdot  \prod_{k=0}^{K-1} 
         \phi\left(z_{b,k} \, | \, 0, 
         (\lambda_{b,k+1}-\lambda_{b,k}) J_k^{(b)} \right) 
         \, 
         dz_{b,0} \cdots dz_{b,K-1} 
          du_b 
          \label{eqn:lemA}
          \\
          & = \int_{[0,1]^{d_b}}  \int \cdots \int 
            g(z_1, \ldots, z_{b-1}, z_b, u_1, \ldots, u_{b-1}, u_b) 
            \cdot 
            \prod_{k=0}^{K-1} 
            \exp\left( h^\prime z_{b,k} - \frac{(\lambda_{b,k+1}-\lambda_{b,k})}{2} 
            h^\prime J^{(b)}_k h  \right) 
            \nonumber 
            \\ 
            & \hskip2.4cm  
            \cdot 
            \prod_{k=0}^{K-1} 
            \exp\left( h^\prime z^-_{b,k}  - \frac{(\bar{\lambda}_b - \lambda_{b,k+1}+\lambda_{b,k})}{2} 
            h^\prime J^{(b)}_k h 
            \right) 
            \nonumber 
            \\ 
            & \hskip2.4cm  
            \cdot 
            \prod_{k=0}^{K-1} \phi\left(z^-_{b,k} \, | \, 0, 
            (\bar{\lambda}_b -\lambda_{b,k+1}+\lambda_{b,k}) J_k^{(b)} \right) 
            \nonumber 
           \\ 
           & \hskip2.4cm  
           \cdot  \prod_{k=0}^{K-1} 
           \phi\left(z_{b,k} \, | \, 0, 
           (\lambda_{b,k+1}-\lambda_{b,k}) J_k^{(b)} \right) 
           \, 
           dz^-_{b,0} \cdots dz^-_{b,K-1} 
           dz_{b,0} \cdots dz_{b,K-1} 
            du_b 
            \nonumber 
            \\ 
          & = \int_{[0,1]^{d_b}}  \int \cdots \int 
            g(z_1, \ldots, z_{b-1}, z_b, u_1, \ldots, u_{b-1}, u_b) 
            \cdot 
            \prod_{k=0}^{K-1} 
            \exp\left( h^\prime (z_{b,k} + z^-_{b,k})- \frac{\bar{\lambda}_b}{2} 
            h^\prime J^{(b)}_k h  \right) 
            \nonumber 
            \\ 
            & \hskip2.4cm  
            \cdot 
            \prod_{k=0}^{K-1} \phi\left(z^-_{b,k} \, | \, 0, 
            (\bar{\lambda}_b -\lambda_{b,k+1}+\lambda_{b,k}) J_k^{(b)} \right) 
            \nonumber 
           \\  
           & \hskip2.4cm  
           \cdot  \prod_{k=0}^{K-1} 
           \phi\left(z_{b,k} \, | \, 0, 
           (\lambda_{b,k+1}-\lambda_{b,k}) J_k^{(b)} \right) 
           \, 
           dz^-_{b,0} \cdots dz^-_{b,K-1} 
           dz_{b,0} \cdots dz_{b,K-1} 
            du_b  
            \nonumber 
    \\ 
  & = 
  E_0\bigg[ 
  g(Z_1, \ldots, Z_b, U_1, \ldots, U_b) 
  \prod_{k=0}^{K-1}  \exp\left( h^\prime \bar{Z}_{b,k} 
  - \frac{\bar{\lambda}_b}{2} h^\prime J^{(b)}_k h 
    \right)  
    \nonumber 
    \\ 
    & \hskip1.5cm 
  \, \bigg| \, 
  Z_1 = z_1, \ldots, Z_{b-1}=z_{b-1}, U_1 = u_1, \ldots, U_{b-1} = u_{b-1}  
  \bigg].  
  \nonumber 
    \end{align}
    \hfill $\Box$

\vspace{1cm}

\section{Alternative Form of Proposition \ref{thm:artfixed}}\label{app:altprop}

We can state Proposition \ref{thm:artfixed} in an alternative way. Let 
  \[ V(\lambda) = W(\lambda) - \frac{\lambda}{\lambda_B} W(\lambda_B), \quad 0 \le \lambda \le \lambda_B, \]
  for $W(\lambda)$ in the construction of $Z(\lambda)$. 
  Then there is a collection of functions $\tilde{T}_1,\dots,\tilde{T}_B$ such that for every $h$, 
  \[ 
    \begin{pmatrix}
      \tilde{T}_1( Z(\lambda_1), U) \\
      \tilde{T}_2( Z(\lambda_2), V(\lambda_1), V(\lambda_2), U) \\
      \vdots \\
      \tilde{T}_{B-1} ( Z(\lambda_{B-1}), V(\lambda_1),\dots,V(\lambda_{B-1}),U) \\ 
      \tilde{T}_D( Z(\lambda_B), V(\lambda_1), V(\lambda_2),\dots,V(\lambda_{B-1}),U)
    \end{pmatrix} \sim \mathcal{L}_h. 
    \] 
The process $V(\lambda)$ is a standard Brownian bridge. 
It does not depend on $h$, and is independent of $Z(\lambda_B)$. However, the processes $Z(\lambda)$ and $V(\lambda)$ are 
dependent over $\lambda < \lambda_B$.  Their inclusion in the functions $\tilde{T}_b$ suffices to preserve the joint dependence structure in the laws $\mathcal{L}_h$.
A related construction was used in \citet{hira:wrig:2017}.

\section{Asymptotic Representation of Thompson Sampling}\label{sec:TSlimit}

We seek the limit of the Thompson sampling rule $\Lambda_{2,n}$ in the setup of Section \ref{sec:experiment}. 
The parameter is $\theta = (\beta_0^\prime, \beta_1^\prime)^\prime \in \Theta \subset \mathbb{R}^{2d_\beta}$ and we consider local alternatives $\theta_n(h) = \theta_0 + h/\sqrt{n}$ for $h = (h_0^\prime,h_1^\prime)^\prime \in \mathbb{R}^{2d_\beta}$. 
The data in Batch 1 consist of $\{ D_{1i},Y_{1i}\}_{i=1}^n$. 
Let the joint distribution of the Batch 1 data under parameter $\theta_n$ be denoted as $P_{\theta_n}^n$.  
Under Assumption \ref{asm:dqm}, and applying Theorem \ref{thm:bandit}, 
the corresponding limit experiment is $Z_1 \sim N(h, J_0^{-1})$ where 
$J_0 = \text{diag}\left\{ \frac{1}{2}\Sigma_0, \frac{1}{2}\Sigma_1 \right\}$. In the setting we consider here, both $\Sigma_0$ and $\Sigma_1$ are nonsingular matrices. 

The following result shows that, under relatively mild conditions, the Thompson sampling rule converges to the posterior rule under a flat prior in the limit experiment for the Batch 1 data. 
\begin{proposition}\label{prop:bvm}
Suppose that Assumption \ref{asm:dqm} holds and that 
    \begin{enumerate}[label=(\alph*)]
        \item The prior distribution on $\theta = (\beta_0,\beta_1)$ is absolutely continuous in a neighborhood of $\theta_0$ with continuous positive density $\pi(\beta_0,\beta_1)$. 
        \item There exists a sequence of tests $\phi_n$ based on the Batch 1 data, such that 
        \[ E_{\theta_0}[\phi_n] \to 0 \quad \text{ and } \quad \sup_{\|\theta-\theta_0\|\ge 0}  E_{\theta}[1-\phi_n] \to 0.\]
        \item The function $g(\cdot)$ is continuously differentiable at $\beta_{0,0}$ and $\beta_{0,1}$. 
    \end{enumerate}
Then, for every $h=(h_0,h_1) \in \mathbb{R}^{2d_{\beta}}$, the Thompson sampling rule satisfies 
\[ \Lambda_{2,n} \stackrel{h}{\leadsto} \Lambda_{2}(Z_1) = \text{Pr}\left( \dot{g}_1'h_1 < \dot{g}_0'h_0 \mid Z_1 \right) = 
  \Phi\left( \frac{\dot{g}_0'Z_{1,0}-\dot{g}_1'Z_{1,1} }{\sqrt{2 \dot{g}_0'\Sigma_0^{-1}\dot{g}_0+ 2 \dot{g}_1'\Sigma_1^{-1}\dot{g}_1}}\right),\]
where $Z_1 = (Z_{1,0},Z_{1,1}) \sim N(h, J_0^{-1})$.
\end{proposition}

\bigskip 
{\bf Proof:}

The posterior after Batch 1 is 
\[ \pi(\beta_0,\beta_1\mid \text{Batch 1}) \propto \pi(\beta_0,\beta_1) \cdot \prod_{i=1}^n p_{\beta_1}(Y_{1,i})^{D_{1,i}} \cdot p_{\beta_0}(Y_{1,i})^{(1-D_{1,i})}. \]
Fix $\theta_0  = (\beta_{0,0},\beta_{0,1})^\prime$ and reparametrize the random quantity $\theta$ as ${h}=({h}_0,{h}_1)^\prime = \sqrt{n}(\theta - \theta_0)$. 
Let $P_{{H}|\text{Batch 1}}$ denote the posterior distribution for ${h}$ given the data in Batch 1.

We first consider the limit of $\Lambda_{2,n}$ under $\theta_0 = (\beta_0,\beta_1)$. 
By conditions (a) and (b), the Bernstein-von Mises Theorem \citep[Theorem 10.1]{vdv:1998} gives
\begin{equation} 
\label{eqn:bvm2} 
\left\| P_{{H}|\text{Batch 1}} - N\left(\Delta_{n,\theta_0}, J_0^{-1}\right)\right\|_{TV} \stackrel{P_{\theta_0}^n}{\rightarrow} 0,
\end{equation}
where $\| \cdot \|_{TV}$ is the total variation norm for probability measures, and 
\[ \Delta_{n,\theta_0} = \frac{1}{\sqrt{n}} \sumin J_0^{-1} s(Y_{1i},D_{1i}),\]
where $s(\cdot)$ is the score function in Assumption \ref{asm:dqm}.
The (finite-sample) Thompson sampling rule is 
\begin{align*}
    \Lambda_{2,n} &= \text{Pr}\left( g(\beta_0)>g(\beta_1)  \mid \text{Batch 1} \right) \\
    &= \int \int {\bf 1}(g(\beta_0)>g(\beta_1)) \pi(\beta_0,\beta_1 \mid \text{Batch 1}) d\beta_0 d\beta_1 \\ 
    &= \int {\bf 1}(h \in B_{n}) dP_{H\mid \text{Batch 1}}(h),
\end{align*}
where $B_{n} = \left\{ h = (h_0,h_1)^\prime \in \mathbb{R}^{2d_{\beta}} : g(\beta_{0,0} + h_0/\sqrt{n})> g(\beta_{0,1} + h_1/\sqrt{n}) \right\}$.

Then the total variation convergence in \eqref{eqn:bvm2} implies 
that 
\[ \left| \Lambda_{2,n} - N\left(B_n | \Delta_{n,\theta_0},J_0^{-1}\right) \right| \stackrel{P_{\theta_0}^n}{\rightarrow} 0 \quad \text{ as } n \to \infty. \]

Moreover, by Le Cam's First Lemma \citep[Lemma 6.4(iv)]{vdv:1998} this convergence in probability also holds for contiguous local parameter sequences, so that for every local parameter $h$, 
\[ \left| \Lambda_{2,n} - N\left(B_n | \Delta_{n,\theta_0},J_0^{-1}\right) \right| \stackrel{P_{\theta_0+h/\sqrt{n}}^n}{\rightarrow} 0 \quad \text{ as } n \to \infty. \]

Next, we consider the limit of $ N\left(B_n | \Delta_{n,\theta_0},J_0^{-1}\right)$ under $P_{\theta_0+h/\sqrt{n}}^n$.  
By the proof of Theorem \ref{thm:bandit}, we have 
$\Delta_{n,\theta_0} \stackrel{h}{\leadsto}  Z_1$, where 
$Z_1 \sim N(h, J_0^{-1})$.  
For $n\ge 1$, define the maps 
\[ m \mapsto f_n(m) := \int {\bf 1}(h \in B_n) dN(h|m,J_0^{-1}).\] 

Then we can rewrite $ N\left(B_n | \Delta_{n,\theta_0},J_0^{-1}\right)$ as $f_n(\Delta_{n,\theta_0})$. 
Lemma \ref{lem:extendedcm} below shows that this sequence of maps satisfies the following condition: for every converging sequence $m_n \to m$, 
\[ 
 f_n(m_n) \to f_\infty(m),
\]
where 
\[ f_{\infty}(m) = \int {\bf 1}(h \in B_{\infty}) dN(h|m,J_0^{-1}) 
\] 
and 
\[ B_{\infty} = \left\{ h \in \mathbb{R}^{2d_{\beta}} : \dot{g}(\beta_{0,0})^\prime h_0 > \dot{g}(\beta_{0,1})^\prime h_1  \right\}.\]
Then, by the extended continuous 
mapping theorem \citep[Theorem 1.11.1]{vdv:well:1996}, 
we have that  $f_n(\Delta_{n,\theta_0}) \stackrel{h}{\leadsto} f_\infty(Z_1)$.  It follows that 
\[ 
\Lambda_{2,n} \, \stackrel{h}{\leadsto} \, 
\int {\bf 1}(h \in B_{\infty}) dN(h|Z_1,J_0^{-1}) 
\, = 
\, 
\Phi\left( \frac{\dot{g}(\beta_{0,0})'Z_{1,0}-\dot{g}(\beta_{0,1})'Z_{1,1} }{\sqrt{2 \dot{g}(\beta_{0,0})'\Sigma_0^{-1}\dot{g}(\beta_{0,0})+ 2 \dot{g}(\beta_{0,1})'\Sigma_1^{-1}\dot{g}(\beta_{0,1})}}\right). 
\]
$\Box$

\bigskip 
\bigskip 
\noindent 
\rule{\textwidth}{.1ex}
\bigskip

\bigskip 

\begin{lemma}\label{lem:extendedcm}
Under the conditions of Proposition \ref{prop:bvm}, the map $f_n$ defined in the proof of that Proposition satisfies 
$f_n(m_n) \to f_{\infty}(m)$ for every sequence $m_n\to m$. 
\end{lemma}

{\bf Proof:} \\ 
Given $\varepsilon >0$, we show there exists $n_0$ such that for $n \ge n_0$,  
$|f_n(m_n) \to f_{\infty}(m)| < \varepsilon$.

Take any $0 < \delta < \infty$.  There exists $n_1$ such that for $n \ge n_1$, $|m_n-m| < \delta$.  

There exists a compact set $A \subset \mathbb{R}^{2d_{\beta}}$ such that 
\[ 
\sup_{\tilde{m}: |\tilde{m}-m| < \delta} 
\int {\bf 1}({h}  \in A^c) dN({h} | \tilde{m},J_0^{-1}) \ < \ \frac{\varepsilon}{5}. 
\]

Define
\begin{align*} 
\nu_n = \sup_{\lambda_0\in[0,1], \lambda_1\in[0,1], {h} \in A} & \Bigg(
\left| \left[ 
\dot{g}\left(\beta_{0,0} + \lambda_0\frac{h_0}{\sqrt{n}}\right) - \dot{g}(\beta_{0,0})\right]^\prime 
h_0
\right| 
\, + \,  
\left| \left[ 
\dot{g}\left(\beta_{0,1} + \lambda_1\frac{h_1}{\sqrt{n}}\right) - \dot{g}(\beta_{0,1})\right]^\prime 
h_1
\right| \Bigg) 
\end{align*} 
By continuous differentiability, $\nu_n \to 0$ as $n \to \infty$.  

Recall that 
\[ B_{\infty} = \left\{  {h}  : \dot{g}(\beta_{0,0})^\prime h_0 - \dot{g}(\beta_{0,1})^\prime h_1  >0 \right\}.\]
By the mean value theorem, for some $\lambda_0$ and $\lambda_1$ (functions of $h_0$ and $h_1$), 
\begin{align*} 
B_{n} = &  \left\{ {h} : g(\beta_{0,0} + h_0/\sqrt{n})> g(\beta_{0,1} + h_1/\sqrt{n}) \right\} 
\\ 
= &  \left\{ {h}  : \sqrt{n} \left[  g\left(\beta_{0,0} + \frac{ h_0}{ \sqrt{n}} \right) - g(\beta_{0,0}) \right]  - 
\sqrt{n} \left[ g\left(\beta_{0,1} + \frac{ h_1}{ \sqrt{n}} \right)  - g(\beta_{0,1} ) 
\right] > 0
 \right\} 
 \\ 
= &  \left\{  {h}  :  \left[  
\dot{g}\left(\beta_{0,0} + \lambda_0\frac{h_0}{\sqrt{n}}\right)^\prime  h_0 
 \right]  - 
 \left[ \dot{g}\left(\beta_{0,1} + \lambda_1\frac{h_1}{\sqrt{n}}\right)^\prime h_1
\right] > 0
 \right\} 
 \\ 
= &  \left\{{h}  :  {
\left[ 
\dot{g}\left(\beta_{0,0} + \lambda_0\frac{h_0}{\sqrt{n}}\right) - \dot{g}(\beta_{0,0})\right]^\prime 
h_0  
-
 \left[ 
\dot{g}\left(\beta_{0,1} + \lambda_1\frac{h_1}{\sqrt{n}}\right) - \dot{g}(\beta_{0,1})\right]^\prime 
h_1
+ \dot{g}(\beta_{0,0})^\prime h_0  - 
\dot{g}(\beta_{0,1})^\prime 
h_1
>  0  
}.
\right\} 
\end{align*} 
So, 
\begin{align*} 
 B_n^c  & \cap B_{\infty} \cap A 
\\ 
 = &
\left\{ {h} \in A :  {
\left[ 
\dot{g}\left(\beta_{0,0} + \lambda_0\frac{h_0}{\sqrt{n}}\right) - \dot{g}(\beta_{0,0})\right]^\prime 
h_0 
-  
 \left[ 
\dot{g}\left(\beta_{0,1} + \lambda_1\frac{h_1}{\sqrt{n}}\right) - \dot{g}(\beta_{0,1})\right]^\prime 
h_1 
+ \dot{g}(\beta_{0,0})^\prime h_0  - 
\dot{g}(\beta_{0,1})^\prime 
h_1
\le   0  
} 
\right. 
\\ 
& \left. \ \mbox{and} \ 
\dot{g}(\beta_{0,0})^\prime h_0  - \dot{g}(\beta_{0,1})^\prime h_1  >0 
\right\} 
\\ 
 = &
\left\{ {h} \in A :  {
 \dot{g}(\beta_{0,0})^\prime h_0 - 
\dot{g}(\beta_{0,1})^\prime 
h_1 
\le   - \left( 
\left[ 
\dot{g}\left(\beta_{0,0} + \lambda_0\frac{h_0}{\sqrt{n}}\right) - \dot{g}(\beta_{0,0})\right]^\prime 
h_0 
-  
 \left[ 
\dot{g}\left(\beta_{0,1} + \lambda_1\frac{h_1}{\sqrt{n}}\right) - \dot{g}(\beta_{0,1})\right]^\prime 
h_1 \right) 
} 
\right. 
\\ 
& \left. \ \mbox{and} \ 
\dot{g}(\beta_{0,0})^\prime h_0  - \dot{g}(\beta_{0,1})^\prime h_1 >0 
\right\} 
\\ 
\subset  &
\left\{  {h} \in A : 
 \dot{g}(\beta_{0,0})^\prime h_0- 
\dot{g}(\beta_{0,1})^\prime 
h_1 
\le   \nu_n 
 \ \mbox{and} \ 
\dot{g}(\beta_{0,0})^\prime h_0  - \dot{g}(\beta_{0,1})^\prime h_1  >0 
\right\}  
\\ 
=  &
\left\{  {h} \in A : 0< 
 \dot{g}(\beta_{0,0})^\prime h_0  - 
\dot{g}(\beta_{0,1})^\prime 
h_1
\le   \nu_n 
\right\}. 
\end{align*} 
Similarly, 
\begin{align*} 
 B_n  & \cap B_{\infty}^c \cap A 
\subset  
\left\{ {h} \in A : \ -  \nu_n < 
 \dot{g}(\beta_{0,0})^\prime h_0 - 
\dot{g}(\beta_{0,1})^\prime 
h_1
\le 0
\right\}.
\end{align*} 
So, $ B_n^c   \cap B_{\infty} \cap A $ and $ B_n   \cap B_{\infty}^c \cap A $ are bounded sets with area converging to zero as $n \to 
\infty$.  Hence, 
\[ 
\sup_{\tilde{m}: |\tilde{m}-m| < \delta} 
\int {\bf 1}({h}  \in B_n^c   \cap B_{\infty} \cap A) dN({h}  | \tilde{m},J_0^{-1}) \to 0
\]
and 
\[ 
\sup_{\tilde{m}: |\tilde{m}-m| < \delta} 
\int {\bf 1}({h}  \in  B_n   \cap B_{\infty}^c \cap A ) dN({h}  | \tilde{m},J_0^{-1}) \to 0.
\]
So there exists $n_2$ such that for $n \ge n_2$, 
\[ 
\sup_{\tilde{m}: |\tilde{m}-m| < \delta} 
\int {\bf 1}({h}  \in B_n^c   \cap B_{\infty} \cap A) dN({h}  | \tilde{m},J_0^{-1})  \ < \ \frac{\varepsilon}{5}
\]
and 
\[ 
\sup_{\tilde{m}: |\tilde{m}-m| < \delta} 
\int {\bf 1}({h}  \in  B_n   \cap B_{\infty}^c \cap A ) dN({h}  | \tilde{m},J_0^{-1})  \ < \ \frac{\varepsilon}{5}.  
\]

By convergence in distribution ($N(m_n, J_0^{-1}) \cvgd N(m, J_0^{-1})$), 
\[ 
 \int {\bf 1}({h} \in B_{\infty} \cap A ) dN({h}|m_n ,J_0^{-1}) 
\to  \int {\bf 1}({h} \in B_{\infty}\cap A) dN({h}|m,J_0^{-1}), 
\] 
so there exists $n_3$ such that for $n \ge n_3$, 
\[ 
\left|  \int {\bf 1}({h} \in B_{\infty} \cap A ) dN({h}|m_n ,J_0^{-1}) 
  -   \int {\bf 1}({h} \in B_{\infty}\cap A) dN({h}|m,J_0^{-1})  \right| < \frac{\varepsilon}{5}.
  \]   
  Let $n_0 = \max\{ n_1, n_2, n_3\}$.   For $n \ge n_0$, 
\begin{align*} 
 |f_n(m_n) & - f_{\infty}(m)| 
  \\ 
  =& \left| \int {\bf 1}({h} \in B_n) dN({h}|m_n ,J_0^{-1}) 
  -   \int {\bf 1}({h} \in B_{\infty}) dN({h}|m,J_0^{-1})  \right| 
   \\ 
  = 
  & \left| \int {\bf 1}({h} \in B_n \cap A^c ) dN({h}|m_n ,J_0^{-1}) 
  -   \int {\bf 1}({h} \in B_{\infty}\cap A^c) dN({h}|m,J_0^{-1})  \right. 
  \\ 
  & +  \int {\bf 1}({h} \in B_n \cap A ) dN({h}|m_n ,J_0^{-1})  - 
  \int {\bf 1}({h} \in B_{\infty} \cap A ) dN({h}|m_n ,J_0^{-1}) 
  \\ 
  & \left. 
  +   \int {\bf 1}({h} \in B_{\infty} \cap A ) dN({h}|m_n ,J_0^{-1}) 
  -   \int {\bf 1}({h} \in B_{\infty}\cap A) dN({h}|m,J_0^{-1})  \right| 
     \\ 
  = 
  & \left| \int {\bf 1}({h} \in B_n \cap A^c ) dN({h}|m_n ,J_0^{-1}) 
  -   \int {\bf 1}({h} \in B_{\infty}\cap A^c) dN({h}|m,J_0^{-1})  \right. 
  \\ 
  & +  \int {\bf 1}({h} \in B_n \cap B_{\infty} \cap A ) dN({h}|m_n ,J_0^{-1})  +  \int {\bf 1}({h} \in B_n  \cap B_{\infty}^c \cap A ) dN({h}|m_n ,J_0^{-1})  
    \\ 
  & - 
  \int {\bf 1}({h} \in B_n \cap B_{\infty} \cap A ) dN({h}|m_n ,J_0^{-1}) -  \int {\bf 1}({h} \in B_n^c \cap B_{\infty} \cap A ) dN({h}|m_n ,J_0^{-1}) 
  \\ 
  & \left. 
  +   \int {\bf 1}({h} \in B_{\infty} \cap A ) dN({h}|m_n ,J_0^{-1}) 
  -   \int {\bf 1}({h} \in B_{\infty}\cap A) dN({h}|m,J_0^{-1})  \right| 
  \\ 
    = 
  & \left| \int {\bf 1}({h} \in B_n \cap A^c ) dN({h}|m_n ,J_0^{-1}) 
  -   \int {\bf 1}({h} \in B_{\infty}\cap A^c) dN({h}|m,J_0^{-1})  \right. 
  \\ 
  & +   \int {\bf 1}({h} \in B_n  \cap B_{\infty}^c \cap A ) dN({h}|m_n ,J_0^{-1})  
 -  \int {\bf 1}({h} \in B_n^c \cap B_{\infty} \cap A ) dN({h}|m_n ,J_0^{-1}) 
  \\ 
  & \left. 
  +   \int {\bf 1}({h} \in B_{\infty} \cap A ) dN({h}|m_n ,J_0^{-1}) 
  -   \int {\bf 1}({h} \in B_{\infty}\cap A) dN({h}|m,J_0^{-1})  \right| 
    \\ 
    \le 
  & \left| \int {\bf 1}({h} \in B_n \cap A^c ) dN({h}|m_n ,J_0^{-1}) \right| 
  + \left|   \int {\bf 1}({h} \in B_{\infty}\cap A^c) dN({h}|m,J_0^{-1}) \right| 
  \\ 
  & + \left|   \int {\bf 1}({h} \in B_n  \cap B_{\infty}^c \cap A ) dN({h}|m_n ,J_0^{-1})  \right| 
+ \left|  \int {\bf 1}({h} \in B_n^c \cap B_{\infty} \cap A ) dN({h}|m_n ,J_0^{-1}) \right| 
  \\ 
  & + \left| 
     \int {\bf 1}({h} \in B_{\infty} \cap A ) dN({h}|m_n ,J_0^{-1}) 
  -   \int {\bf 1}({h} \in B_{\infty}\cap A) dN({h}|m,J_0^{-1})  \right| 
  \\ 
  \le & 
  \sup_{\tilde{m}: |\tilde{m}-m| < \delta} 
\int {\bf 1}({h} \in A^c) dN({h}|\tilde{m},J_0^{-1})  + \sup_{\tilde{m}: |\tilde{m}-m| < \delta} 
\int {\bf 1}({h} \in A^c) dN({h}|\tilde{m},J_0^{-1}) 
\\ 
& + 
\sup_{\tilde{m}: |\tilde{m}-m| < \delta} 
\int {\bf 1}({h} \in B_n^c   \cap B_{\infty} \cap A) dN({h}|\tilde{m},J_0^{-1}) + \sup_{\tilde{m}: |\tilde{m}-m| < \delta} 
\int {\bf 1}({h} \in  B_n   \cap B_{\infty}^c \cap A ) dN({h}|\tilde{m},J_0^{-1}) 
  \\ 
  & + \left| 
     \int {\bf 1}({h} \in B_{\infty} \cap A ) dN({h}|m_n ,J_0^{-1}) 
  -   \int {\bf 1}({h} \in B_{\infty}\cap A) dN({h}|m,J_0^{-1})  \right|  
   \\ 
< & 
  \frac{\varepsilon}{5} +  \frac{\varepsilon}{5} +  \frac{\varepsilon}{5} +  \frac{\varepsilon}{5} +  \frac{\varepsilon}{5}  = \varepsilon.
 \end{align*} 
 $\Box$
 
\section{Batchwise Difference in Means Test}\label{app:bdm}

For each batch $b$, let 
\[
 \hat{S}_{b,n} = \hat{\beta}_{b,1} - \hat{\beta}_{b,0}
 = 
 \frac{ \sum_{i=1}^n Y_{b,i} D_{b,i}}{ \sum_{i=1}^n  D_{b,i} } - 
 \frac{  \sum_{i=1}^n Y_{b,i} (1-D_{b,i})}{  \sum_{i=1}^n (1- D_{b,i}) }
 \] 
denote the difference in means between the treatment and control arms. 
We can construct a finite-sample batchwise difference in means statistic as 
\[ 
{S}_n^{BDM} = \sum_{b=1}^B \omega_{b,n} 
\hat{S}_{b,n} 
\]
where 
\begin{align*}
\omega_{b,n} & = \frac{1}{\sqrt{B}} 
\cdot 
\frac{ 1 }{\sqrt{ \frac{\hat{\sigma}_{b,1}^2}{N_{b,1}} +  \frac{\hat{\sigma}_{b,0}^2}{N_{b,0 } }}}
\\ 
\hat{\sigma}_{b,t}^2 & = \frac{1}{N_{b,t}}  \sum_{i=1}^n 
(Y_{b,i} - \hat{\beta}_{b,t})^2 {\bf 1}\{ D_{b,i} = t \} 
\\ 
N_{b,t} & = \sum_{i=1}^n 
{\bf 1}\{ D_{b,i} = t \}. 
\end{align*} 
It is straightforward to show that $S_n^{BDM} \leadsto T^{BDM}$ with the representation given in Section \ref{subsec:powerenv} under every local parameter $h$.

\section{Additional Figures}\label{app:fig}

\begin{figure}[ht]
 \caption{PDM Statistic Null Distribution} 
 \label{fig:appPDM}
     \begin{subfigure}[]{0.5\textwidth}
             \caption{Two Batches ($B=2$): $\sigma_0=\frac{1}{2}$, $\sigma_1=\frac{1}{2}$}
       \includegraphics[width=.95\linewidth]{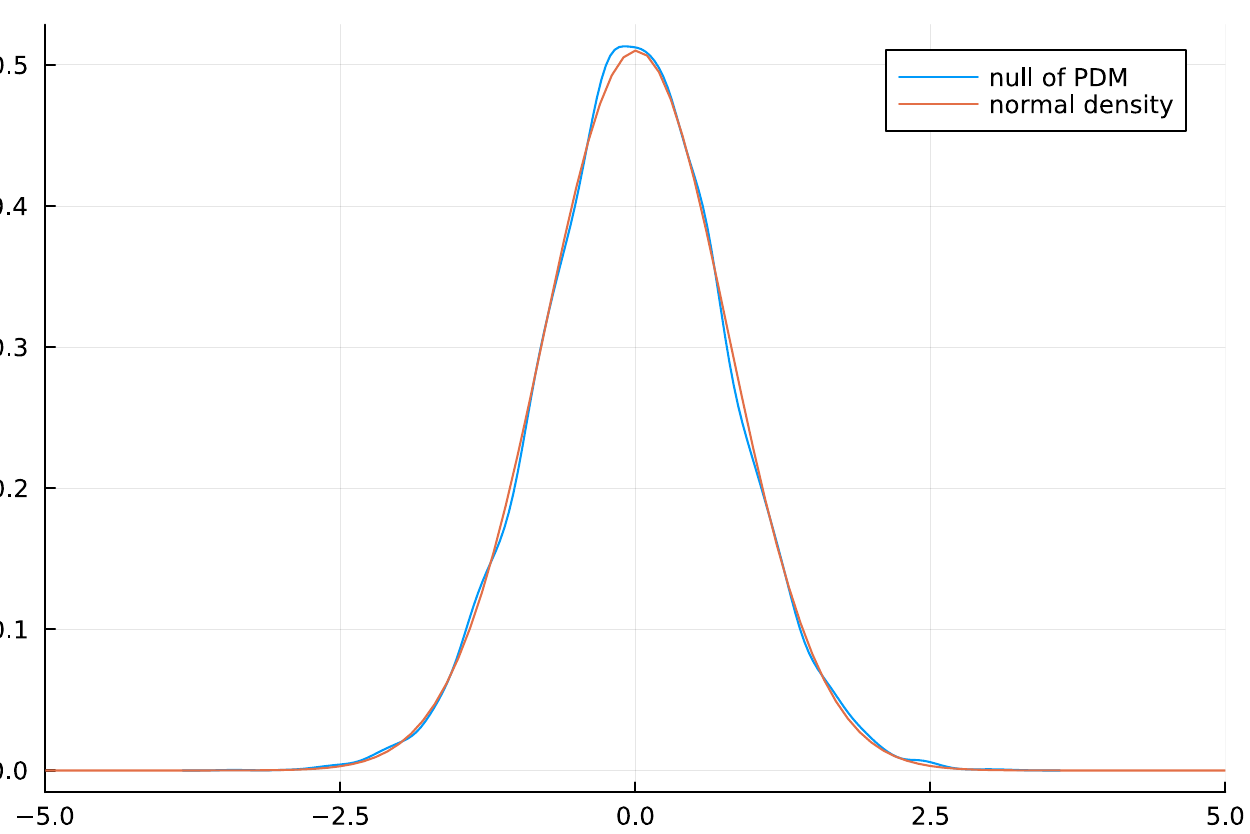}
    \end{subfigure} 
         \begin{subfigure}[]{0.5\textwidth}
             \caption{Five Batches ($B=5$): $\sigma_0=\frac{1}{2}$, $\sigma_1=\frac{1}{2}$ }
       \includegraphics[width=.95\linewidth]{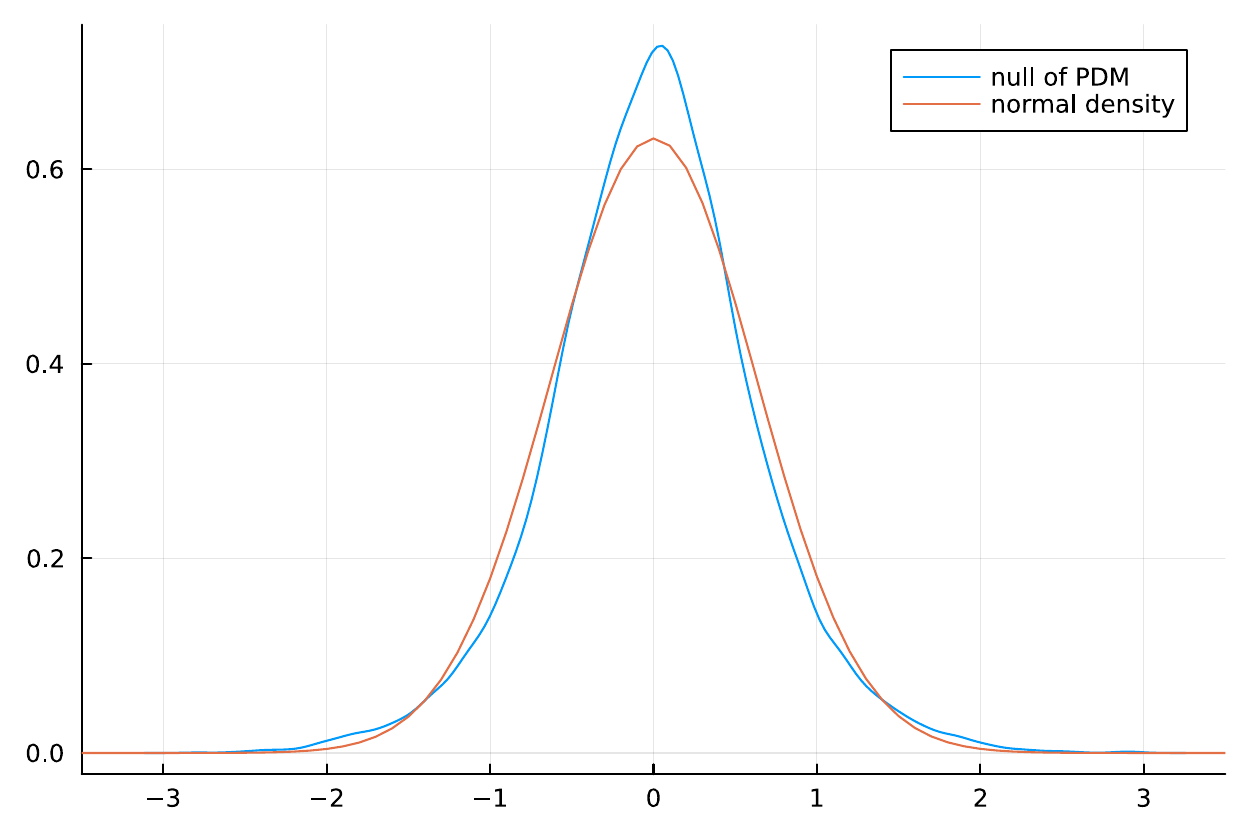}
    \end{subfigure} 
    \\ 
    \medskip 
    
         \begin{subfigure}[]{0.5\textwidth}
             \caption{Two Batches ($B=2$): $\sigma_0=\frac{1}{2}$, $\sigma_1=\frac{1}{16}$}
       \includegraphics[width=.95\linewidth]{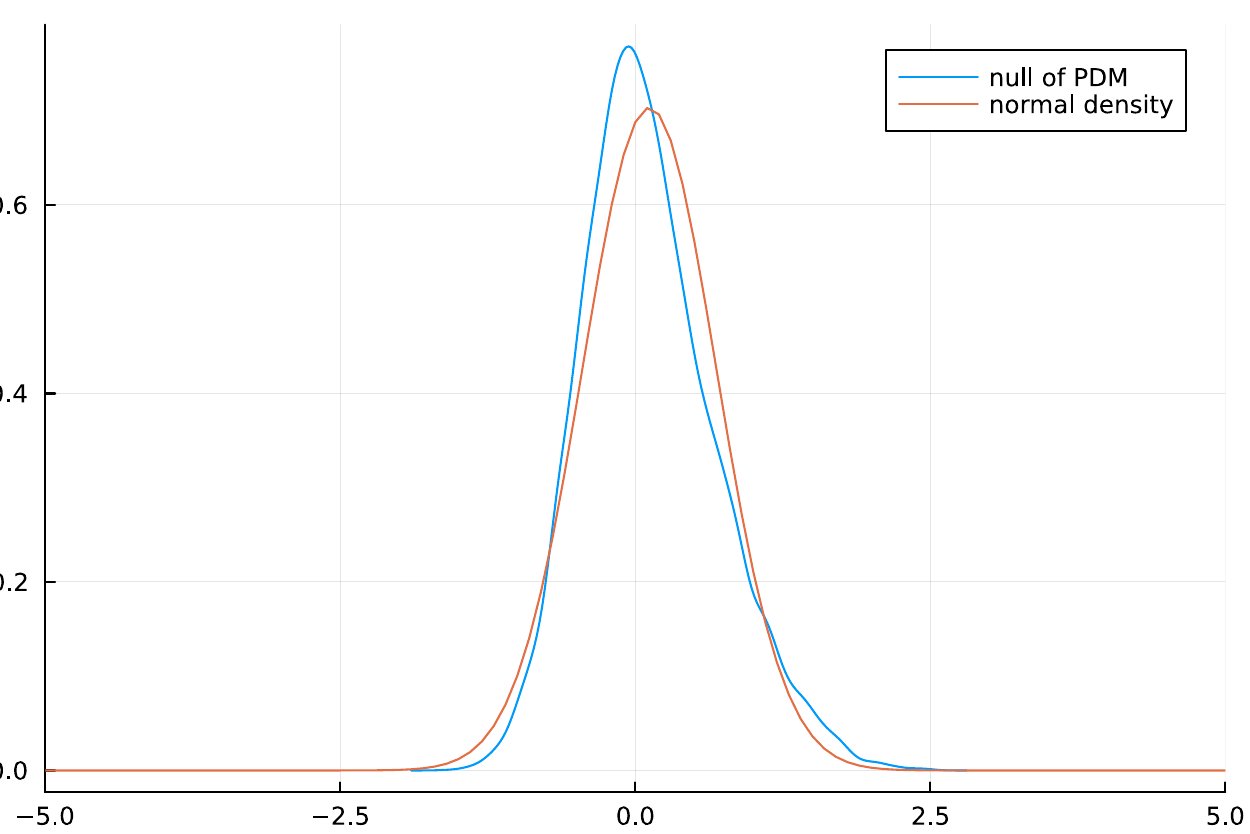}
    \end{subfigure}%
         \begin{subfigure}[]{0.5\textwidth}
             \caption{Five Batches ($B=5$): $\sigma_0=\frac{1}{2}$, $\sigma_1=\frac{1}{16}$ }
       \includegraphics[width=.95\linewidth]{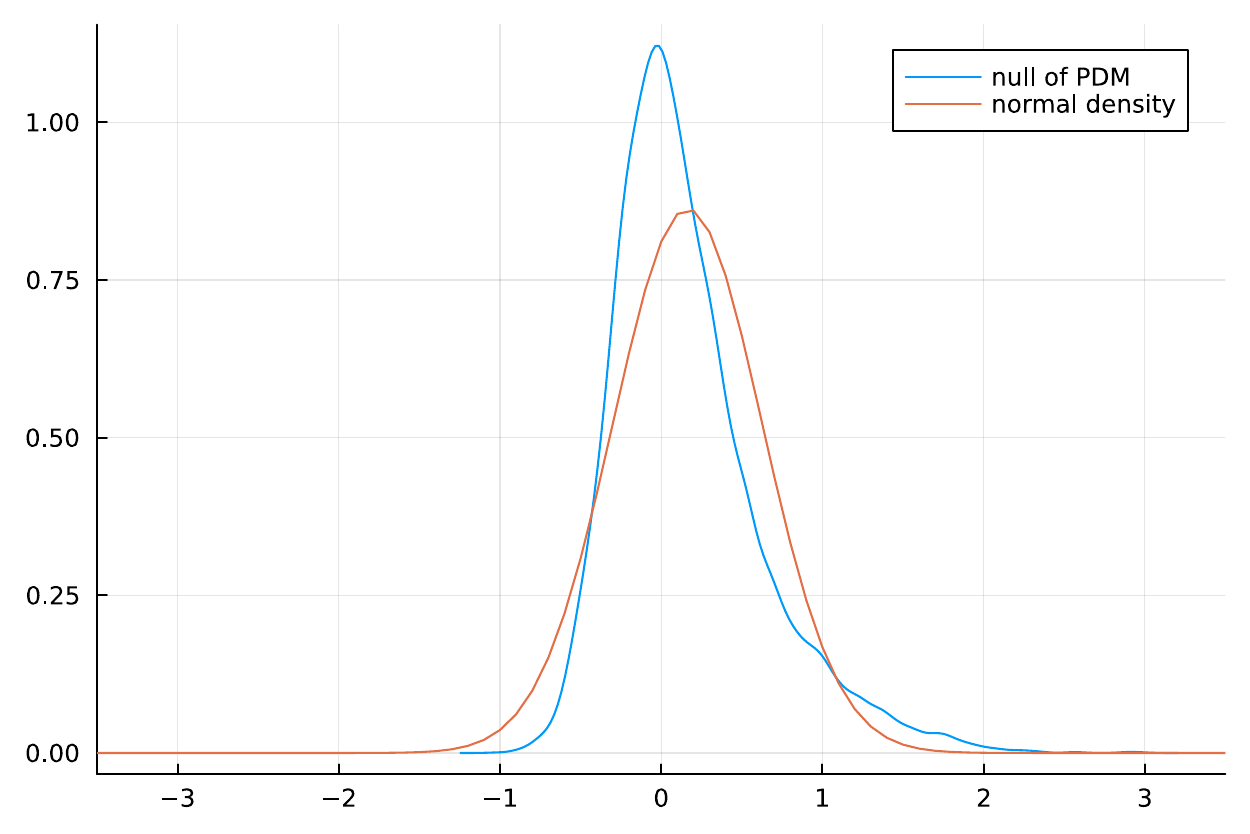}
    \end{subfigure}
        \\ 
    \medskip 
    
         \begin{subfigure}[]{0.5\textwidth}
             \caption{Two Batches ($B=2$): $\sigma_0=\frac{1}{16}$, $\sigma_1=\frac{1}{2}$}
       \includegraphics[width=.95\linewidth]{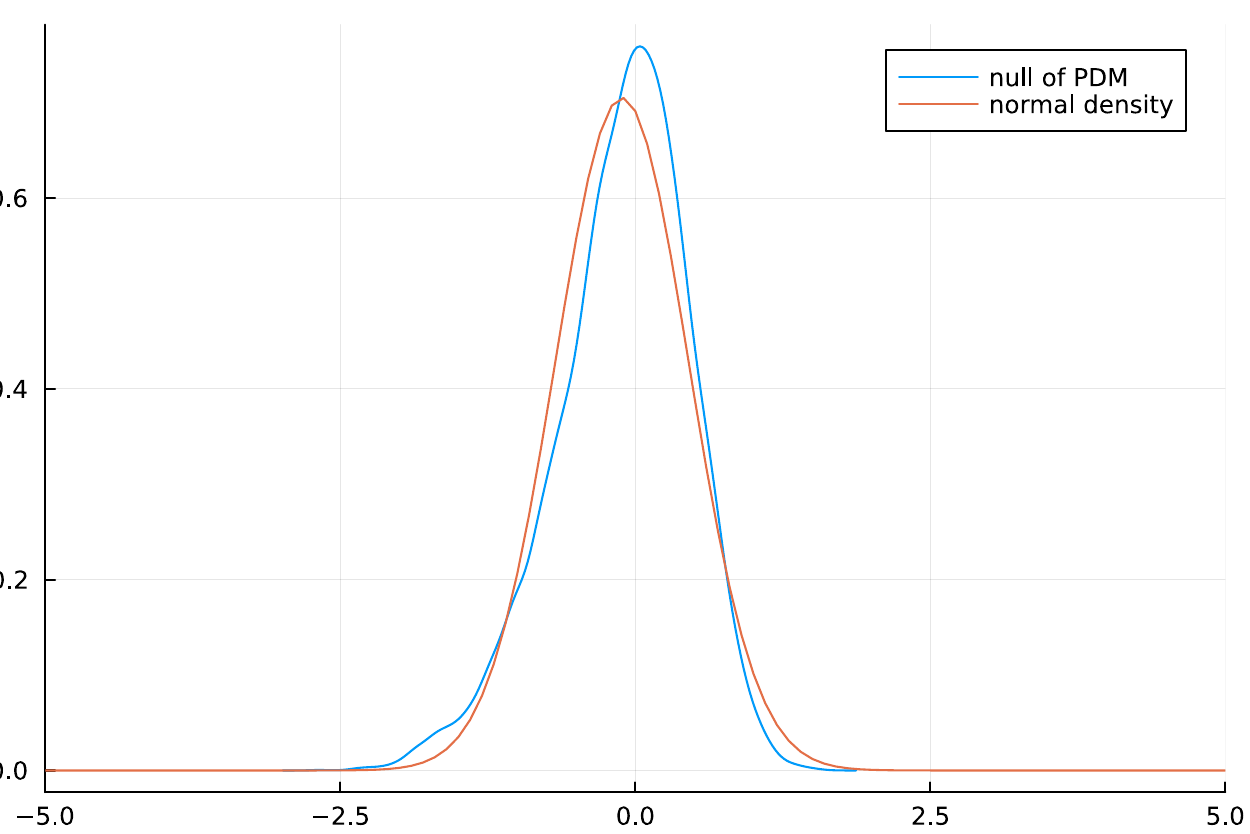}
    \end{subfigure}
         \begin{subfigure}[]{0.5\textwidth}
             \caption{Five Batches ($B=5$): $\sigma_0=\frac{1}{16}$, $\sigma_1=\frac{1}{2}$ }
       \includegraphics[width=.95\linewidth]{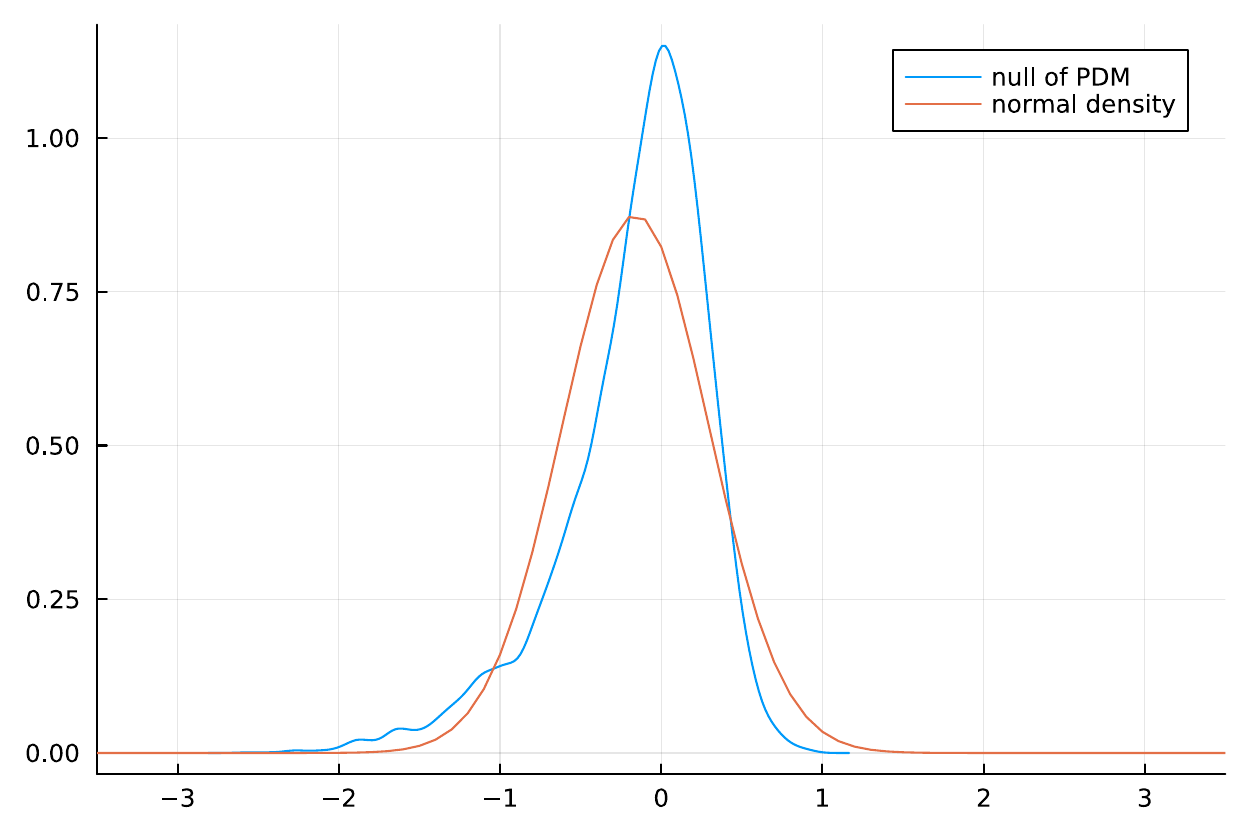}
    \end{subfigure}
\end{figure}

\begin{figure}[ht]
 \caption{Local Asymptotic Power Curves under Thompson Sampling: Equal Variances} 
 \label{fig:app1}
     \begin{subfigure}[]{0.5\textwidth}
             \caption{Three Batches ($B=3$): $\sigma_0=\sigma_1=\frac{1}{2}$}
       \includegraphics[width=.95\linewidth]{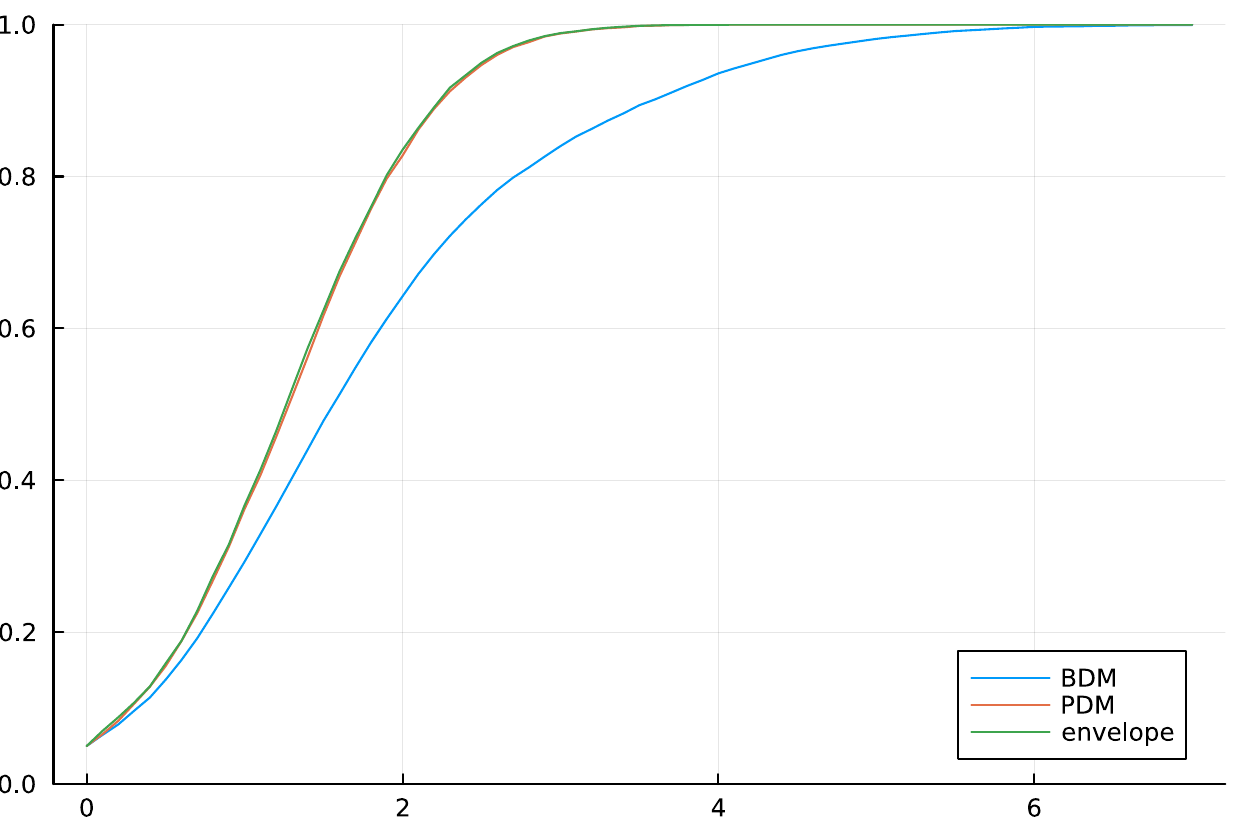}
    \end{subfigure}%
         \begin{subfigure}[]{0.5\textwidth}
             \caption{Four Batches ($B=4$): $\sigma_0=\sigma_1=\frac{1}{2}$}
       \includegraphics[width=.95\linewidth]{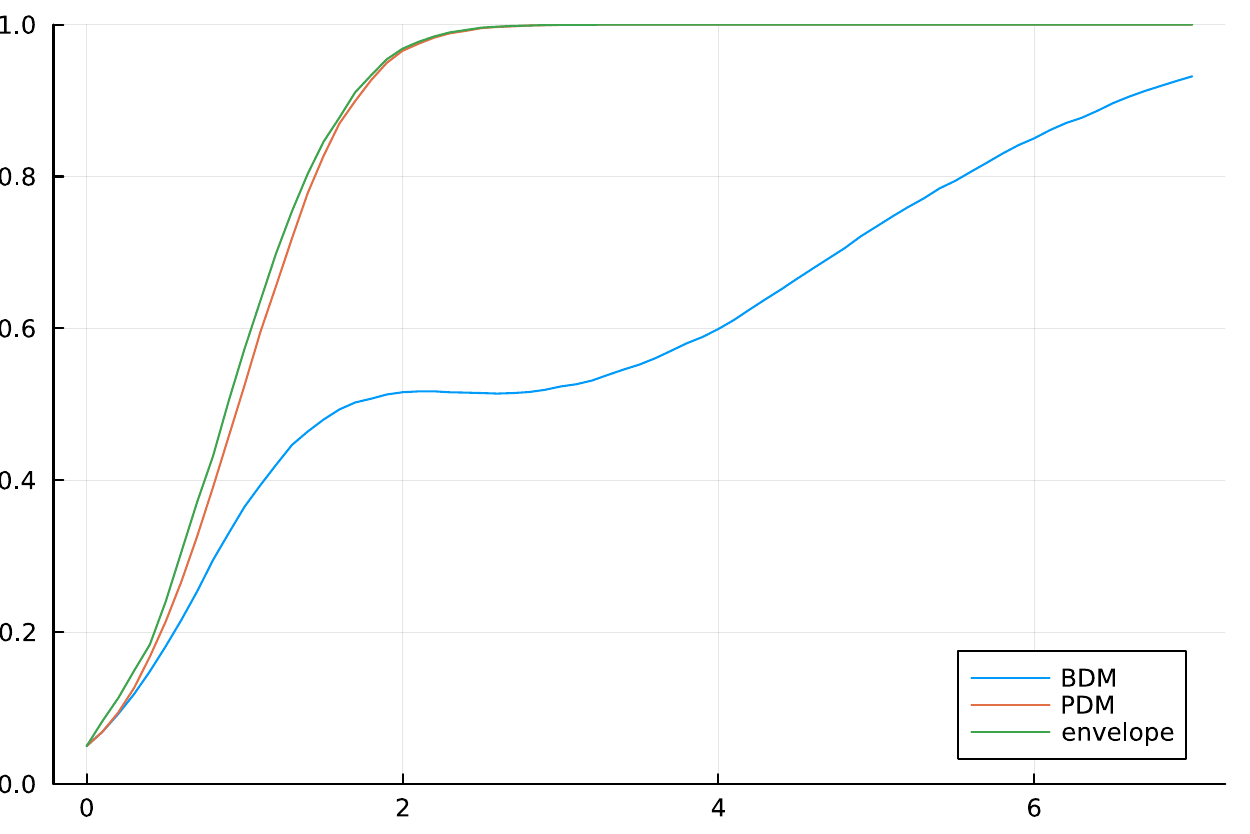}
    \end{subfigure}
\end{figure}

\begin{figure}[ht]
 \caption{Local Asymptotic Power Curves under Thompson Sampling: Unequal Variances} 
 \label{fig:app2}
     \begin{subfigure}[]{0.5\textwidth}
             \caption{Two Batches ($B=2$): $\sigma_0=\frac{1}{2}$, $\sigma_1=\frac{1}{16}$}
       \includegraphics[width=.95\linewidth]{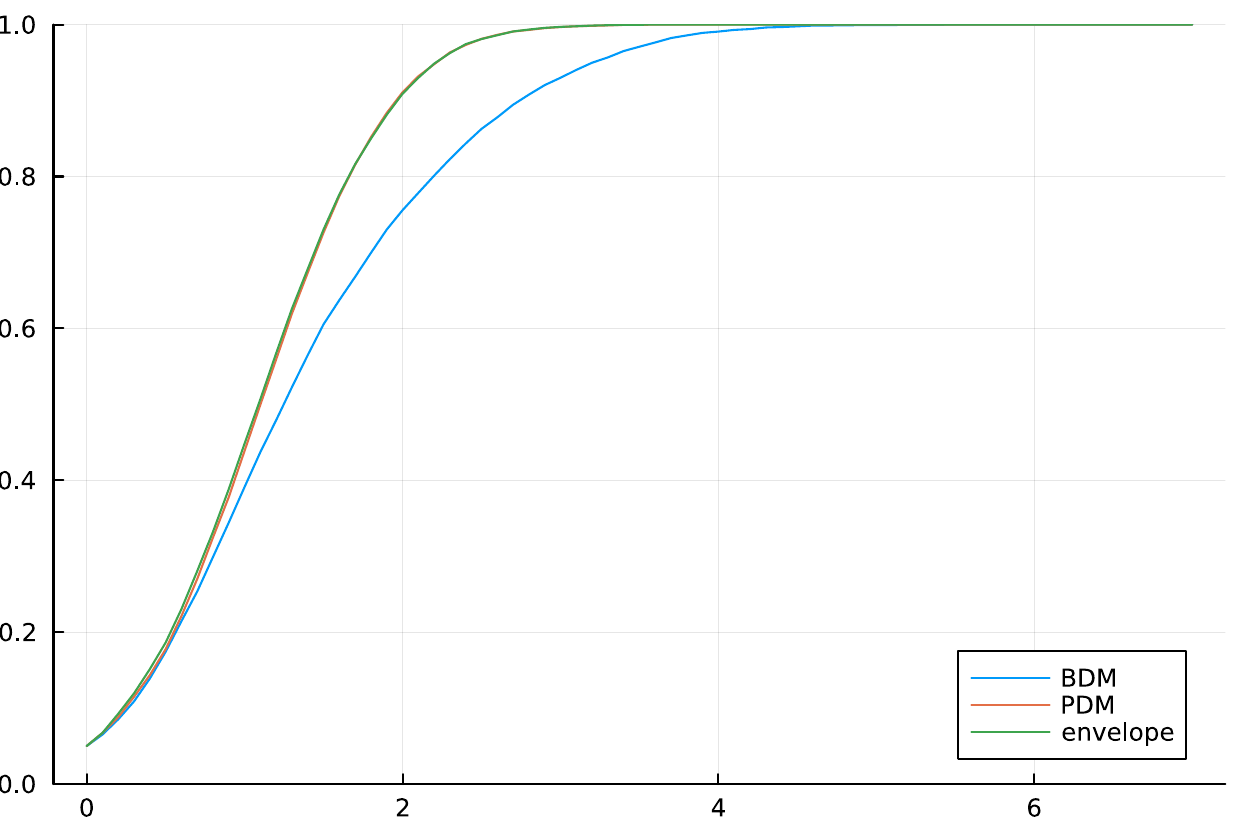}
    \end{subfigure}
         \begin{subfigure}[]{0.5\textwidth}
             \caption{Two Batches ($B=2$): $\sigma_0=\frac{1}{16}$, $\sigma_1=\frac{1}{2}$ }
       \includegraphics[width=.95\linewidth]{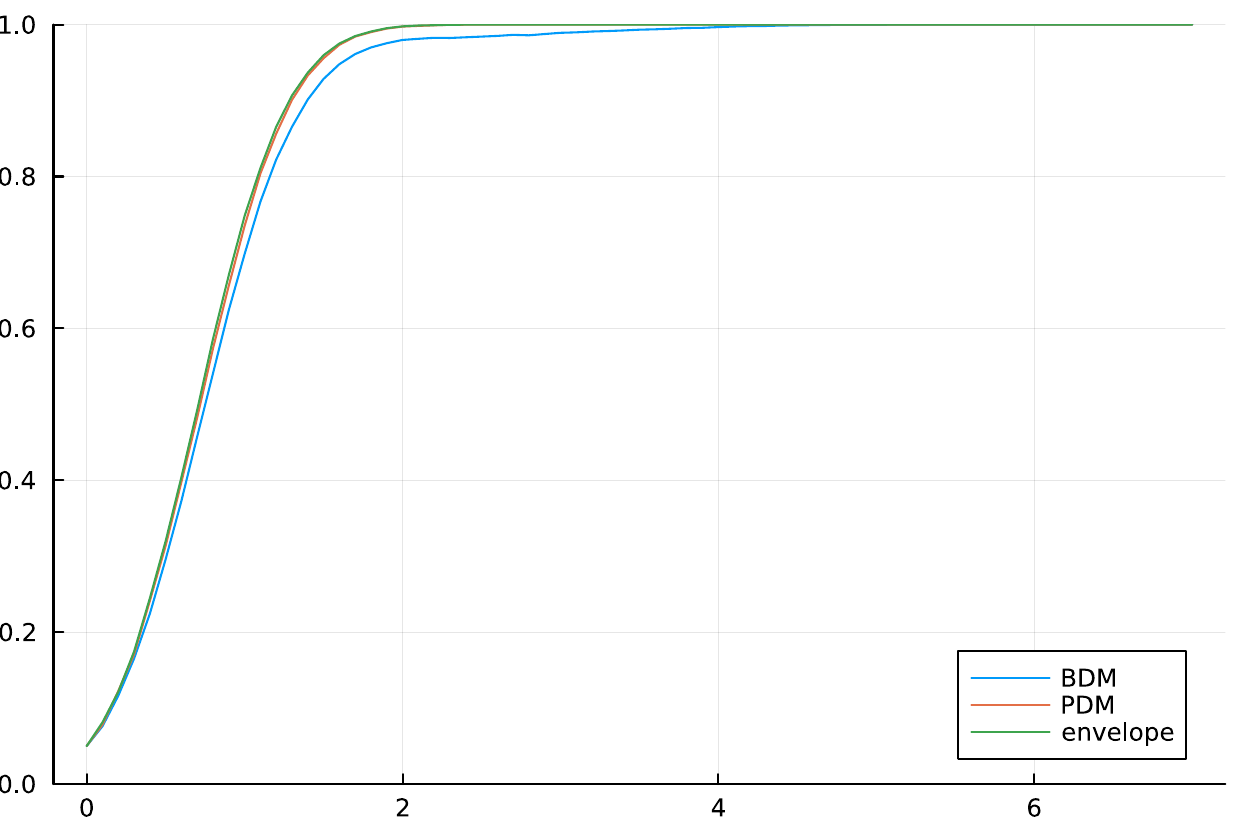}
    \end{subfigure} 
    \\ 
    \medskip 
    
         \begin{subfigure}[]{0.5\textwidth}
             \caption{Five Batches ($B=5$): $\sigma_0=\frac{1}{2}$, $\sigma_1=\frac{1}{16}$}
       \includegraphics[width=.95\linewidth]{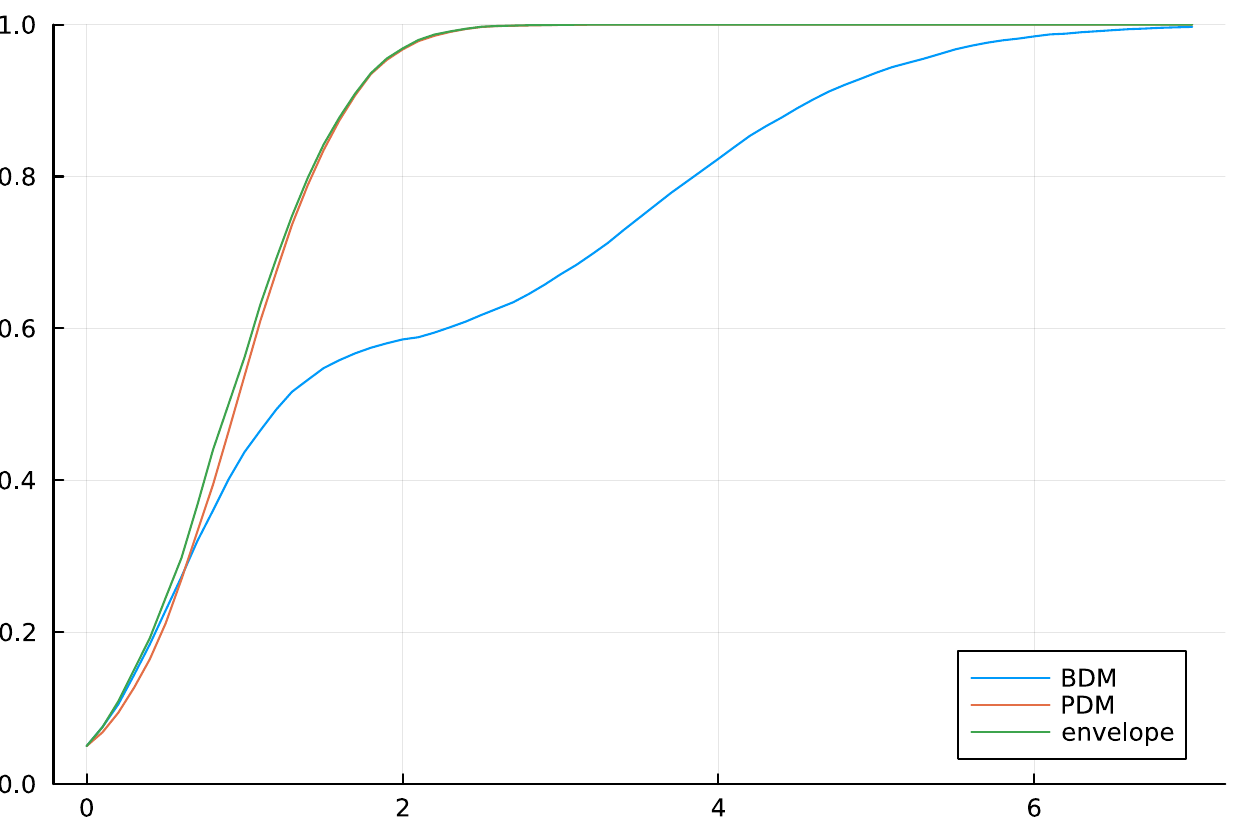}
    \end{subfigure}
         \begin{subfigure}[]{0.5\textwidth}
             \caption{Five Batches ($B=5$): $\sigma_0=\frac{1}{16}$, $\sigma_1=\frac{1}{2}$ }
       \includegraphics[width=.95\linewidth]{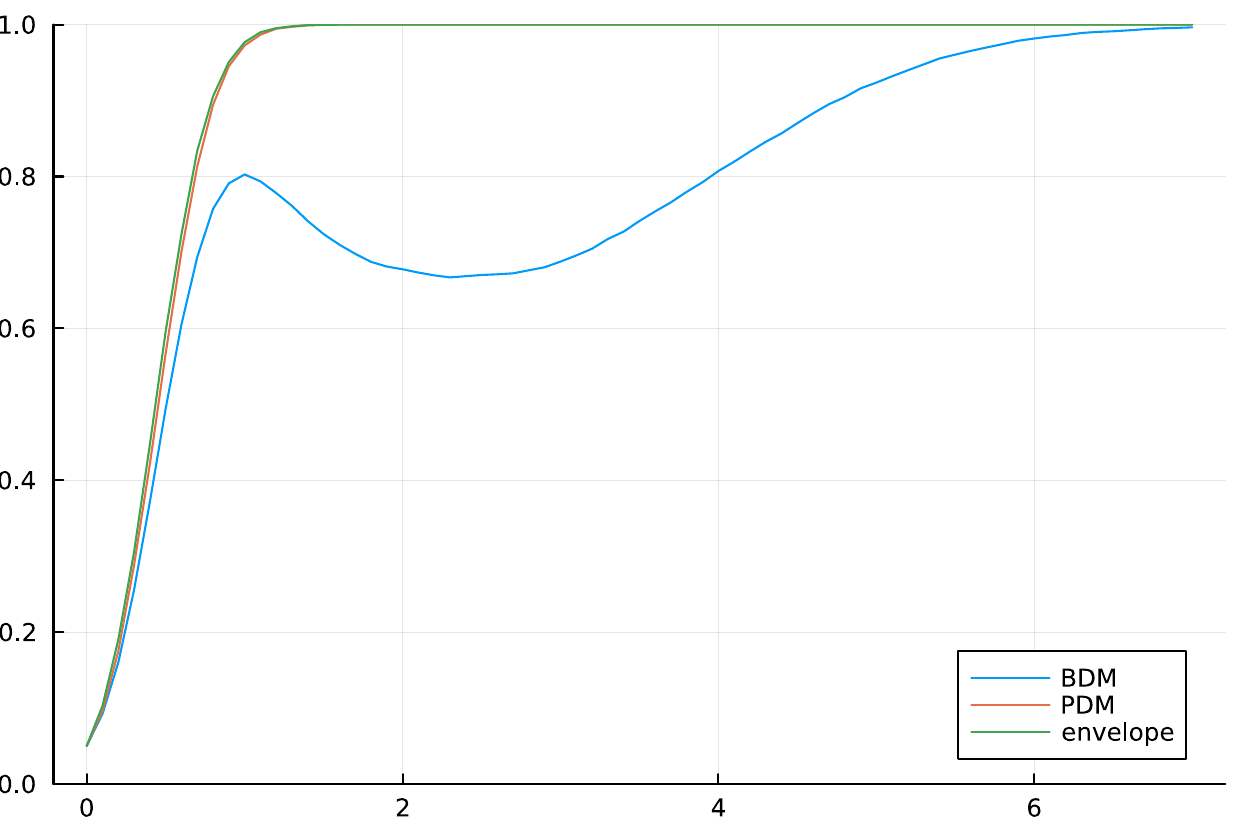}
    \end{subfigure}
\end{figure}

\begin{figure}[ht]
 \caption{Local Asymptotic Power Curves under Thompson Sampling: Unequal Variances} 
 \label{fig:app2a}
     \begin{subfigure}[]{0.5\textwidth}
             \caption{Two Batches ($B=2$): $\sigma_0=\frac{1}{2}$, $\sigma_1=\frac{1}{4}$}
       \includegraphics[width=.95\linewidth]{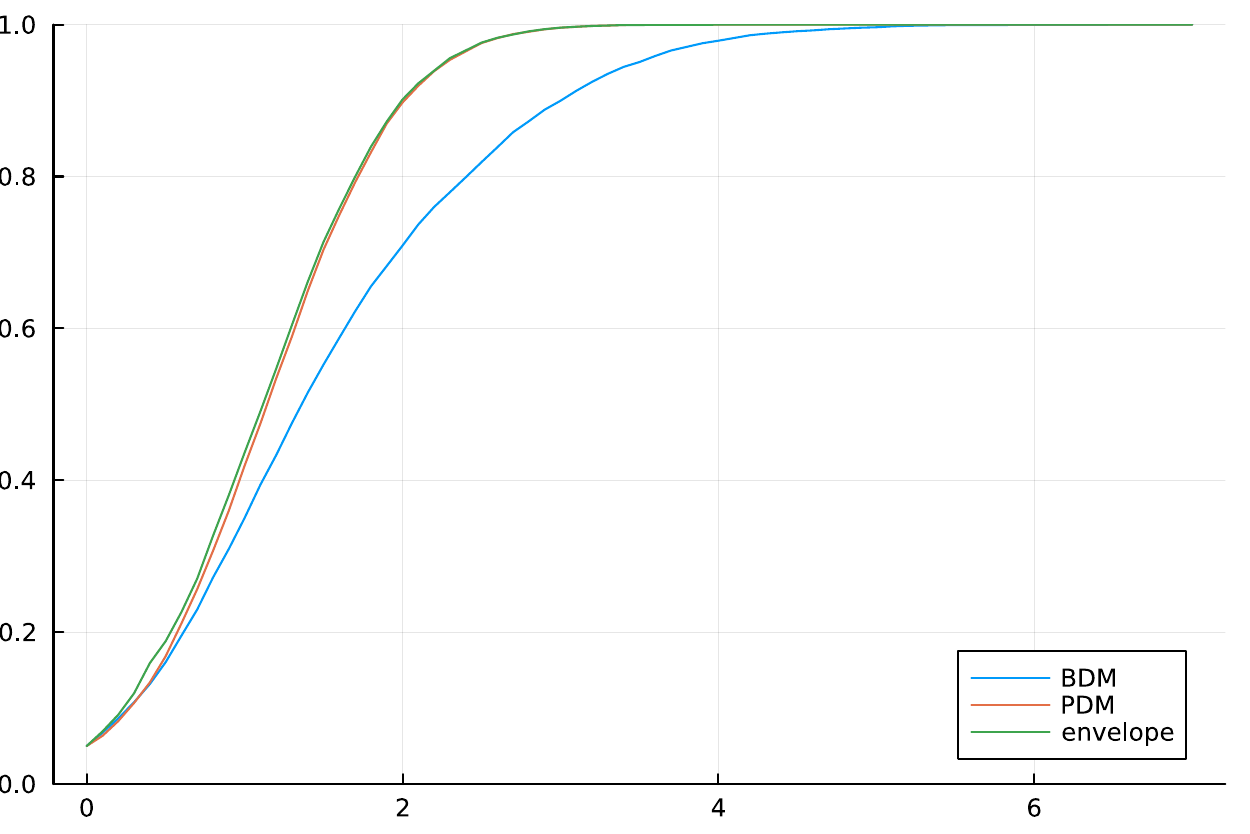}
    \end{subfigure}
         \begin{subfigure}[]{0.5\textwidth}
             \caption{Two Batches ($B=2$): $\sigma_0=\frac{1}{4}$, $\sigma_1=\frac{1}{2}$ }
       \includegraphics[width=.95\linewidth]{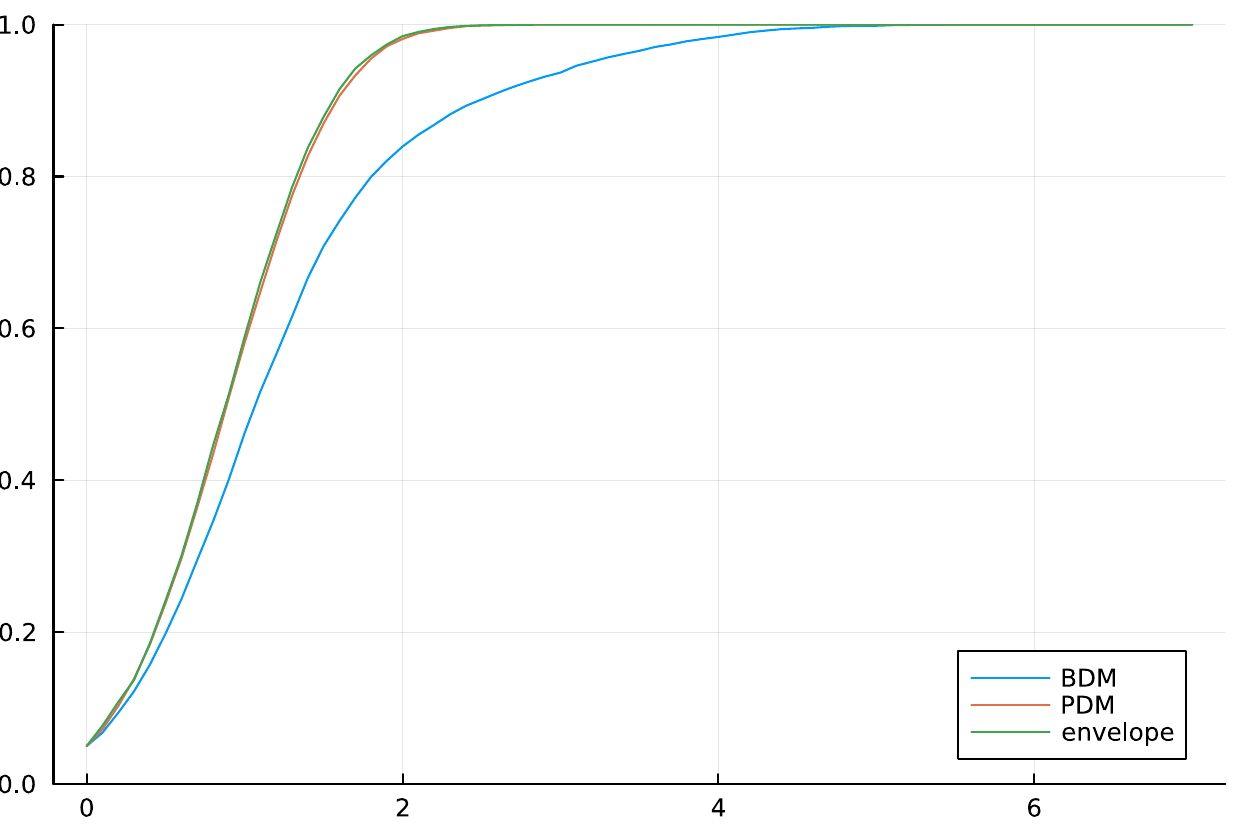}
    \end{subfigure} 
    \\ 
    \medskip 
    
         \begin{subfigure}[]{0.5\textwidth}
             \caption{Three Batches ($B=3$): $\sigma_0=\frac{1}{2}$, $\sigma_1=\frac{1}{4}$}
       \includegraphics[width=.95\linewidth]{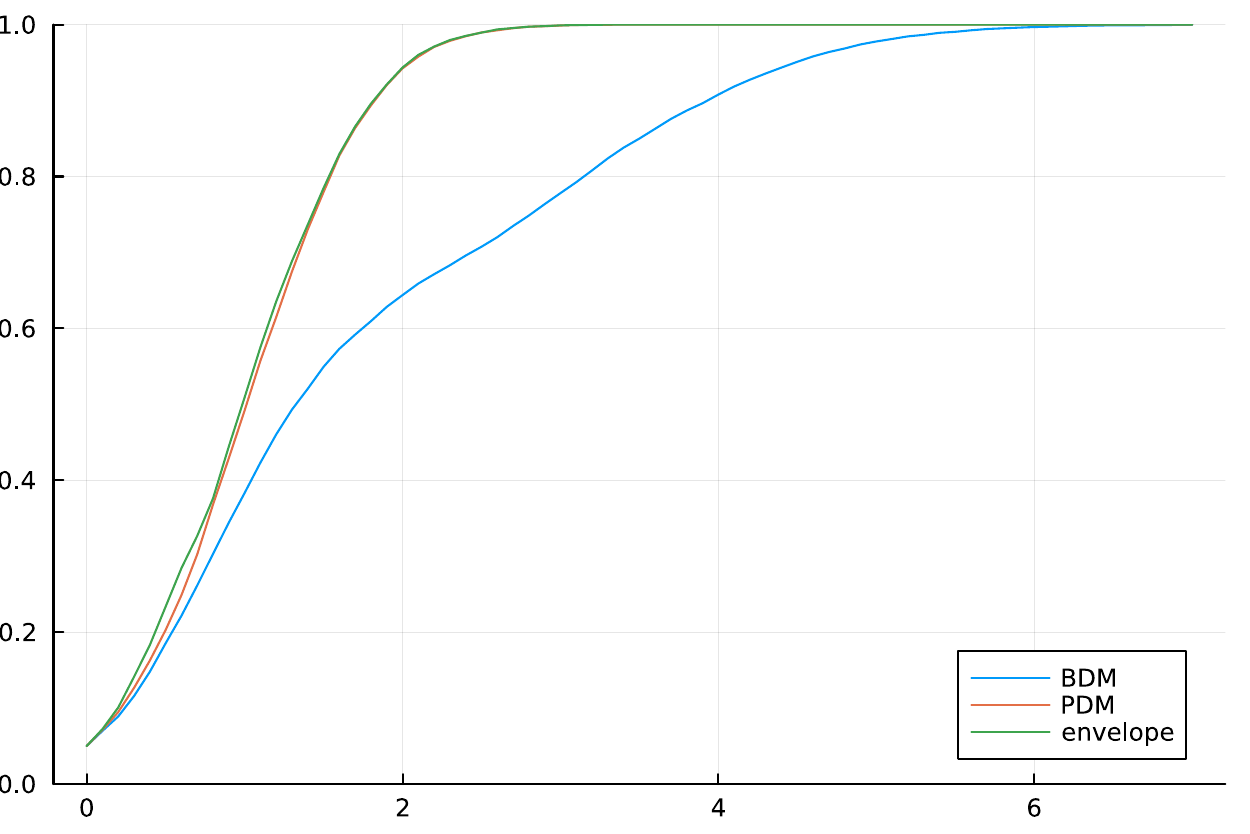}
    \end{subfigure} 
         \begin{subfigure}[]{0.5\textwidth}
             \caption{Three Batches ($B=53$): $\sigma_0=\frac{1}{4}$, $\sigma_1=\frac{1}{2}$ }
       \includegraphics[width=.95\linewidth]{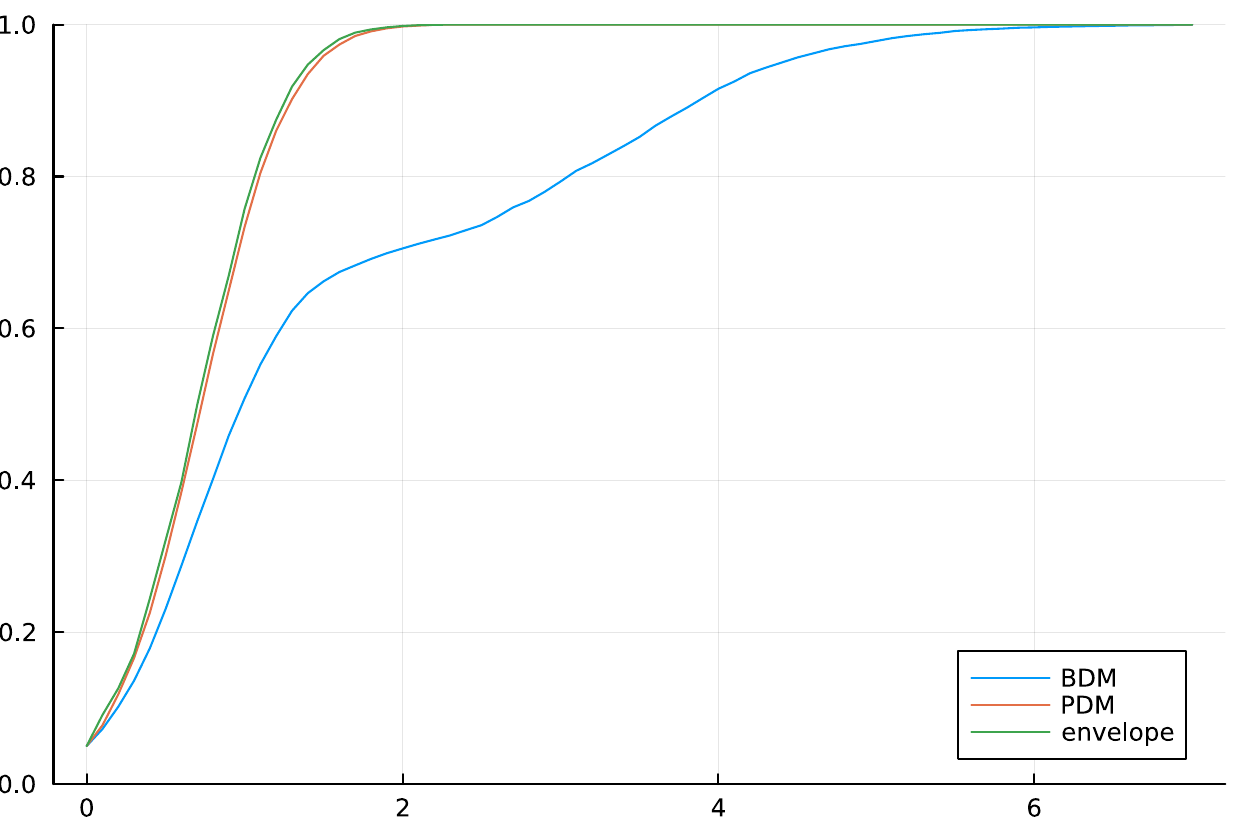}
    \end{subfigure}
        \\ 
    \medskip 
    
         \begin{subfigure}[]{0.5\textwidth}
             \caption{Four Batches ($B=4$): $\sigma_0=\frac{1}{2}$, $\sigma_1=\frac{1}{4}$}
       \includegraphics[width=.95\linewidth]{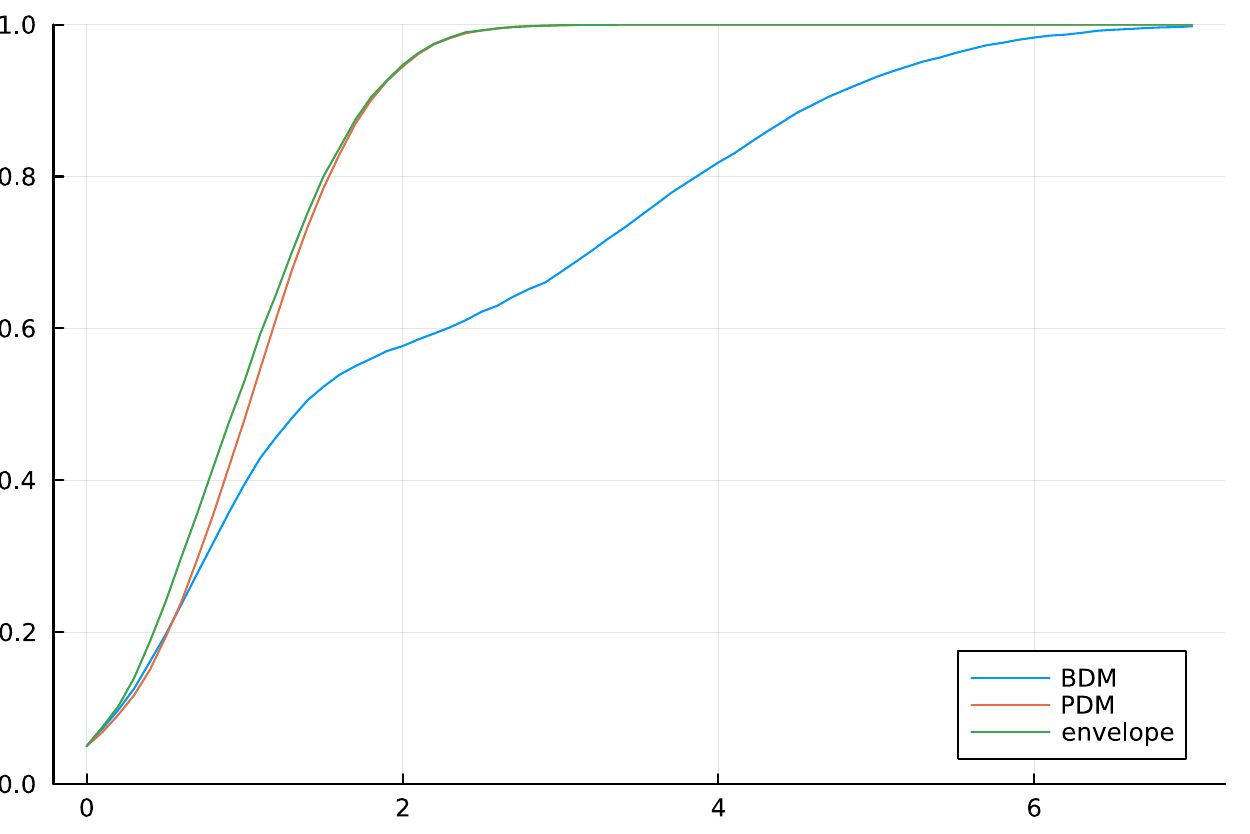}
    \end{subfigure}%
         \begin{subfigure}[]{0.5\textwidth}
             \caption{Four Batches ($B=4$): $\sigma_0=\frac{1}{4}$, $\sigma_1=\frac{1}{2}$ }
       \includegraphics[width=.95\linewidth]{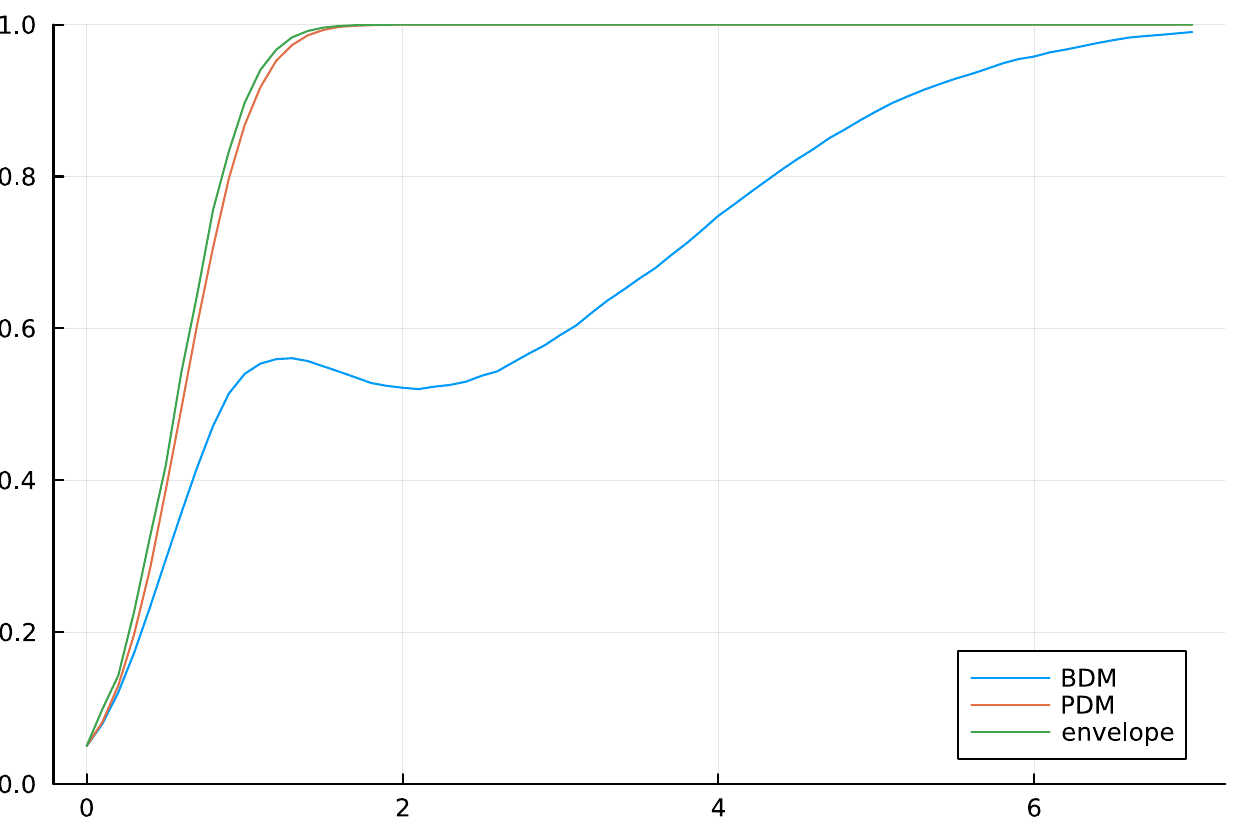}
    \end{subfigure}
        \\ 
    \medskip 
    
         \begin{subfigure}[]{0.5\textwidth}
             \caption{Five Batches ($B=5$): $\sigma_0=\frac{1}{2}$, $\sigma_1=\frac{1}{4}$}
       \includegraphics[width=.95\linewidth]{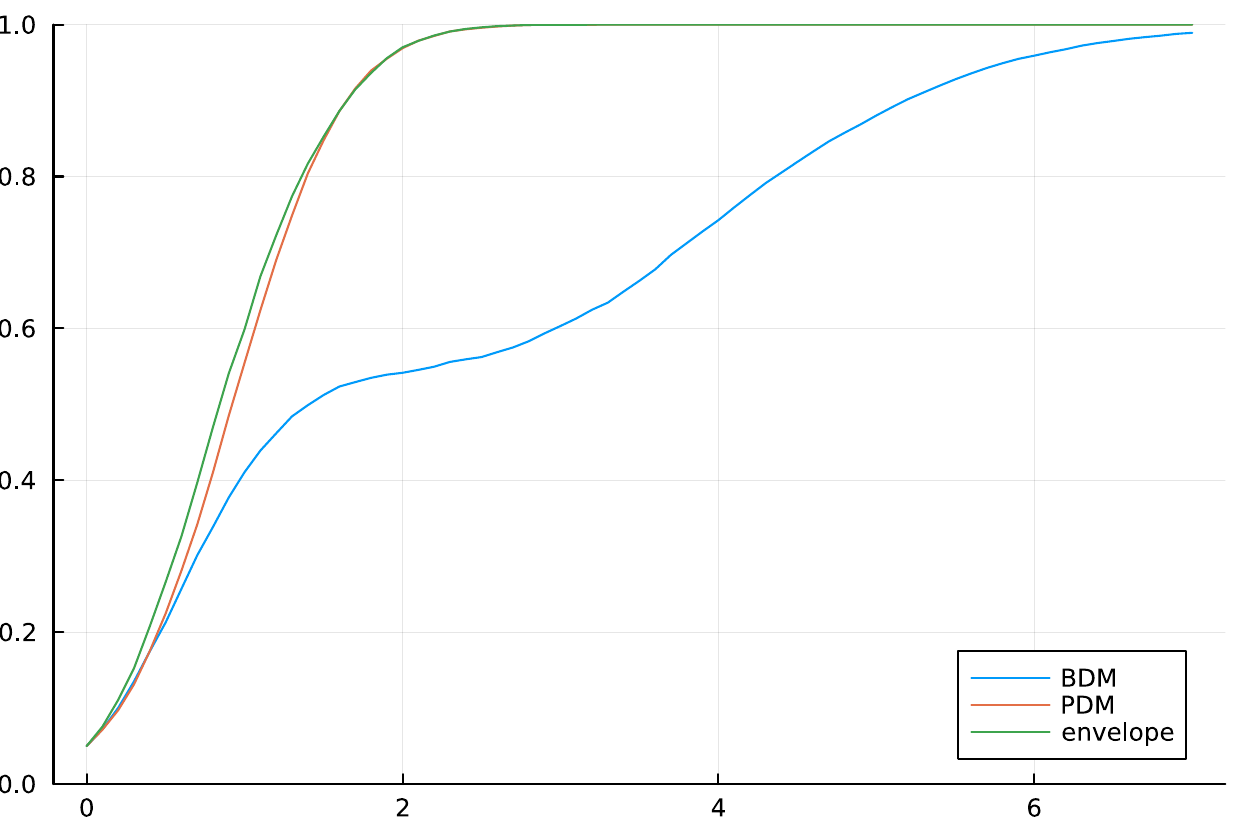}
    \end{subfigure}
         \begin{subfigure}[]{0.5\textwidth}
             \caption{Five Batches ($B=5$): $\sigma_0=\frac{1}{4}$, $\sigma_1=\frac{1}{2}$ }
       \includegraphics[width=.95\linewidth]{std1div4and1div2_5batch.pdf}
    \end{subfigure}
\end{figure}

\begin{figure}[ht]
 \caption{Local Asymptotic Power Curves under Thompson Sampling: Unequal Variances} 
 \label{fig:app2b}
     \begin{subfigure}[]{0.5\textwidth}
             \caption{Two Batches ($B=2$): $\sigma_0=\frac{1}{2}$, $\sigma_1=\frac{1}{8}$}
       \includegraphics[width=.95\linewidth]{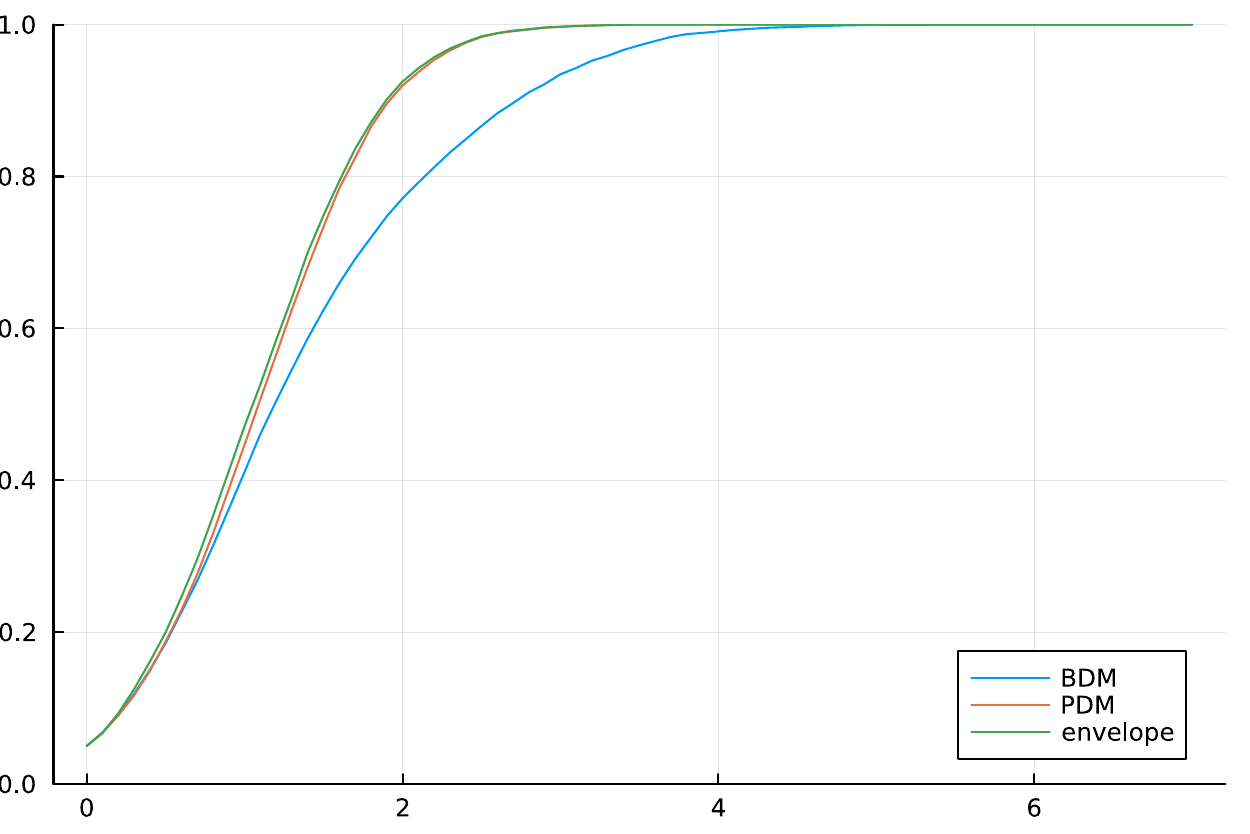}
    \end{subfigure}
         \begin{subfigure}[]{0.5\textwidth}
             \caption{Two Batches ($B=2$): $\sigma_0=\frac{1}{8}$, $\sigma_1=\frac{1}{2}$ }
       \includegraphics[width=.95\linewidth]{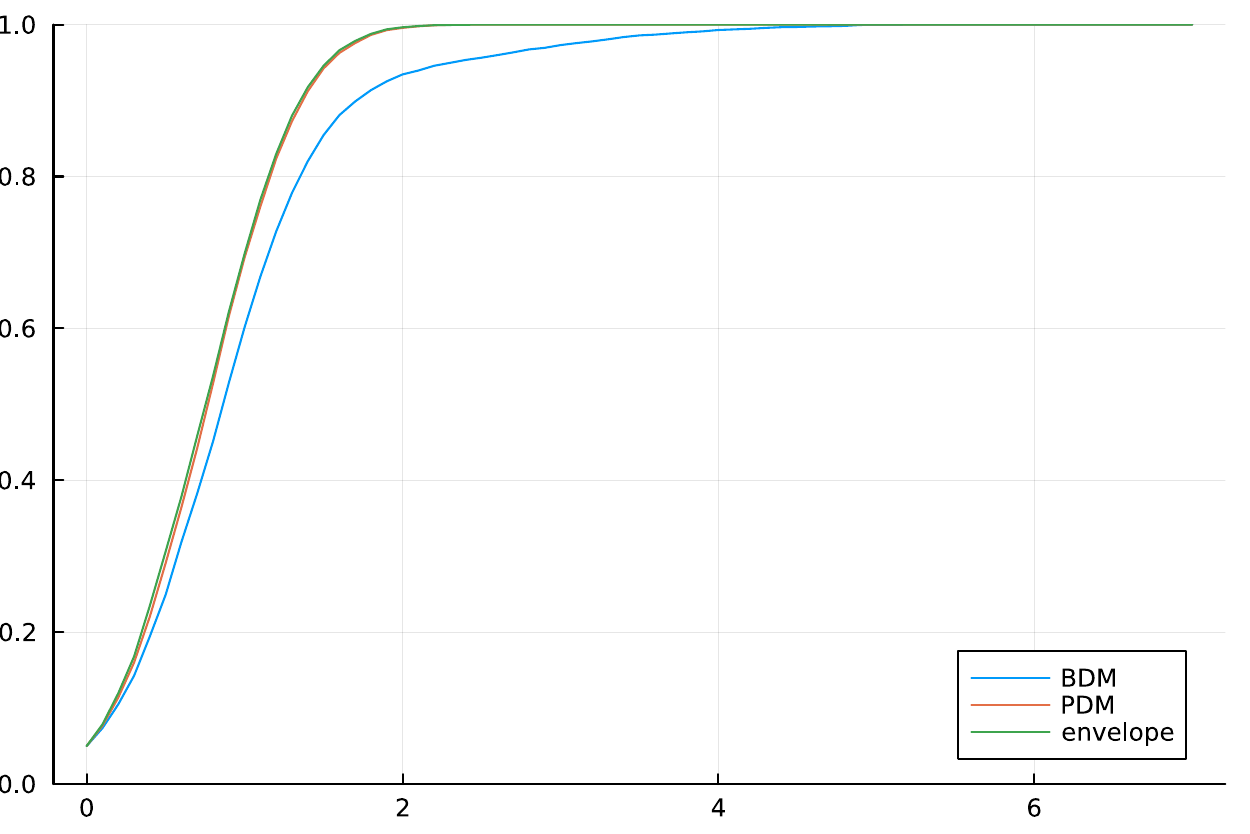}
    \end{subfigure} 
    \\ 
    \medskip 
    
         \begin{subfigure}[]{0.5\textwidth}
             \caption{Three Batches ($B=3$): $\sigma_0=\frac{1}{2}$, $\sigma_1=\frac{1}{8}$}
       \includegraphics[width=.95\linewidth]{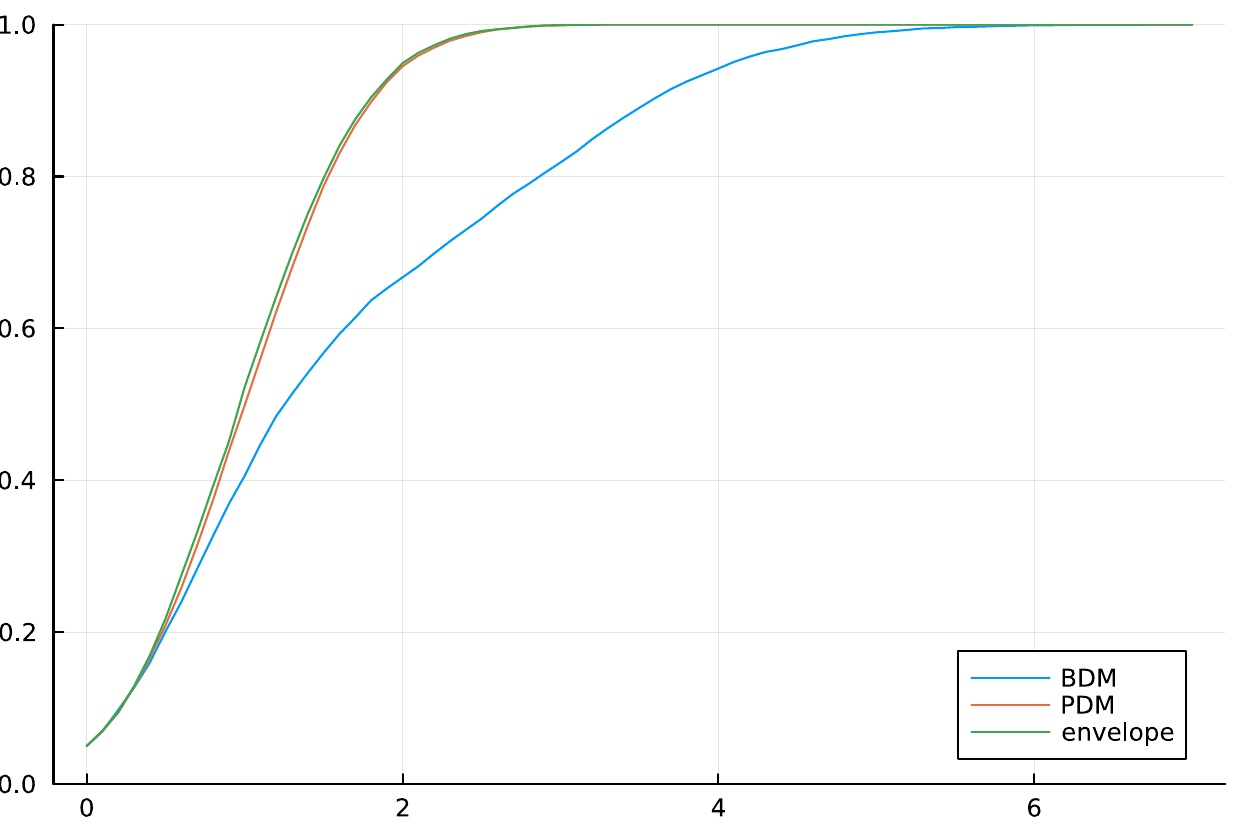}
    \end{subfigure} 
         \begin{subfigure}[]{0.5\textwidth}
             \caption{Three Batches ($B=53$): $\sigma_0=\frac{1}{8}$, $\sigma_1=\frac{1}{2}$ }
       \includegraphics[width=.95\linewidth]{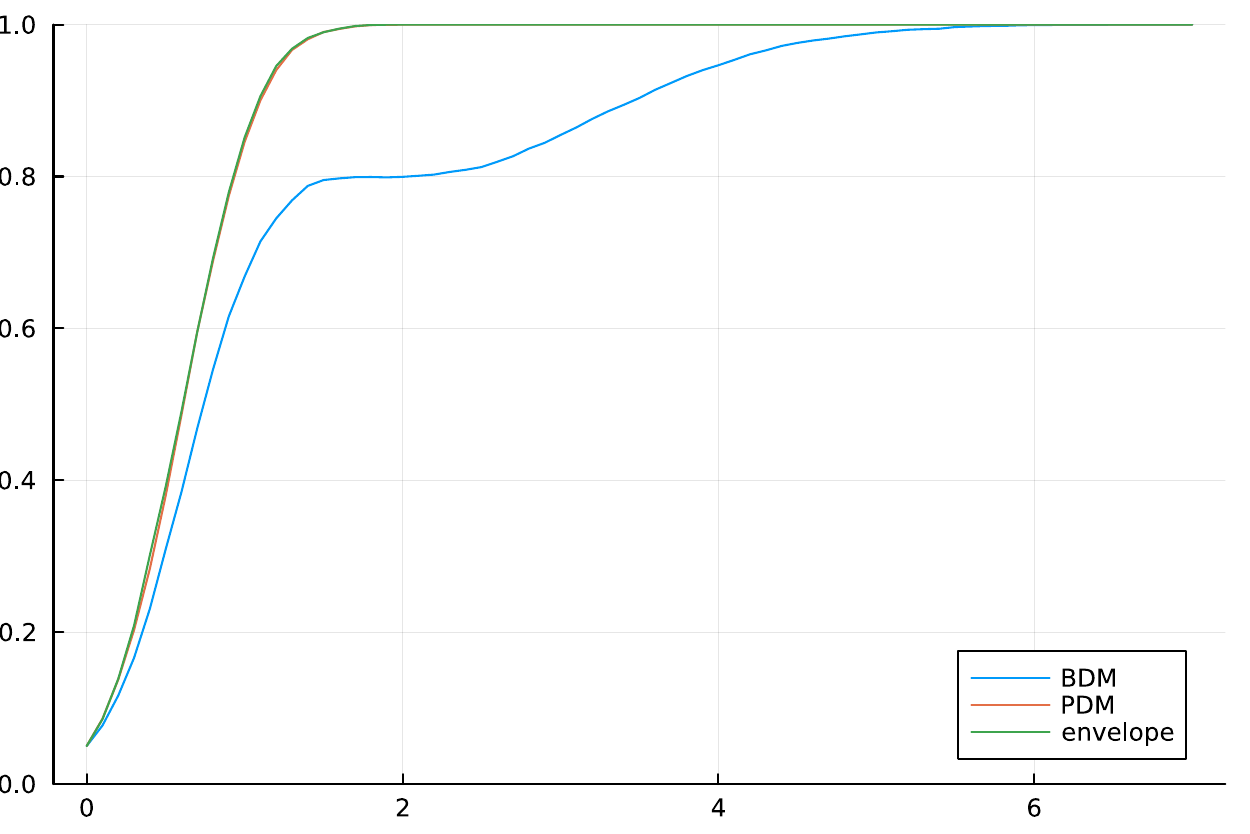}
    \end{subfigure} 
        \\ 
    \medskip 
    
         \begin{subfigure}[]{0.5\textwidth}
             \caption{Four Batches ($B=4$): $\sigma_0=\frac{1}{2}$, $\sigma_1=\frac{1}{8}$}
       \includegraphics[width=.95\linewidth]{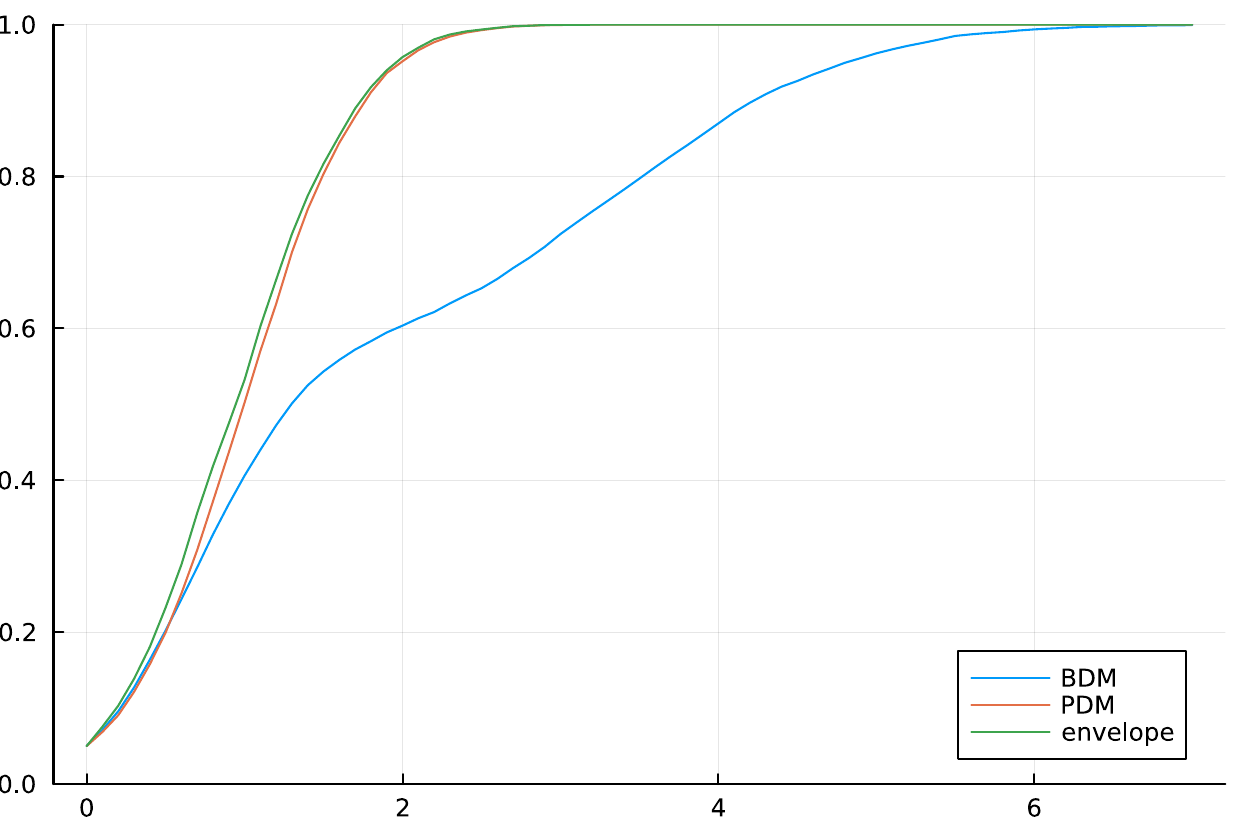}
    \end{subfigure}
         \begin{subfigure}[]{0.5\textwidth}
             \caption{Four Batches ($B=4$): $\sigma_0=\frac{1}{8}$, $\sigma_1=\frac{1}{2}$ }
       \includegraphics[width=.95\linewidth]{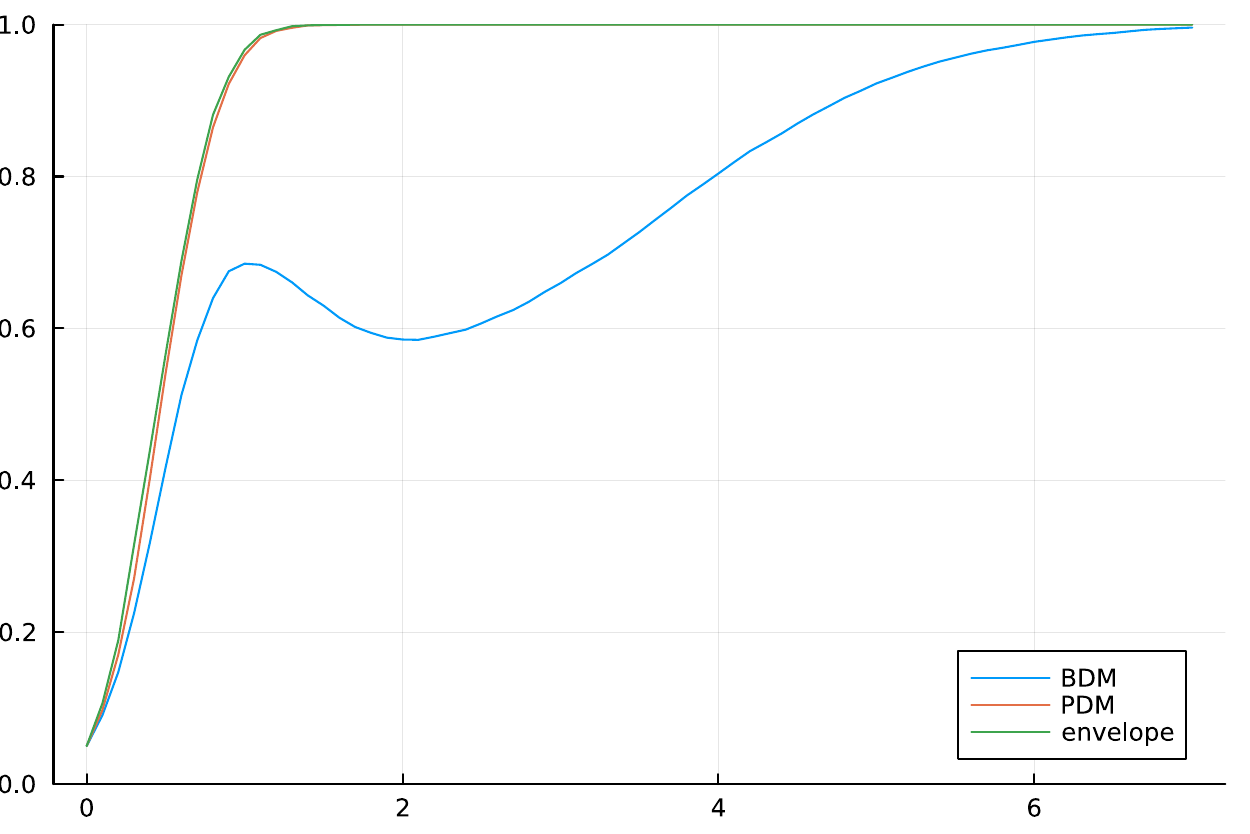}
    \end{subfigure}
        \\ 
    \medskip 
    
         \begin{subfigure}[]{0.5\textwidth}
             \caption{Five Batches ($B=5$): $\sigma_0=\frac{1}{2}$, $\sigma_1=\frac{1}{8}$}
       \includegraphics[width=.95\linewidth]{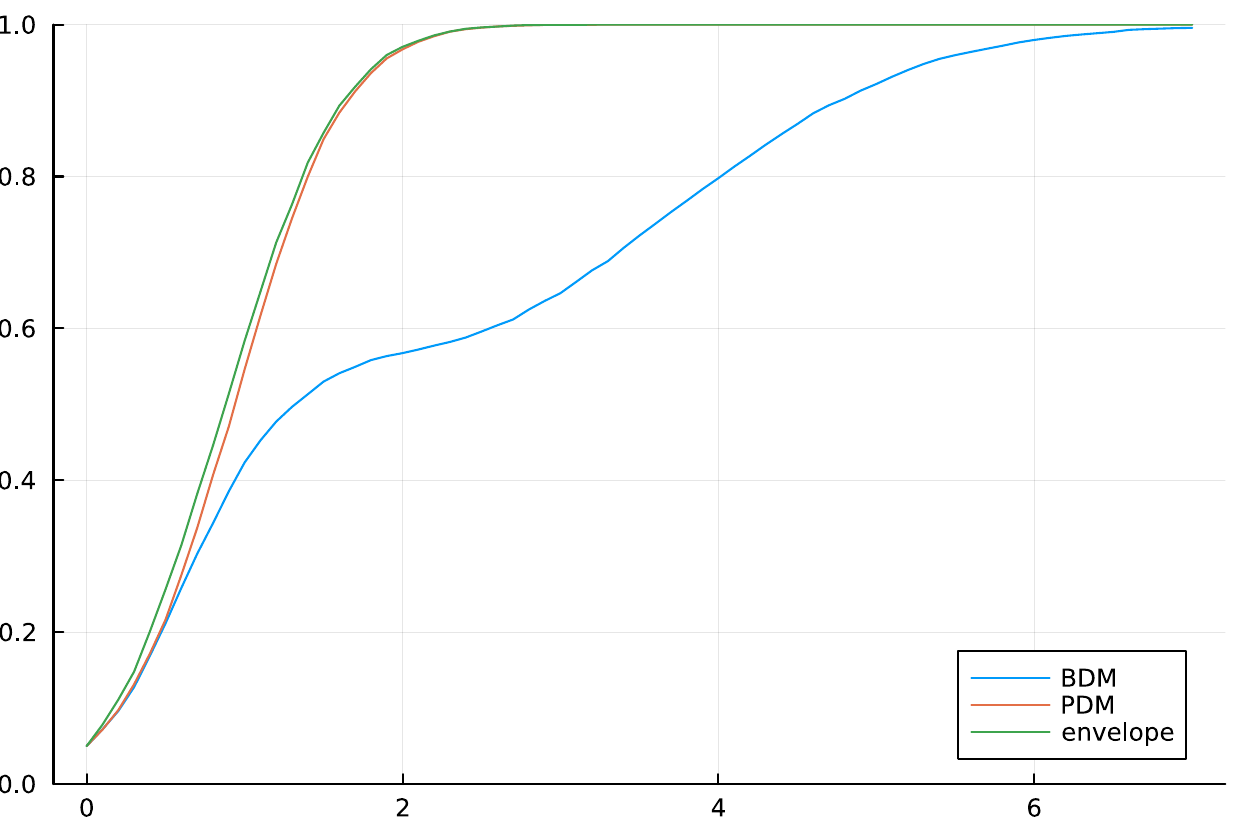}
    \end{subfigure} 
         \begin{subfigure}[]{0.5\textwidth}
             \caption{Five Batches ($B=5$): $\sigma_0=\frac{1}{8}$, $\sigma_1=\frac{1}{2}$ }
       \includegraphics[width=.95\linewidth]{std1div8and1div2_5batch.pdf}
    \end{subfigure}
\end{figure}

\end{document}